\documentclass[superscriptaddress,twocolumn,floatfix,prx]{revtex4-2}

\usepackage{amsmath}
\usepackage{amssymb}
\usepackage{mathtools}
\usepackage{graphicx}
\usepackage[dvipsnames]{xcolor}
\usepackage[colorlinks=true,linkcolor=RoyalBlue,citecolor=ForestGreen,urlcolor=RoyalPurple]{hyperref}
\usepackage[capitalise]{cleveref}
\usepackage{dsfont}
\usepackage{placeins}
\usepackage{cancel}
\usepackage[normalem]{ulem}
\usepackage{dcolumn}
\usepackage{bm}

\newcommand{\bvec}[1]{\boldsymbol{#1}}
\newcommand{\unit}[1]{\,\mathrm{#1}}
\renewcommand{\Im}{\mathrm{Im}}

\newcommand{\approptoinn}[2]{\mathrel{\vcenter{
  \offinterlineskip\halign{\hfil$##$\cr
    #1\propto\cr\noalign{\kern2pt}#1\sim\cr\noalign{\kern-2pt}}}}}
\newcommand{\appropto}{\mathpalette\approptoinn\relax}
\newcommand{\unsim}{{\sim}} 
\newcommand{\Edens}{\mathcal{E}}
\newcommand{\Estacking}{\mathcal{E}_\mathrm{stacking}}
\newcommand{\Eelastic}{\mathcal{E}_\mathrm{elastic}}
\newcommand{\captiontitle}[1]{{\bf #1}}
\newcommand{\captionlabel}[1]{{\bf(#1)}}

\usepackage{makerobust}

\begin{document}

\title{Multi-layered atomic relaxation in van der Waals heterostructures}

\author{Dorri Halbertal}
\altaffiliation{These authors contributed equally}
\affiliation{Department of Physics, Columbia University, New York, New York 10027, United States}
\author{Lennart Klebl}
\altaffiliation{These authors contributed equally}
\affiliation{Institut für Theorie der Statistischen Physik, RWTH Aachen University and JARA-Fundamentals of Future Information Technology, 52056 Aachen, Germany}
\author{Valerie Hsieh}
\affiliation{Department of Physics, Columbia University, New York, New York 10027, United States}
\author{Jacob Cook}
\affiliation{Department of Physics and Astronomy, University of Missouri, Columbia, MO 65211}
\author{Stephen Carr}
\affiliation{Physics Department, Brown University, Providence, Rhode Island 02912, United States}
\author{Guang Bian}
\affiliation{Department of Physics and Astronomy, University of Missouri, Columbia, MO 65211}
\author{Cory R.~Dean}
\affiliation{Department of Physics, Columbia University, New York, New York 10027, United States}
\author{Dante M.~Kennes}
\affiliation{Institut für Theorie der Statistischen Physik, RWTH Aachen University and JARA-Fundamentals of Future Information Technology, 52056 Aachen, Germany}
\affiliation{Max Planck Institute for the Structure and Dynamics of Matter, Center for Free Electron Laser Science, 22761 Hamburg, Germany}
\author{Dmitri N.~Basov}
\affiliation{Department of Physics, Columbia University, New York, New York 10027, United States}


\date{\today}

\begin{abstract}
When two-dimensional van der Waals materials are stacked to build heterostructures, moir\'e patterns emerge from twisted interfaces or from mismatch in lattice constant of individual layers. Relaxation of the atomic positions is a direct, generic consequence of the moir\'e pattern, with many implications for the physical properties. Moir\'e driven atomic relaxation may be naively thought to be restricted to the interfacial layers and thus irrelevant for multi-layered heterostructures. However, we provide experimental evidence for the importance of the three dimensional nature of the relaxation in two types of van der Waals heterostructures: First, in multi-layer graphene twisted on graphite at small twist angles ($\theta\approx0.14^\circ$) we observe propagation of relaxation domains even beyond 18 graphene layers. Second, we show how for multi-layer PdTe\textsubscript{2} on Bi\textsubscript{2}Se\textsubscript{3} the moir\'e lattice constant depends on the number of PdTe\textsubscript{2} layers. Motivated by the experimental findings, we developed a continuum approach to model multi-layered relaxation processes based on the generalized stacking fault energy functional given by \emph{ab-initio} simulations. Leveraging the continuum property of the approach enables us to access large scale regimes and achieve agreement with our experimental data for both systems. Furthermore it is well known that the electronic structure of graphene sensitively depends on local lattice deformations. Therefore we study the impact of multi-layered relaxation on the local density of states of the twisted graphitic system. We identify measurable implications for the system, experimentally accessible by scanning tunneling microscopy. Our multi-layered relaxation approach is not restricted to the discussed systems, and can be used to uncover the impact of an interfacial defect on various layered systems of interest.
\end{abstract}

\maketitle

\begin{figure}%
    \centering
    \includegraphics[width=\columnwidth]{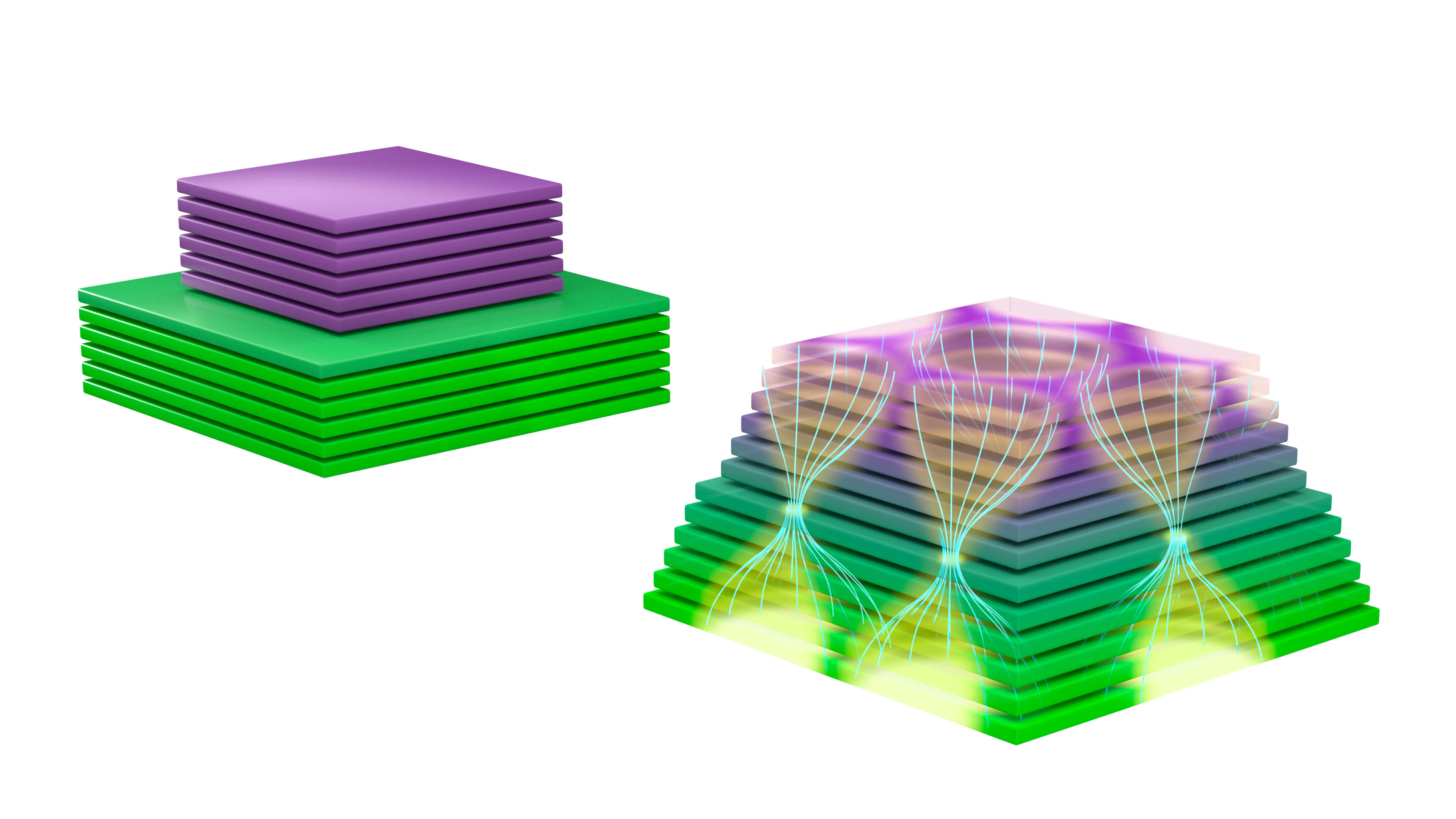}
    \caption{Two-dimensional heterostructure composed of two mismatched (green and purple) materials. Left panel: Lego-brick-like cartoon picture in which no atomic relaxation occurs~\cite{geim2013van}. The two materials have different lattice constants visualized by the width of the stacks. Right panel: Cartoon picture including the main finding of this work: layers relax continuously away from the interface. The extent of relaxation is qualitatively encoded in the yellow color. The emergent moir\'e pattern slowly decays towards the outer layers of the stack.}
    \label{fig:sketch}
\end{figure}%

\section{Introduction}
Moir\'e van der Waals (vdW) heterostructures have drawn considerable interest in the past few years. Whenever two layers with relative twist or lattice mismatch are stacked, large-scale periodic perturbations of stacking configurations emerge. So called moir\'e superlattices hence universally appear in any kind of vdW heterostructure, including manually assembled as well as chemical vapour deposition grown~\cite{wang2013platform, wang2014two, niu2015two, novoselov20162d} stacks. In twisted moir\'e heterostructures, various (unconventional) superconducting and correlated \cite{cao2018unconventional, lu2019superconductors, yankowitz2019tuning, cao2020nematicity, liu2021tuning, stepanov2020untying, arora2020superconductivity, oh2021evidence, park2021tunable, cao2021large, hao2021electric, kim2021spectroscopic, cao2018mott, cao2020strange, polshyn2019large, zondiner2020cascade, wong2020cascade, xie2019spectrosopic, kerelsky2019maximized, jiang2019charge, choi2019correlations, chen2021electrically, shi2020tunable, liu2020tunable, shen2020correlated, cao2020tunable, wang2020correlated, tang2020simulation, regan2020mott}, excitonic \cite{nayak2017probing, rivera2018interlayer, alexeev2019resonantly, andersen2021excitons, jin2019observation}, and topological \cite{burg2019correlated, rubio2021moire, chen2019signatures, chen2019evidence, chen2020tunable, nuckolls2020strongly, xie2021fractional, pierce2021unconventional, stepanov2021competing, choi2020tracing, sharpe2019emergent, wu2021chern, das2021symmetry, park2021flavour, saito2020independent} phases have been discovered -- demonstrating that engineering heterostructures of atomically thin layers of vdW materials opens up a vast space to design properties of quantum materials~\cite{kennes2021moire}. However, the atomic layers in vdW materials are not rigid in general, but instead behave as deformable membranes. When two layers with an interfacial defect (i.e. lattice mismatch or twist angle) come in contact, the atomic positions relax to minimize the total energy~\cite{carr2020electronic}.

Naturally, the relaxation itself is strongest at the defect interface, but we demonstrate here that it may also extent beyond the two atomic layers at the interface. Therefore, the ``Lego bricks" analogy put forward by Geim and Grigorieva~\cite{geim2013van}, where heterostructures are composed of mechanically independent layers, though extremely helpful for illustration purposes, needs to be refined considerably to incorporate the flexible nature of each layer as well as the interplay of mechanical relaxation between different layers. We contrast the two paradigms in \cref{fig:sketch}, where the left panel illustrates how to understand a stack of two multi-layered, mismatched materials in the picture disregarding relaxation. As no relaxation occurs there is a sharp transition from the bottom stack (green) to the top stack (purple) visualized by the difference in width. In contrast, a multi-layered relaxation leads to a smooth deformation of layers from the bottom to the top (right panel), with signatures of the moir\'e superlattice slowly decaying away from the interface (yellow vortices).

\begin{figure*}
    \centering
    \includegraphics[width=\textwidth]{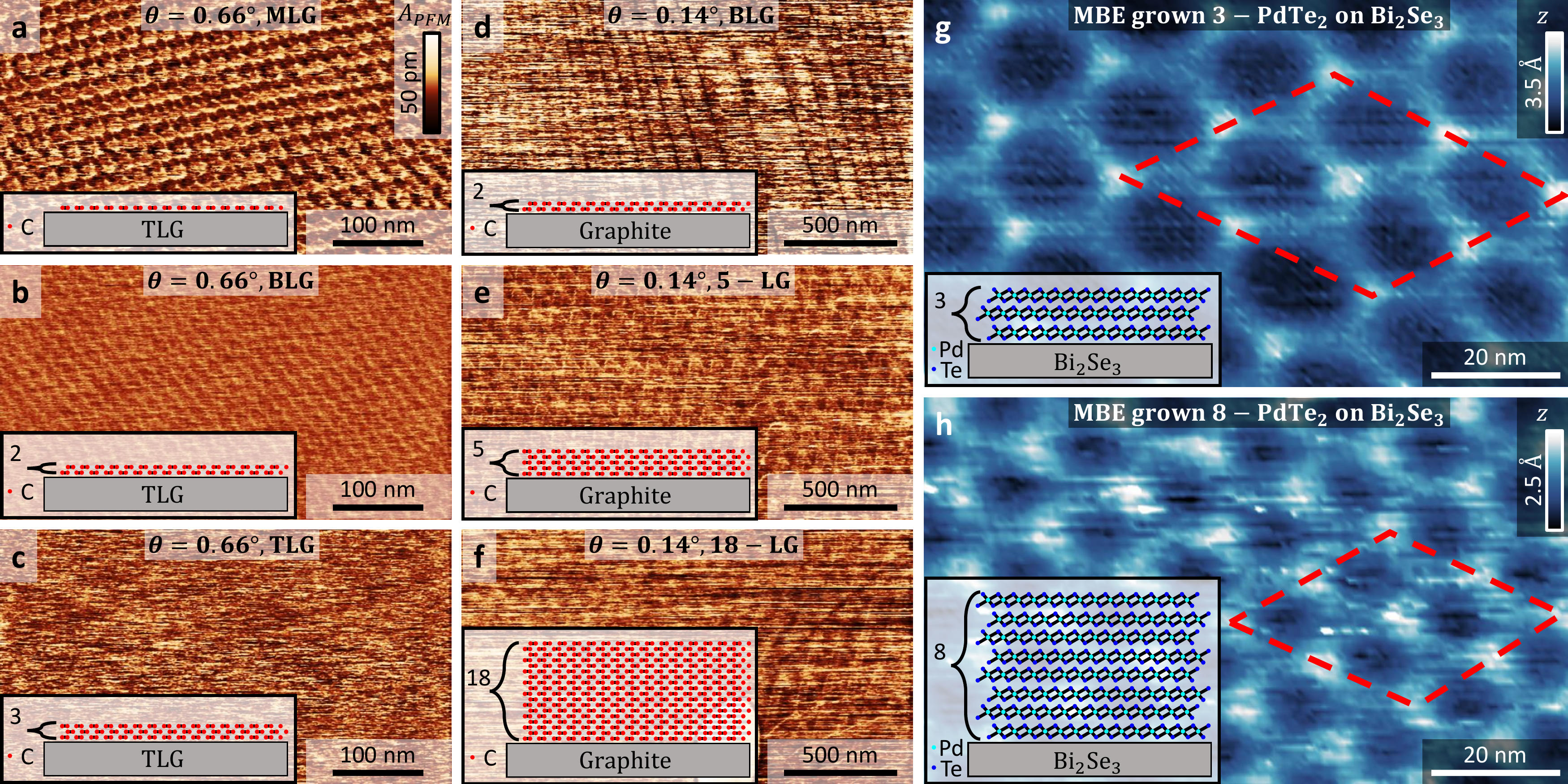}
    \caption{\captiontitle{Experimental motivation: Evidence for multi-layered relaxation.}
    \captionlabel{a-c} Piezoelectric force microscopy (PFM) amplitude mapping of \captionlabel{a} monolayer (MLG), \captionlabel{b} bilayer (BLG) and \captionlabel{c} trilayer (TLG) graphene twisted by $0.65^{\circ}$ relative to a TLG substrate. \captionlabel{d-f} PFM amplitude mapping of \captionlabel{d} bilayer, \captionlabel{e} 5-layered and \captionlabel{f} 18-layered graphene twisted by $0.14^{\circ}$ relative to a graphite substrate. The smaller the twist angle the thicker the layer through which the moir\'e superlattice can be resolved. \captionlabel{g-h} Topography maps by scanning tunneling microscopy (STM) of molecular beam epitaxy (MBE) grown PdTe$_2$ on a Bi$_2$Se$_3$ substrate. Two thicknesses are presented 3-layers \captionlabel{g} and 8-layers \captionlabel{h}, showing a thickness dependence of the domain structure and the moir\'e period indicating a layer dependent relaxation process. The dashed red diamond marks four moir\'e unit cells ($2\times2$) in each case. Inset in each panel presents a schematic of the system under study. The experimental details are presented in \cref{App:graphene_exp,App:PdTe2_exp}}
    \label{fig:experimental-data}
\end{figure*}

In fact, we show, elaborate domain formation even far away from the interface, stressing the problem  is truly three dimensional in nature.
From this follows that material properties and the electronic structure~\cite{uchida2014atomic, lucignano2019crucial, guinea2019continuum, cantele2020structural} may be substantially influenced as soon as domain formation occurs, not only in direct vicinity of the interface. \Cref{fig:experimental-data} summarizes experimental evidence for the fact that the domain formation picture is rather complex, and for its three dimensional nature. \Cref{fig:experimental-data}\captionlabel{a-c} explore piezoresponse force microscopy (PFM) maps of a terraced few layer graphene (FLG) structure on top of a trilayer graphene (TLG) substrate at a twist angle of 0.66$^{\circ}$ between the TLG and the FLG part of the stack. PFM is an atomic force microscopy (AFM) contact-mode variant which has been shown to be a powerfull large area imaging tool of twisted moir\'e superlattices \cite{McGilly2020c}. It measures the deflection of the cantilever of a conductive tip at a given point on the sample, due to an oscillating voltage between the metallic tip and a grounded sample (more details in \cref{App:graphene_exp}). Simplistically, the amplitude of the signal can be interpreted as a measure of local deformation resulted by the electric field, i.e. a local piezoelectric or flexoelectric effect (See Ref.~\onlinecite{McGilly2020c} for a dedicated study on the use of PFM for visualization of moir\'e superlattices in in van der Waals heterostructures). In the context of this work we consider the PFM contrast as a qualitative measure of the strength of the relaxation at the sample surface. The terraced FLG has monolayer graphene (MLG), bilayer graphene (BLG), and TLG segments, (highlighted over the optical image in \cref{App:graphene_exp}), and studied by PFM in \cref{fig:experimental-data}\captionlabel{a-c} respectively. As the top layer thickness increases the PFM contrast quickly decays. On the other hand, when a similar measurement is performed for a smaller twist angle of 0.14$^{\circ}$ (\cref{fig:experimental-data}\captionlabel{d-f}) the moir\'e pattern can be visualized by PFM at the surface of a  considerably higher stack, even up to 18 atomic layers above the twist interface (\cref{fig:experimental-data}\captionlabel{f}). A different consequence of the layered atomic relaxation emerges in atomically mismatched heterostructures. Such a case is presented in \cref{fig:experimental-data}\captionlabel{g,h} for an epitaxially grown PdTe$_2$ over a Bi$_2$Si$_3$ substrate~\cite{cook2022moire}. In this system a moir\'e superlattice emerges from the lattice mismatch between the two materials. Surprisingly, the domain shapes and even the moir\'e period itself seem to depend on the layer thickness of the PdTe$_2$ stack. In fact, the observed moir\'e period exceeds the maximal (zero twist angle) predicted moir\'e period, solely considering the lattice mismatch. All of the above suggest that the relaxation process and the resulting domain structures are more involved than considered thus far. The spatial extent of the relaxation process and even the moir\'e period depend on the twist angle and material properties.

In this work we first develop the multi-layered relaxation framework in \cref{sec:multi-layered-relaxation-approach}. The framework allows a description of large systems (both in terms of thickness as well as moir\'e period) which cannot be accessed by atomistic models \cite{Naik2018,Gargiulo2018,Haddadi2020,van2015relaxation,Maity2021,jain2016structure}.
We will then use this framework to reproduce the data in \cref{fig:experimental-data} (cf.~\cref{sec:mismatched-systems}), explore the three dimensional nature of the relaxation process (cf.~\cref{sec:structural-study}), and study its consequences for the electronic structure using graphitic stacks as a prototypical example in \cref{sec:impact-electronic}.

\section{Multi-layered relaxation - the approach}
\label{sec:multi-layered-relaxation-approach}

In this section we develop the mathematical framework to allow the description of the atomic relaxation in a multi-layered stack of vdW materials. We follow a continuous relaxation approach, conceptually similar to Refs.~\onlinecite{Carr2018,Zhu2020,Cazeaux2020,Halbertal2021}. By continuous relaxation, we assume that the in-plane displacement fields change on a length scale larger than the atomic unit cell spacing.
Using the above assumption for continuous models, the atomic relaxation can be formulated in terms of the energy functional being a sum of a stacking mismatch and an elastic energy term.
The sought after solution is the displacement field that minimizes the total energy. Ref.~\onlinecite{Carr2018} formulated the problem in configuration space for twisted bilayer systems. Ref.~\onlinecite{Cazeaux2020} introduced a generalized formalism to potentially treat multiple twisted interfaces, later implemented by Ref.~\onlinecite{Zhu2020} for a specific system of double twisted trilayer graphene. In Ref.~\onlinecite{Halbertal2021} we solved the problem in real space, thus allowing an omission of the periodicity requirement imposed by the previous treatments in configuration space. It allowed treating experimentally realistic strain induced relaxation patterns.
The derivation follows the lines and notation of  Ref.~\onlinecite{Halbertal2021}, but instead of treating each flake of the heterostructure as a single elastic membrane,  here we assume a stack of $m$ layered materials labelled by $j\in\{1,2,\dots,m\}$ with a thickness of $n_j$ layers, respectively, where $j=1$ denotes the lowermost and $j=m$ the uppermost flake of the heterostructure. 
Crucially, we allow for individual relaxation in each layer with  the in-plane displacement fields of layer $k$ of flake $j$ being described by
\begin{align}
    \bvec{u}^{(j,k)} (\bvec{r}) = \big(u_{x}^{(j,k)}(x,y),u_{y}^{(j,k)}(x,y)
    \big)^\mathrm{T}\,.
\end{align}
The interface between flakes $j$ and $j+1$ is formed between layer $k=n_j$ of flake $j$ and layer $k=1$ of flake $j+1$. The energy functional in our model is composed of the elastic energy density and the stacking energy density,
\begin{equation}
\Edens(\bvec{r},\bvec{u},\nabla\bvec{u})=\Eelastic(\nabla\bvec{u})+ \Estacking(\bvec{r},\bvec{u}) \,.
\end{equation}
We assume that the $n_1+n_2+\dots+n_m$ displacement fields change on a length scale larger than the lattice constants $\alpha_{j}$ in accordance with prior literature~\cite{Carr2018,Cazeaux2020,Halbertal2021}.
However, Ref.~\onlinecite{Halbertal2021} was restricted to a single interlayer interface, where now we consider one interface from the twist or lattice constant mismatch between each set of flakes $j$ and $j+1$, which after relaxation induces   $n_{j}-1$ interfaces within flake $j$ and similarly $n_{j+1}-1$ interfaces within flake $j+1$.
We show here that only such a modelling captures the important, finite penetration depth of the relaxation induced by stacking vdW heterostructures from one to a few to a large number of layers in the flakes. To this end, we will therefore assume that the stacking energy term can be described by two sets of GSFEs: one set covering the initial interfaces {\it between} flakes and another set describing the relaxation-induced interfaces {\it within} each of the $m$ separate flakes.
Note that the separation of the system into a sum over the different interfaces makes a further approximation. It neglects contributions to the stacking energy from neighboring layers beyond the nearest neighbor. Such an approximation would fail to capture subtle differences like the energy imbalance between the Bernal and rhombohedral phases in twisted double bilayer graphene~\cite{Halbertal2021} and could be relaxed in principle in the future.

First, we write the elastic energy density $\Eelastic(\nabla\bvec u)$ as a function of the gradients of the displacement fields $\nabla\bvec u$:
\begin{widetext}
\begin{equation}
  \Eelastic(\nabla\bvec{u}) =\!\!\!\sum_{j,1\leq k \leq n_j}\!\!\! \varepsilon^{(j,k)}(\nabla\bvec u) = \!\!\!\sum_{j,1\leq k \leq n_j} \Bigg[
    \frac{K_j \big( \nabla \cdot \bvec u^{(j,k)} \big)^{2}}2 + \frac{G_j}2\bigg( \big( \partial_{x}u_{x}^{(j,k)}-
      \partial_{y}u_{y}^{(j,k)} \big)^{2}+ \big(\partial_{x}u_{y}^{(j,k)}+
      \partial_{y}u_{x}^{(j,k)} \big)^{2} \bigg) \Bigg] \,,
\end{equation}
\end{widetext}
with $\varepsilon^{(j,k)}(\nabla\bvec u)$ the layer resolved elastic energy density, and $K_j$ and $G_j$ the bulk and shear moduli of flake $j$, respectively.

The stacking energy density, $\Estacking(\bvec r,\bvec u)$, explicitly depends on position and displacement fields. The general expression is a function of the stacking configuration of each interface.
As the stacking configuration is defined in terms of the underlying Bravais lattice, we denote the lattice vectors of flake $j$ with $\alpha_j \bvec a_{1,2}^j$
where $\bvec a_{1,2}^j$
are normalized lattice vectors. Here, we choose the normalization factors $\alpha_j$ to be the lattice constants. For the general case, the relative length of lattice vectors can be included in $\bvec a_{1,2}^j$%
, and $\alpha_j$ acts as an overall length scale of flake $j$.
Further, we can always define a two-dimensional matrix $A_j$ that transforms to the lattice vectors such that
\begin{equation}
    A_j = \begin{bmatrix}
    \bvec a_1^j & \bvec a_2^j
    \end{bmatrix}
\end{equation}
We then proceed to define the stacking configuration $\bvec\Omega_0$ of each layer $k$ in flake $j$ as
\begin{align}
    \bvec\Omega_0^{(j,k)}(\bvec r,\bvec u) &{}= \frac{2\pi}{\alpha_j}\,(A_j)^{-1} \big(\bvec r-\bvec u^{(j,k)}(\bvec r)\big) \,,
\end{align}
The origins of the unit cells of the layers within the $m$ rigid flakes, respectively, can be chosen at $\bvec r_j=\alpha_j(l\bvec a_1^j + m\bvec a_2^j)$ with integers $l$ and $m$, such that for zero displacement field $\bvec \Omega_0^{(j,k)}(\bvec r_j,\bvec u=0)=2\pi (l\,,\,m)^\mathrm{T}$.

We assume that each unit area on each interface between two layers contributes to the stacking energy density according to the GSFE for that interface with the relative stacking configuration as input. So we define the relative stacking configuration $\bvec\Omega_*$ for the intra-flake interfaces $k_j=1,\dots,n_j-1$ as
\begin{align}
    \bvec\Omega_*^{(j,k_j)} &{}=-\frac{2\pi}{\alpha_j} (A_j)^{-1} \big(\bvec{u}^{(j,k_{j}+1)}-\bvec{u}^{(j,k_{j})} \big) \,.
\end{align}
The relative stacking configuration at the inter-flake interfaces can similarly be expressed as
\begin{equation}
    \bvec\Omega_*^{(j\rightarrow j+1)} = \bvec \Omega_0^{j+1,1} - \bvec \Omega_0^{j,n_j}\,.
\end{equation}
As we also want to explicitly treat isotropic expansion of each flake and constant stacking registry shifts of each layer (such as Bernal (AB) stacking in graphene), we define the total displacement of flake $j$ and layer $k$ as $\bvec U^{(j,k)} = \bvec u^{(j,k)} + \epsilon_j\bvec r$ with $\epsilon_j$ the expansion parameter and add the space independent constant $\bvec\Delta\bvec\Omega^{(j,k)}$ to the stacking configuration of each layer. These refinements yield the following expressions for the stacking configurations of all interfaces:
\begin{widetext}
\begin{subequations}
\label{eqn:stackingConfigurations}
\begin{align}
  \bvec{\Omega}^{(j,k)}(\bvec{U}) &{}=
    -\frac{2\pi}{\alpha_{j}} \,(A_j)^{-1}
    \Big(\bvec{u}^{(j,k_j+1)}-\bvec{u}^{(j,k_j)}\Big) +
    \bvec{\Delta\Omega}^{(j,k_{j}+1)}-\bvec{\Delta\Omega}^{(j,k_{j})} \,,\\
  \bvec{\Omega}^{(j\rightarrow j+1)}(\bvec{U}) &{}=
  \frac{2\pi}{\alpha_{j+1}} (A_{j+1})^{-1} \Big( (1-\epsilon_{j+1}) \bvec r
  - \bvec u^{(j+1,1)} \Big) - \frac{2\pi}{\alpha_{j}} (A_j)^{-1} \Big( (1-\epsilon_{j}) \bvec r - \bvec u^{(j,n_j)} \Big)
      + \bvec{\Delta\Omega}^{(j+1,1)}-\bvec{\Delta\Omega}^{(j,n_{j})} \,.
\end{align}
\end{subequations}
In general, we formulate the stacking energy density of the system as a function of the relative stacking configurations $\bvec\Omega$:
\begin{equation}
    \Estacking(\bvec U) =
    \sum_{j=1}^m \sum_{k_j=1}^{n_j-1} V_\mathrm{GSFE}^{(j)}\big(
            \bvec\Omega^{(j,k_j)}(\bvec U)
        \big)
    +
    \sum_{j=1}^{m-1} V_\mathrm{GSFE}^{(j\rightarrow j+1)}\big(
            \bvec\Omega^{(j\rightarrow j+1)}(\bvec U)
        \big) \,.
\end{equation}
\end{widetext}

For the specific physical systems under consideration in this work, we we can use the known form of the GSFE functional \cite{Carr2018,Cazeaux2020,Zhu2020,Halbertal2021} for each of the interfaces:
\begin{multline}
\label{eqn:Vgsfe}
V_\mathrm{GSFE}\Big(\bvec{\Omega}, c_{0,\dots,5} \Big)=c_{0} \\
  \begin{aligned}[b]
    &+c_{1}\big(\cos v+\cos w+\cos(v+w)        \big) \\
    &+c_{2}\big(\cos(v+2w)+\cos(v-w)+\cos(2v+w)\big) \\
    &+c_{3}\big(\cos(2v)+\cos(2w)+\cos(2v+2w)  \big) \\
    &+c_{4}\big(\sin v+\sin w-\sin(v+w)        \big) \\
    &+c_{5}\big(\sin(2v+2w)-\sin(2v)-\sin(2w)  \big) \,,
  \end{aligned}
\end{multline}
where we set $(v,w)=\bvec{\Omega}$ as the components of the relative stacking configurations. The parameters $c_{0,\dots,5}$ depend on the type of interface. As the systems considered in this work consist of $m=2$ flakes, we need three sets of parameters to calculate the stacking energy density, namely $c_{0,\dots,5}^{(j)}$ for each of the interfaces \emph{within} flakes $j=1,2$ and $c_{0,\dots,5}^{(1\rightarrow2)}$ for the interface \emph{between} flakes $1$ and $2$ (see \cref{tab:material_param} for numerical values for the systems under consideration in this work).

To find the solution $\bvec u^{(j,k)}(\bvec r)$ and $\epsilon_j$, we minimize the total energy
\begin{equation}
    E_\mathrm{tot} = \int\mathrm d^2r\,\Big(\Estacking(\bvec U) + \Eelastic(\nabla \bvec U)\Big)
\end{equation}
using standard unconstrained minimization techniques (trust region algorithm). Note that changes to $\epsilon_j$ affect the size of the moir\'e unit cell, so one has to properly account for this effect in defining the grid and the strain terms.

\section{Mismatched system test case}
\label{sec:mismatched-systems}
\begin{figure}
    \centering
    \includegraphics[width=\columnwidth]{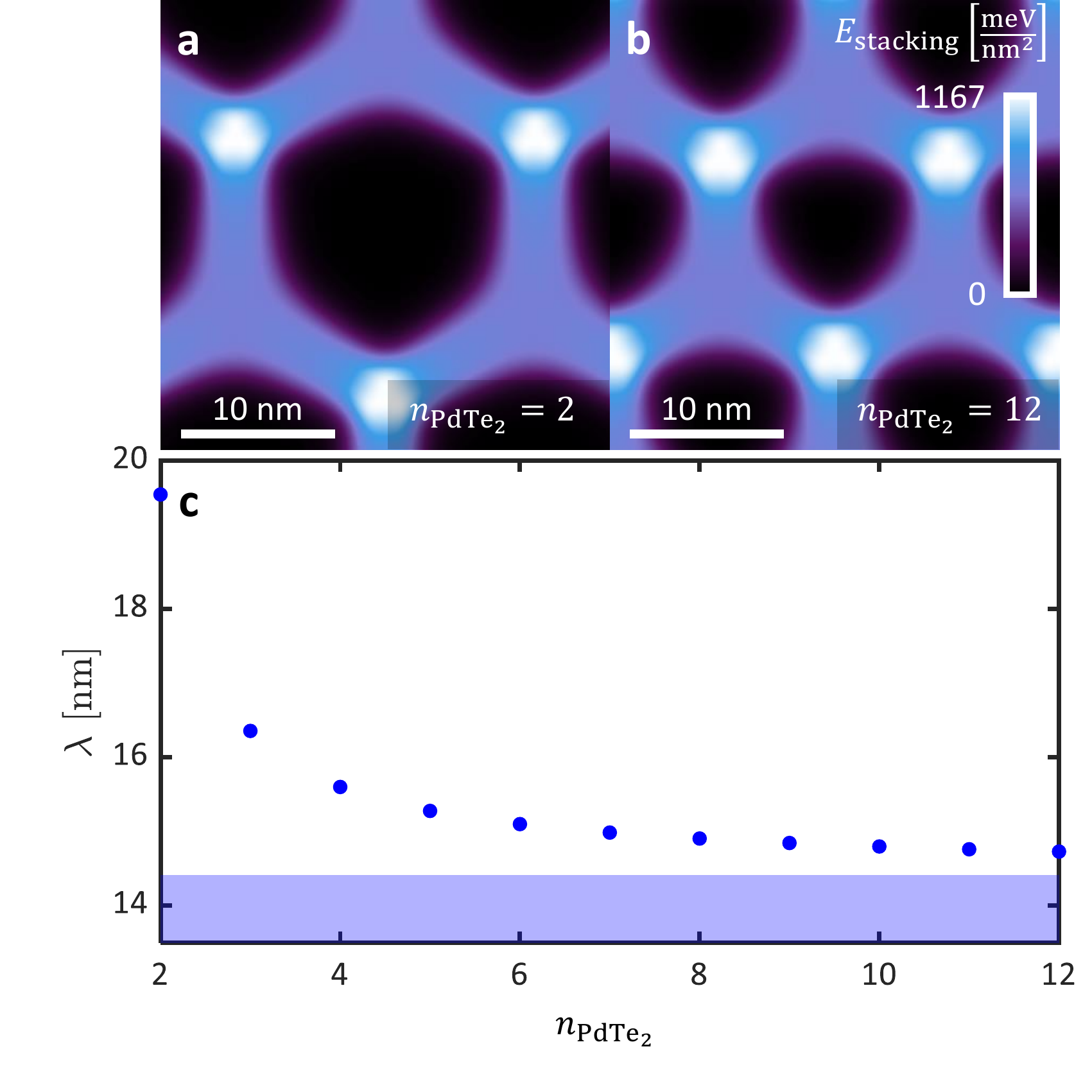}
    \caption{\captiontitle{Modeling of thickness dependent moir\'e superlattice in PdTe\textsubscript{2}/Bi\textsubscript{2}Se\textsubscript{3}.} \captionlabel{a-b} Moir\'e superlattice after mechanical relaxation visualized by the stacking energy density for 2 \captionlabel{a} and 12 \captionlabel{b} $\mathrm{PdTe_2}$ layers on bulk $\mathrm{Bi_2Se_3}$. \captionlabel{c} Moir\'e periodicity $\lambda$ as a function of number of $\mathrm{PdTe_2}$ layers. The blue region marks the range of accessible moir\'e period values without isotropic global strain. The relaxation calculations reveal enhanced relaxation and expansion of the moir\'e periodicity in the thin $\mathrm{PdTe_2}$ layer limit. All considered cases assumed no twist between the $\mathrm{PdTe_2}$ and the $\mathrm{Bi_2Se_3}$. See calculation details in \cref{app:material-parameters}.}
    \label{fig:PdTe2_theory}
\end{figure}

Motivated by the experimental findings summarized in \cref{fig:experimental-data}\captionlabel{g,h} we consider a $\mathrm{PdTe_2/Bi_2Se_3}$ heterostructure as a specific example of a mismatched heterostructure.
The underlying Bravais lattice of both materials is triangular, so we set the lattice vectors as
\begin{align}
    \label{eqn:lattice-vectors-mismatched}
    \bvec a_1 &{}= \begin{pmatrix} 1 \\ 0 \end{pmatrix} \,,&
    \bvec a_2 &{}= \begin{pmatrix} 1/2 \\ \sqrt3/2 \end{pmatrix} \,.
\end{align}
The GSFE function parameters $c_{1,\dots,5}$ for each interface of  this multi-layered PdTe$_2$ on a Bi$_2$Se$_3$ substrate, and the bulk and shear moduli for both materials are taken from DFT (see \cref{app:material-parameters}).
For simplicity we fix the substrate and allow all $\mathrm{PdTe_2}$ layers to relax individually. We further let the $\mathrm{PdTe_2}$ flake expand isotropically via $\epsilon_2$. This degree of freedom is crucial in order to explain the observation of \cref{fig:experimental-data}\captionlabel{g,h}. \Cref{fig:PdTe2_theory} summarizes the multi-layered relaxation results for different numbers of $\mathrm{PdTe_2}$ layers. \Cref{fig:PdTe2_theory}\captionlabel{a,b} show how as the layer number drops the relaxation becomes stronger, leading to hexagonal domains and domain expansion. The domain expansion is further quantified in \cref{fig:PdTe2_theory}\captionlabel{c} covering the transition between the thick and thin $\mathrm{PdTe_2}$ regimes. It is important to note that all the considered cases have shown a moir\'e period $\lambda$ in excess of the accessible moir\'e periods as expected without including the isotropic expansion term (blue shaded region in \cref{fig:PdTe2_theory}\captionlabel{c}). Such an excess moir\'e period is in agreement with the experimental results. Our results highlight the multi-layered trade-off in search of an energy minimizing solution by the $\mathrm{PdTe_2}$ flake: By conforming to the lattice constant of the $\mathrm{Bi_2Se_3}$ substrate the stacking energy is reduced, at the expense of a penalty of an elastic energy. As the number of layers increases the elastic energy term becomes too costly and the relaxation is suppressed. However, at the thin $\mathrm{PdTe_2}$ limit due to the relative softness of $\mathrm{PdTe_2}$ (see \cref{tab:material_param}) the elastic energy penalty due to stretching the $\mathrm{PdTe_2}$ is not severe in comparison with the stacking energy benefit, and the result is an enhanced moir\'e period. \Cref{fig:PdTe2_theory}\captionlabel{c} also highlights the penetration of the relaxation field (here probed by the layer dependent moir\'e periodicity $\lambda$) induced by the $\mathrm{PdTe_2/Bi_2Se_3}$ interface into the stack which decays exponentially with the number of layers $n_{\mathrm{PdTe_2}}$. To visualize the layer dependent moir\'e period, we assembled a video of the pattern in \cref{fig:PdTe2_theory}\captionlabel{a,b} varying $n_\mathrm{PdTe_2}$ over time~\cite{supplement}.

Finally, we note that the layer-dependent moir\'e wavelength in $\mathrm{PdTe_2}$/$\mathrm{BiSe_2}$ is \emph{quantitatively} reproduced by our multi-layered relaxation framework, considering our data (\cref{fig:experimental-data}\captionlabel{g-h}) as well as the data presented in Ref.~\onlinecite{cook2022moire}.

\section{Structural study of a graphene based system}
\label{sec:structural-study}
\begin{figure*}
    \centering
    \includegraphics[width=\textwidth]{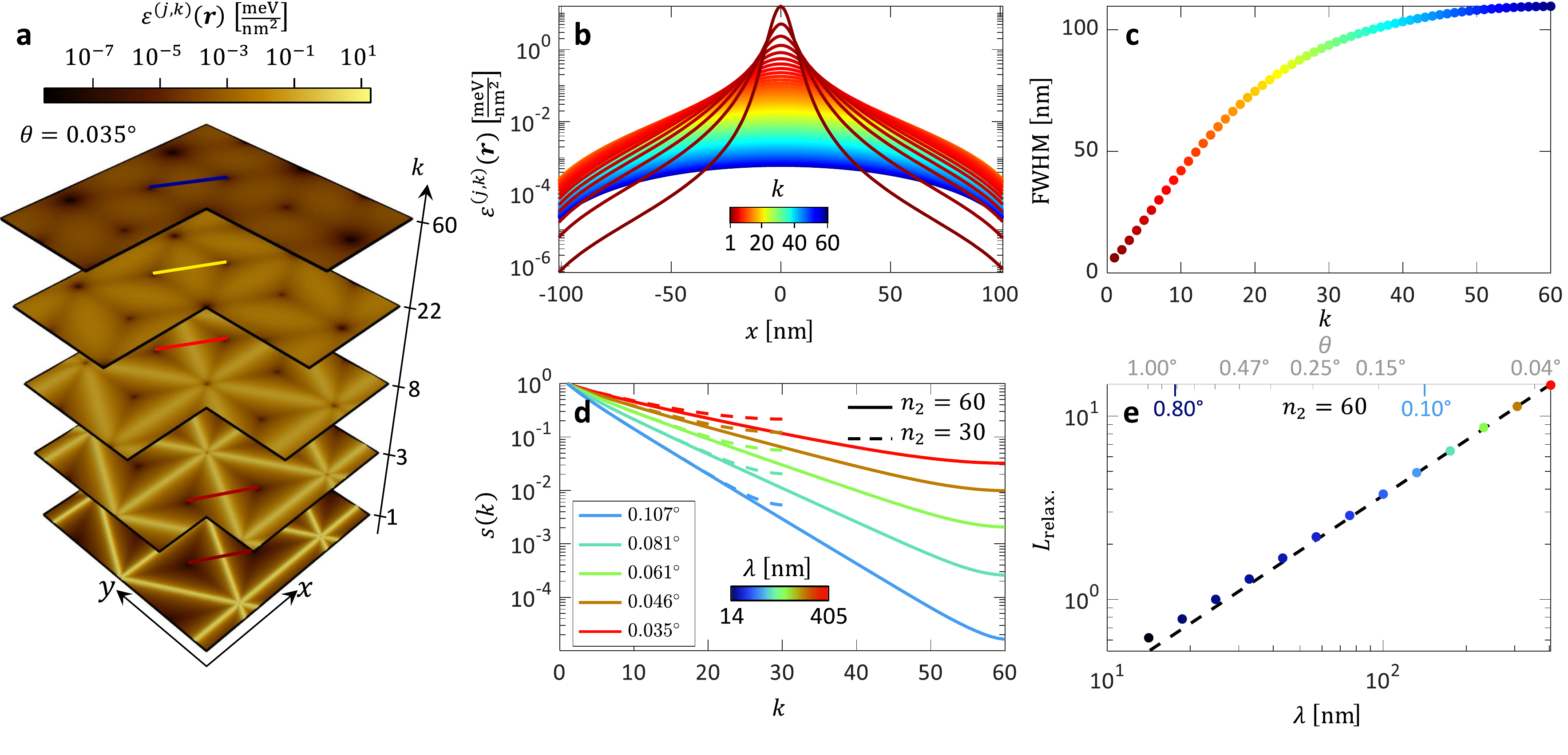}
    \caption{\captiontitle{Multi-layered relaxation in twisted graphite interfaces} \captionlabel{a} Elastic energy density in representative layers for a $n_1=n_2=60$ graphite on graphite structure with a twist angle $\theta=0.035^{\circ}$. \captionlabel{b} Domain wall (DW) profiles and \captionlabel{c} width as full width at half maximum (FWHM) for different layers showing the broadening and weakening of the relaxation away from the interface. Layer number indicated by color-bar in \captionlabel{b}. \captionlabel{d} Decay of the strength of the relaxation process away from the interface for different twist angles as captured by the displacement field scaling factor (see text for details). Two layer thickness cases are considered. \captionlabel{e} The penetration depth extracted from \captionlabel{d} as a function of the moir\'e period $\lambda$. The penetration depth is shown to be proportional to the moir\'e period (dashed line).}
    \label{fig:multilayered-relaxation}
\end{figure*}

Next, we highlight the impact of employing the multi-layered relaxation scheme by the analysis of a graphene based system compared to the experimental findings reported in \cref{fig:experimental-data}\captionlabel{a-f}. To this end, we study the structural properties of two flakes of Bernal stacked multi-layered graphene stacked with a twist at the interface of the two flakes. As in the previous section, the Bravais lattice of graphene is also triangular. However, we need to implement the twist angle $\theta$ as a two-dimensional rotation matrix in the transformation from flake $1$ to $2$:
$A_2 = R(\theta) A_1$.
Moreover, the GSFE parameters and shear and bulk moduli of graphene are presented in \cref{app:material-parameters}.

As a primary result, we obtain the layer- and space-resolved elastic energy density $\varepsilon^{(j,k)}(\bvec r)$. This can be viewed as a layer-resolved elastic energy tomograph scanning through the height of the stacks. \Cref{fig:multilayered-relaxation}\captionlabel{a} displays this result for a small twist angle system ($\theta=0.035^\circ$) and $n_1=n_2=60$. The layer-resolved elastic energy tomograph emphasizes two main points: First, the relaxation pattern decays exponentially as a function of distance to the interface. Second, the system forms triangular relaxation domains separated by domain walls connecting the AB/BA regions of the interfacial moir\'e pattern. In \cref{fig:multilayered-relaxation}\captionlabel{b}, we plot the elastic energy density along the line cut across a domain wall (as indicated in the tomograph of \cref{fig:multilayered-relaxation}\captionlabel{a}) for all layers. At the interface, these domains have very sharp domain walls. The width of the domain wall grows as the distance to the interface increases, while at the same time its height declines. In \cref{fig:multilayered-relaxation}\captionlabel{c} we quantify this behaviour by plotting the full width at half maximum (FWHM) as a function of layer index $k$. We color-code the value of $k$ consistently across panels \captionlabel{a-c}.

Upon varying twist angle, we can examine how the relaxation behaviour evolves. To this end, we analyze the scaling of the atomic displacements $\bvec u^{(j,k)}(\bvec r)$ as a function of $k$ in the second flake $j=2$. To do so we quantify the strength of the relaxation in layer $k$ of flake $2$ by defining a layer-dependent scaling factor $s(k)$ which is chosen such that the in-plane integrated squared difference
\begin{equation}
  D(k) = \int\mathrm d^2r\,\,\big\|\bvec u^{(2,1)}(\bvec r)-s(k)\bvec u^{(2,k)}(\bvec r) \big\|^2 \,,
\end{equation}
of displacement fields  of the $k$-th  layer $\bvec u^{(2,k)}(\bvec r)$ and the first layer $\bvec u^{(2,1)}(\bvec r)$  is minimal. 

\Cref{fig:multilayered-relaxation}\captionlabel{d} shows the scaling factor $s(k)$ for five twist angles $0.035^\circ \leq \theta \leq 0.107^\circ$, with red curves corresponding to the small twist angle (large moir\'e wavelength $\lambda$) case and blue curves to the large twist angle (small $\lambda$) case. For each of the twist angles considered, we additionally set the number of layers of the second flake to $n_2=30$ (dashed lines) and $n_2=60$ (solid lines). Upto a cutoff when approaching the upper end of the  stack (dependent on the height $n_2$),  we find penetration of the relaxation fields into the stack which decays exponentially. The exponent decreases with the twist angle, yielding longer-ranged relaxation effects for smaller twist angle. We quantify the exponential decay of the relaxation effects by determining the penetration depth $L_\mathrm{relax.}$ of the relaxation pattern. Here, we define $L_\mathrm{relax.}$ through
\begin{align}
    s(k) \appropto e^{-k/L_\mathrm{relax.}}
\end{align}
and calculate it using a fit for the case $n_2=60$ as a function of $\lambda$. The result is shown in \cref{fig:multilayered-relaxation}\captionlabel{e} with consistent color-coding of $\lambda$ and $\theta$ across panels \captionlabel{d-e}. We observe that the penetration depth $L_\mathrm{relax.}(\lambda)$ scales linearly with $\lambda$ in the small-$\theta$ and large-$n_2$ limit.

For the two angles observed in the experiment (\cref{fig:experimental-data}\captionlabel{a-f}), we thus expect a penetration depth that is $0.66^\circ/0.15^\circ\approx4.7$ times as large in the $0.15^\circ$ system than in the $0.66^\circ$ system. As the $0.66^\circ$ case cannot be assumed to be in the large-$n_2$ limit ($n_2=3$), we conclude that the theory \emph{qualitatively} agrees with the experimental data --- $18\unit{layers}/3\unit{layers}$ is the same order of magnitude as $4.7$.

\section{Multi-layered relaxation's impact on the electronic structure of a graphitic system}
\label{sec:impact-electronic}

As suggested by our experimental data and confirmed by using the multi-layered relaxation framework, the in-plane relaxation can affect layers far from the interface of the flakes. For twisted graphitic systems this becomes particularly prominent  at small twist angle (cf.~\cref{fig:experimental-data,fig:multilayered-relaxation}). Such in-plane relaxations generate effective magnetic fields~\cite{morozov2006strong,vozmediano2010gauge,katsnelson2012graphene,amorim2016novel} and therefore may have a substantial impact on the electronic structure. Furthermore, using relaxed atomic positions has been shown to be crucial to explain low-energy properties of twisted graphitic systems~\cite{uchida2014atomic,Nam2017,lucignano2019crucial,guinea2019continuum,liang2020effect}. Therefore, we analyze the effect of multi-layered relaxation on the electronic structure of twisted many-layer graphene systems in the following.

We focus on the case of several layers of Bernal stacked graphene on bulk graphite (corresponding to the small angle experimental PFM data shown in \cref{fig:experimental-data}). We approximate the bulk by $n_1=20$ graphene layers and vary the number of layers in the top (twisted) flake from $n_2=3$ to $n_2=20$. We set the twist angle to $\theta=0.1^\circ$ (the $\theta=0.8^\circ$ case can be found in \cref{app:large-angle-electronic-structure}) which implies a unit cell of $\unsim150\unit{nm}$ in size and $\unsim650,\!000$ carbon atoms per layer. For a multi-layer geometry, the number of atoms is intractable for a tight-binding framework, and so we adopt the Bistritzer-MacDonald continuum model of twisted bilayer graphene~\cite{bistritzer2011moire,lopes2012continuum,lopes2007graphene,koshino2018maximally,cea2019twists} to include arbitrary in-plane relaxations (see Refs.~\onlinecite{balents2019general,koshino2020effective}) and Bernal multilayer hopping terms~\cite{gruneis2008tightbinding,koshino2019band,Haddadi2020,samajdar2020microscopic}.

For our analysis, we make use of the Bistritzer-MacDonald Hamiltonian, which can be
viewed as a reciprocal moir\'e lattice vector expansion of the full tight-binding Hamiltonian of the system.
The expansion as presented below is valid for SE-odd moir\'e structures~\cite{lopes2012continuum}, which applies to the graphitic systems studied in this work.
With $\bvec B_{1,2}$ the reciprocal moir\'e lattice vectors, we expect
the multi-layered Hamiltonian to be a matrix in the ``indices" $\bvec G=p\bvec B_1 + q\bvec B_2$, $p,q\in\mathbb Z$. The expansion is then truncated at $\|\bvec G\|<G_c$, where $G_c$ must be chosen large enough to ensure convergence. We further need to account for the sublattice (A, B), flake, layer, and valley $\xi=\pm1$ degrees of freedom. For the sake of clarity of notation we drop flake indices and continuously index the layers with $l=1,\dots,n_1+n_2$, where $l\leq n_1$ corresponds to flake $j=1$ and $l>n_1$ to flake $j=2$.

Our starting point is the intralayer continuum Hamiltonian, which is composed of the following three parts:
\begin{equation}
    H_\mathrm{intra}^{\bvec G\bvec G',\xi,l}(\bvec k) = H_\mathrm{bernal}^{\bvec G,\xi,l}(\bvec k)\delta_{\bvec G,\bvec G'} + 
    H_\mathrm{mag}^{\bvec G\bvec G',\xi,l} + H_\mathrm{stat}^{\bvec G\bvec G',\xi,l} \,.
    \label{eqn:intralayer-continuum-overview}
\end{equation}
Only the standard Bernal intralayer Hamiltonian explicitly depends on momentum $\bvec k$ (in the moir\'e Brillouin zone). As the low-energy states all are close to the individual graphene layers' $K_\xi$-points, we shift and rotate the momentum vector:
\begin{equation}
    \bvec K_v = R(\theta^l) \,\big(\bvec k - \bvec K_\xi^l + \bvec G\big),
\end{equation}
with $R(\theta^{l>n_1})=R(\theta)$ for the twisted layers in flake $2$ and $R(\theta^{l\leq n_1})=\mathds1$ for the non-twisted layers in flake $1$. The $K_\xi$-points of a given layer can be expressed by the reciprocal moir\'e lattice vectors as
\begin{align}
    \bvec K_\xi^{l\leq n_1} &{}= -\xi/3\,\big(2\bvec B_2 + \bvec B_1\big) \,, \\
    \bvec K_\xi^{l>n_1} &{}= -\xi/3\,\big( \bvec B_2 - \bvec B_1 \big) \,.
\end{align}
Using $K_v = \xi \bvec K_{v,x} + i \bvec K_{v,y}$, the Bernal intralayer Hamiltonian as a matrix in (A, B) sublattice space is then expressed as
\begin{align}
    H_\mathrm{bernal}^{\bvec G,\xi,l}(\bvec k) = \begin{pmatrix}
        \delta_{p,\mathrm{A}} + \delta^l_\mathrm{isp} & -v_F\,K_v^* \\ -v_F\,K_v & \delta_{p,\mathrm{B}} + \delta^l_\mathrm{isp}
    \end{pmatrix}\,.
\end{align}
Here, $v_F$ denotes the Fermi velocity of graphene. We further include the intrinsic symmetric polarization $\delta^l_\mathrm{isp}$ that is nonzero only in the central (twisted) layers: $\delta^{n_1}_\mathrm{isp}=\delta^{n_1+1}_\mathrm{isp}=\delta_\mathrm{isp}$. Additionally, we account for sublattice polarization with $\delta_{p,\mathrm{A}}=\delta_p$, $\delta_{p,\mathrm{B}}=0$ in all layers.

The other two parts of the intralayer continuum Hamiltonian \cref{eqn:intralayer-continuum-overview} depend on the in-plane displacement field. As the in-plane displacement field is periodic in moir\'e lattice vectors and varies within the moir\'e unit cell, we apply a Fourier transform and obtain a representation in reciprocal moir\'e lattice vector space:
\begin{equation}
    \bvec u_{\bvec G}^l = \sum_{\bvec r} e^{-i\bvec G\cdot\bvec r} \bvec u^{l=(j,k)}(\bvec r) \,.
\end{equation}
First, we can write down the vector potential of the effective magnetic field generated by the displacement field as a function of $\bvec G$:
\begin{equation}
    \bvec A^{\bvec G,l} = \begin{pmatrix}
        i G_x & -iG_y \\
        -i G_y & -i G_x
    \end{pmatrix} \cdot \bvec u_{\bvec G}^l \,.
\end{equation}
This potential is added to the Hamiltonian with a coupling strength $g_2$ by standard substitution:
\begin{align}
    H_\mathrm{mag}^{\bvec G\bvec G',\xi,l} &{}= g_2 \big[ \sigma_x A_x^{\bvec G-\bvec G',l} + \xi\sigma_y A_y^{\bvec G-\bvec G',l} \big]\,,
\end{align}
with $\sigma_x$ and $\sigma_y$ Pauli matrices in sublattice space. Note that the initial dependency of the displacement field on the position vector $\bvec r$ is translated to a (reciprocal moir\'e lattice) momentum transfer $\bvec G-\bvec G'$.

Second, the displacement field also generates a static potential that leads to the following contribution to the intralayer Hamiltonian:
\begin{align}
    H_\mathrm{stat}^{\bvec G\bvec G',\xi,l} = i\mathds1 g_1 \big[ (\bvec G-\bvec G') \cdot \bvec u^l_{\bvec G-\bvec G'} \big] \,.
\end{align}
Here, $g_1$ is the static potential's coupling strength.

The intralayer Hamiltonian is accompanied by several interlayer coupling terms. For all non-twisted interfaces, we employ an effective graphite Hamiltonian. Its coupling from one to the next layer reads (as a matrix in (A$_l$B$_l$; A$_{l+1}$B$_{l+1}$) sublattice space) \cite{Haddadi2020,samajdar2020microscopic}
\begin{align}
    H_\mathrm{inter}^{\bvec G\bvec G',\xi,l\,l+1}(\bvec k) &{}= \begin{pmatrix}
    \gamma_4 K_v^* & \gamma_1 \\ \gamma_3 K_v^* & \gamma_4 K_v
    \end{pmatrix}\,\delta_{\bvec G,\bvec G'}.
\end{align}
The coupling to second-next layers is taken into account as shown in Ref.~\onlinecite{gruneis2008tightbinding} and reads as a matrix in (A$_l$B$_l$; A$_{l+2}$B$_{l+2}$):
\begin{align}
    H_\mathrm{inter}^{\bvec G\bvec G',\xi,l\,l+2}(\bvec k) &{}= \begin{pmatrix}
    \gamma_5 & 0 \\
    0 & \gamma_2
    \end{pmatrix}\,\delta_{\bvec G,\bvec G'}.
\end{align}
For the values of the hopping parameters $\gamma_{1,\dots,5}$ see \cref{tab:cnt-params}.

The interlayer Hamiltonian of the twisted interface between layers $l=n_1$ and $l=n_1+1$ encodes the interfacial moir\'e pattern and is therefore a matrix in $\bvec G$. The coupling function $U^\xi$ depends on the \emph{difference} of in-plane displacement and is most easily formulated in real-space:
\begin{multline}
    U^\xi(\bvec r) = -\sum_{j=1}^3 M^\xi_j\,t_0\,\exp\big[i\bvec Q_j^\xi\cdot
    \big(\bvec u^{(1,n_1)}(\bvec r)-\bvec u^{(2,1)}(\bvec r)\big) \\
    + i\delta\bvec k_j\cdot\bvec r\big]\,.
\end{multline}
Here, $t_0$ is the out-of plane hopping and $M^\xi_j$ are matrices in sublattice (A$_{n_1}$,B$_{n_1}$; A$_{n_1+1}$,B$_{n_1+1}$) space:
\begin{align}
    M_1^\xi = \begin{pmatrix}
    1 & c \\ c & 1
    \end{pmatrix}, \ 
    M_2^\xi = \begin{pmatrix}
    1 & c\,w^{-\xi} \\
    c\,w^\xi & 1
    \end{pmatrix}, \ 
    M_3^\xi = \big(M^\xi_2\big)^* ,
\end{align}
with $c=1.22333$ the ratio of AB-stacking to AA-stacking interlayer tunneling due to out-of plane corrugations and $w=\exp(2\pi i/3)$. The unperturbed momentum transfer vectors $\delta\bvec k_j$ read
\begin{align}
    \delta\bvec k_1 = 0 \,, &&
    \delta\bvec k_2 = \xi\bvec B_1 \,, &&
    \delta\bvec k_3 = \xi(\bvec B_1+\bvec B_2) \,,
\end{align}
and the displacement induced momentum transfer vectors $\bvec Q_j^\xi$ are the center of the twisted and non-twisted $K_\xi$-points of the original graphene lattices:
\begin{alignat}{3}
    \bvec Q_j^\xi
    &{}=
    R(2\pi/3\,j)\,\big[\bvec K^0_\xi + \tilde{\bvec K}^0_\xi\big]/2 \,, \hspace{-25pt}
    \\
    \bvec K_{+1}^0 &{}= -1/3\big(2\bvec b_1+\bvec b_2\big)\,,
    &
    \bvec K_{-1}^0 &{}= R(2\pi/6) \bvec K_{+1}^0\,,
    \\
    \tilde{\bvec K}_{+1}^0 &{}= -1/3\big(2\tilde{\bvec b}_1+\tilde{\bvec b}_2\big)\,,
    &
    \tilde{\bvec K}_{-1}^0 &{}= R(2\pi/6) \tilde{\bvec K}_{+1}^0\,,
\end{alignat}
where $\bvec b_{1,2}$ ($\tilde{\bvec b}_{1,2}$) denote the reciprocal lattice vectors of the non-twisted (twisted) graphene lattice. To arrive at a consistent description we further take the Fourier transform of the real-space interlayer tunneling $U^\xi(\bvec r)$ and obtain a coupling function $U^{\bvec G\bvec G',\xi}$.

The full single-particle continuum Hamiltonian for valley $\xi$ follows as
\begin{multline}
    H_\xi^{\bvec G\bvec G',ll'}(\bvec k) = H_\mathrm{intra}^{\bvec G\bvec G',\xi,l}(\bvec k) \, \delta_{l,l'} + H_\mathrm{inter}^{\bvec G\bvec G',\xi,ll'}(\bvec k) \, \delta_\mathrm{bulk}^{ll'} + \\
    U^{\bvec G\bvec G'} \,\delta_\mathrm{twist}^{ll'}\,,
    \label{eqn:full-continuum}
\end{multline}
where $\delta_\mathrm{bulk}^{ll'} = 1$ for $l$ and $l'$ in the same flake and $\delta_\mathrm{twist}^{ll'} = 1$ for $l$ and $l'$ corresponding to the twisted interface coupling ($l=n_1$ to $l'=n_1+1$). For all other combinations of layers, these ``delta functions" are zero. As the full Hamiltonian requires several numerical parameters, we provide an overview of their values in \cref{tab:cnt-params}.

\begin{table}[t]
\centering
\caption{Parameters for setup of multi-layered continuum Hamiltonian. The length of the graphene lattice vectors $\bvec a_{1,2}$ and the lattice constant $\alpha$ is needed for proper normalization of several hopping parameters.}
\label{tab:cnt-params}
\vspace{0.5em}
\begin{ruledtabular}
\def\arraystretch{1.5}
\def\eqspace{=}
\begin{tabular}{rcl|rcl}
    $a                  $&$\eqspace$&$\alpha||\bvec a_1|| = \alpha||\bvec a_2||$ &
    $t_0                $&$\eqspace$&$0.0797$\,eV \\\hline
    $\gamma_1           $&$\eqspace$&$0.3513$\,eV &
    $\gamma_2           $&$\eqspace$&$-0.0105$\,eV \\\hline
    \hspace{7pt}
    $\gamma_3           $&$\eqspace$&$0.2973\,\mathrm{eV} \cdot a \cdot \sqrt{3}/2$
    \hspace{7pt} &
    \hspace{7pt}
    $\gamma_4           $&$\eqspace$&$0.1954\,\mathrm{eV} \cdot a \cdot \sqrt{3}/2$
    \hspace{7pt} \\\hline
    $\gamma_5           $&$\eqspace$&$0.0187$\,eV &
    $\delta_p           $&$\eqspace$&$0.0540$\,eV \\\hline
    $\delta_\mathrm{isp}$&$\eqspace$&$-0.03$\,eV &
    $g_2                $&$\eqspace$&$-6.3585$\,eV \\\hline
    $g_1                $&$\eqspace$&$4.0$\,eV &
    $v_F                $&$\eqspace$&$2.1354\,\mathrm{eV} \cdot a$ \\
\end{tabular}
\end{ruledtabular}
\end{table}

The application of the continuum description \cref{eqn:full-continuum} to the $\theta=0.1^\circ$ twisted many-layer graphene on graphite system is presented in the following. We inspect the local spectral density in the top layer of the twisted flake as an observable experimentally accessible by scanning tunneling microscopy. To isolate the effect of the relaxation we compare the rigid system (where $\bvec u^{(j,k)}(\bvec r)\equiv0$) to the relaxed one where we apply the method of \cref{sec:multi-layered-relaxation-approach} to find $\bvec u^{(j,k)}(\bvec r)$.

\begin{figure*}
    \centering
    \includegraphics[width=0.95\textwidth]{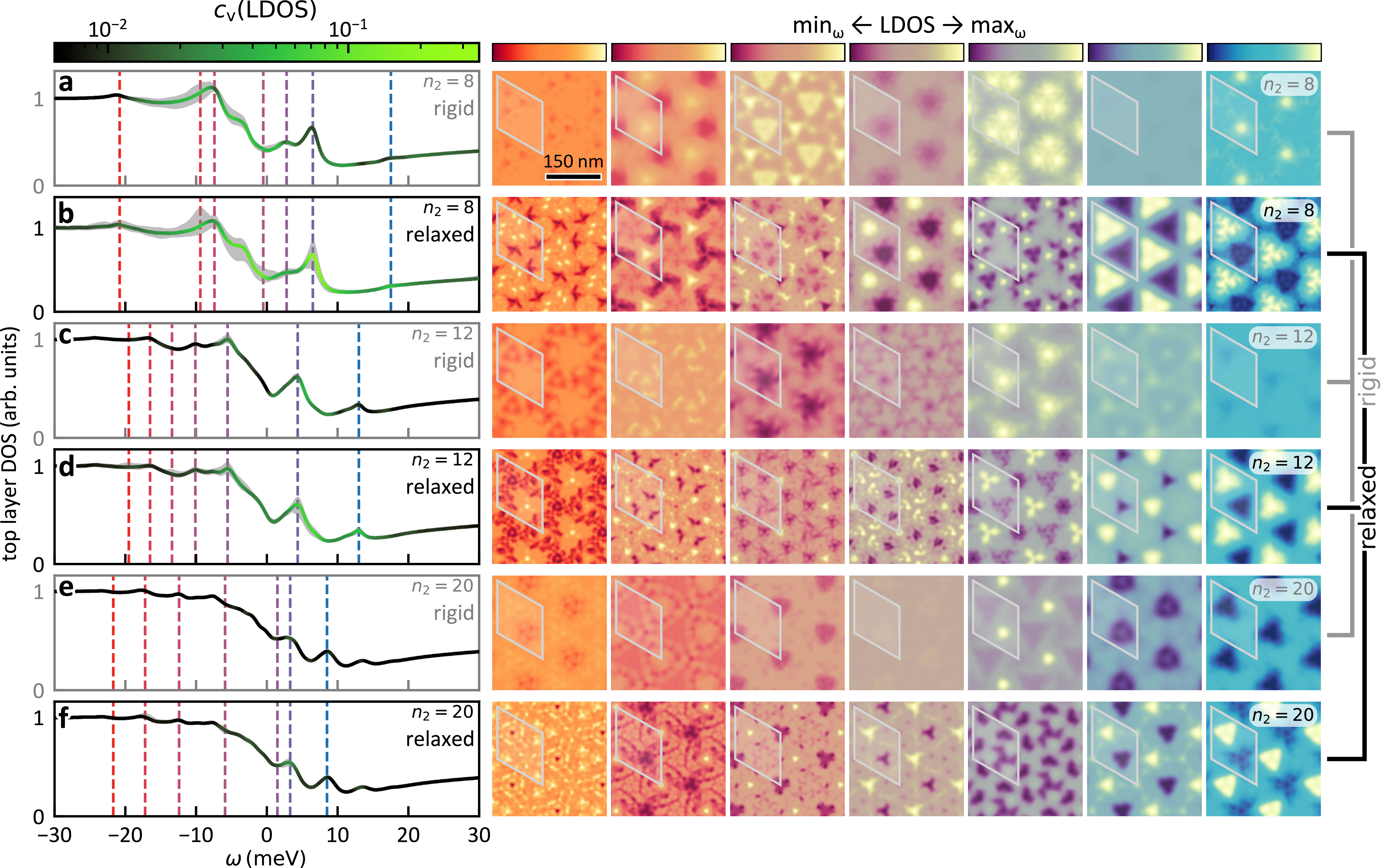}
    \caption{%
    \captiontitle{Electronic structure properties for graphite on graphite at twist angle of $\theta=0.1^\circ$.} Each row \captionlabel{a-f} corresponds to the same system where we vary the number of graphene layers in the top flake ($n_2$) and relaxed/rigid atomic structures across rows.
    Left subpanels: Frequency ($\omega$) resolved density of states (DOS) of the outermost layer of the top graphite flake. Panels~\captionlabel{a,c,e} represent simulations using rigid structures and panels~\captionlabel{b,d,f} relaxed structures. The number of layers of the bottom flake is fixed at $n_1=20$. We color-code the coefficient of variation ($c_\mathrm{v}=\sigma/\mu$) of the local density of states (LDOS) in the top layer as a function of $\omega$. Light green indicates strong spatial variations in the top layer LDOS, whereas black corresponds to a uniform LDOS. The gray area in the background of the curves indicates the minimum and maximum values of the LDOS.
    Right subpanels: Spatial distribution of the LDOS in the outermost layer at a given frequency. The frequencies are indicated in the left subpanels as dashed, colored vertical lines varying from red (low frequencies) to blue (high frequencies). The color-maps for the LDOS maps are chosen to be consistent with this coloring. Light (yellow) regions correspond to high intensity and dark (red $\rightarrow$ blue) regions to low intensity. For each configuration, the rigid (gray) and relaxed (black) panels in two subsequent rows share a color scale to make differences visible. The system's unit cell is indicated in each of the small subpanels. All plots on the right hand side share the scale bar plotted in the top left subpanel.}
    \label{fig:continuum-modeling}
\end{figure*}

The left subpanels in \cref{fig:continuum-modeling} present the integrated (over space) top layer density of states (DOS) as a function of energy $\omega$. Over all panels \captionlabel{a-f}, the top flake layer number $n_2$ is varied from $n_2=8$ to $n_2=20$ (the $n_2=3$ to $n_2=6$ cases are shown in \cref{app:large-angle-electronic-structure}). We subdivide the rigid \captionlabel{a,c,e} and the relaxed \captionlabel{b,d,f} cases. To highlight the spatial variations, we color the curves with the coefficient of variation of the top layer LDOS (over space) that is defined as
\begin{align}
    c_\mathrm{v}(l,\omega) &{}= \frac{
    {\mathrm{std}}_{\bvec r}\big(\mathrm{LDOS}(l,\omega,\bvec r)\big)
    }{
    {\mathrm{mean}}_{\bvec r}\big(\mathrm{LDOS}(l,\omega,\bvec r)\big) } \,.
\end{align}
Additionally, we add a gray background that corresponds to the minimum/maximum values taken by the LDOS of the outermost layer. By the DOS  and its spatial variations we can identify regions in $\omega$ (for each $n_2$) where the effect of relaxation is most pronounced. We highlight these regions by vertical dashed lines colored from red to blue. The right subpanels in each row \captionlabel{a-f} show the corresponding spatial map of the outermost layer LDOS for these specific $\omega$ and $n_2$ values. Note that two subsequent rows [e.g.~\captionlabel{a} and~\captionlabel{b}] share a color-map for each $\omega$ and $n_2$ such that differences from the relaxed to the rigid case become visible. The color-maps of the LDOS maps are chosen to correspond to the colors indicating $\omega$ in the DOS plot on the left. We provide the LDOS maps for all frequencies as videos in the supplemental material~\cite{supplement}. For $n_2=8$ and $n_2=12$, the LDOS maps for rigid atomic structures \captionlabel{a,c} clearly show less variation than the ones for relaxed atomic structures \captionlabel{b,d} for all values of $\omega$ chosen. Likewise, the $c_\mathrm{v}$ value is much larger over a substantial range of frequencies [see left subpanels for \captionlabel{a-d}]. The local density of states shows strong differences between the rigid and relaxed calculations in these cases. At a twist angle of $\theta=0.1^\circ$ this highlights the relevance of relaxation effects for a description of the electronic properties. At $n_2=20$, over a broad range of frequencies the value of $c_\mathrm{v}$ is minor in both the relaxed \captionlabel{f} and rigid \captionlabel{e} case. Only at around $\omega\approx3\unit{meV}$ there is a slightly more pronounced spatial variation showing that for the twist angle of $\theta=0.1^\circ$ the effects of relaxation do not penetrate through $n_2=20$ layers in accord with the analysis of $s(k)$ above suggesting $L_\mathrm{relax.}\sim5$ layers.

\section{Outlook}
Our work highlights the effects of multi-layered relaxation in various vdW heterostructures. We have shown experimentally using PFM that for multi-layered graphene stacks with a twisted interface the atomic displacement can propagate deep into the flake for small twist angles [see \cref{fig:experimental-data}~\captionlabel{a-f}]. We complement this experimental observation of the relevance of relaxation in twisted vdW heterostructures with an untwisted mismatched system: multi-layered PdTe\textsubscript{2} on a Bi\textsubscript{2}Se\textsubscript{3} substrate. In this system, multi-layered relaxation is of great importance as well and it becomes distinctively visible as a change in lattice constant as the number of PdTe\textsubscript{2} layers is increased [cf. \cref{fig:experimental-data}~\captionlabel{g,h}]. We thus establish that multi-layered relaxation is highly relevant for the engineering of stacked vdW heterostructures in the twisted and non-twisted context. On the one hand this emphasizes the need of deepened understanding of relaxation and a refinement of the lego-like picture often employed in the field~\cite{geim2013van}. On the other hand, it adds yet another tuning knob to the thriving field of vdW heterostructure engineering. As exemplified here for PdTe\textsubscript{2} on a Bi\textsubscript{2}Se\textsubscript{3} substrate, the effective lattice constant can be tuned over a rather versatile range by multi-layered relaxation effects.

In tandem, we develop a machinery that allows us to simulate and understand these relaxation properties given the generalized stacking fault energy functional from \emph{ab-initio} calculations. Within this approach, we treat each layer of the heterostructure stack as a separate membrane. Our simulations fully support the experimental evidence of three dimensional, layered relaxation patterns in both systems: First, we reproduce the stacking height dependent lattice constant evolution in a stack of PdTe\textsubscript{2} on Bi\textsubscript{2}Se\textsubscript{3} (\cref{fig:PdTe2_theory}). Second, we observe that the penetration depth of relaxation domains in twisted graphite stacks linearly increases with moir\'e wavelength (\cref{fig:multilayered-relaxation}). These two examples emphasize how the three dimensional nature of multi-layered heterostructures needs to be taken into account when modeling their atomic structure.

Ultimately we predict differences in electronic properties arising from the atomic relaxation in multi-layered graphene twisted on graphite. Similarly to the notion of penetration depth of the relaxation pattern, we see that the local density of states in the outermost layer of the graphene stack sensitively depends on whether the atomic positions of the crystal are relaxed or not. For a small twist angle ($\theta=0.1^\circ$), we demonstrate that the local variation of the electron density in the outermost graphene layer is much stronger when the multi-layered relaxation pattern is taken into account properly (\cref{fig:continuum-modeling}). We propose to analyze such multi-layered (small angle) twisted graphene on graphite stacks using scanning tunneling microscopy in the future, as the local electronic density as a function of frequency can be used to probe the here predicted effects of relaxation on the electronic structure in precise manner. The strong effects of relaxation on the electronic properties of the stack highlight that also these properties, including their collective behavior, can be engineered by the relaxation of stacked vdW heterostructure even beyond the few layer paradigm.

\begin{acknowledgments}
Nano-imaging research at Columbia is supported by DOE-BES grant DE-SC0018426. Research on atomic relaxation is supported by W911NF2120147. We  acknowledge  funding by the Deutsche Forschungsgemeinschaft (DFG, German Research Foundation) under RTG 1995 and RTG 2247, within the Priority Program SPP 2244 ``2DMP'', under Germany's Excellence Strategy - Cluster of Excellence Matter and Light for Quantum Computing (ML4Q) EXC 2004/1 - 390534769 and - Cluster  of  Excellence and Advanced Imaging of Matter (AIM) EXC 2056 - 390715994. We acknowledge computational resources provided by the Max Planck Computing and Data Facility and RWTH Aachen University under project number rwth0716. This work was supported by the Max Planck-New York City Center for Nonequilibrium Quantum Phenomena. DNB is Moore Investigator in Quantum Materials EPIQS GBMF9455. DH was supported by a grant from the Simons Foundation (579913).
We thank Oxford Instruments Asylum Research for performing the large scale tapping mode scan presented in \cref{App:graphene_exp} (\cref{fig:graphene-afm}).
\end{acknowledgments}

\appendix

\section{Graphene experiment}
\label{App:graphene_exp}

\begin{figure}
    \centering
    \includegraphics[width=\columnwidth]{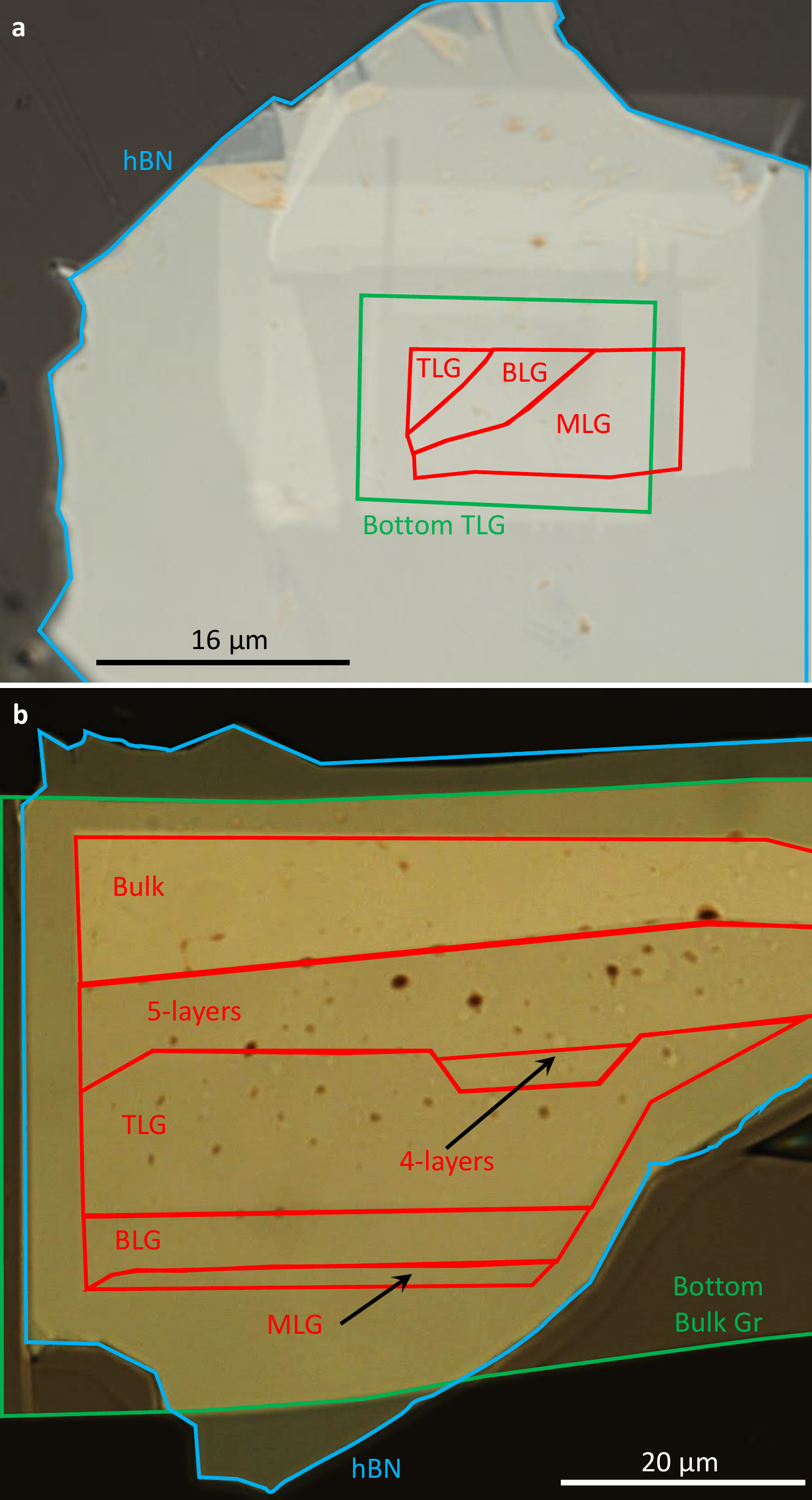}
    \caption{\captiontitle{Optical images of the studied terraced graphene samples.}  \captionlabel{a} The sample measured in \cref{fig:experimental-data}\captionlabel{a-c} (twist angle of $\unsim0.66^\circ$) with hBN, bottom trilayer graphene, and top mono/bi/tri-layer graphene regions indicated by blue, green and red contours respectively. \captionlabel{b} The sample measured in \cref{fig:experimental-data}\captionlabel{d-f} (twist angle of approximately $0.1^\circ$) with regions indicated by similar colors. The top layer has bulk (estimated to be 18 layers), 5-layers, 4-layers, TLG, BLG and MLG regions as indicated.}
    \label{fig:graphene-optical}
\end{figure}

\begin{figure}
    \centering
    \includegraphics[width=\columnwidth]{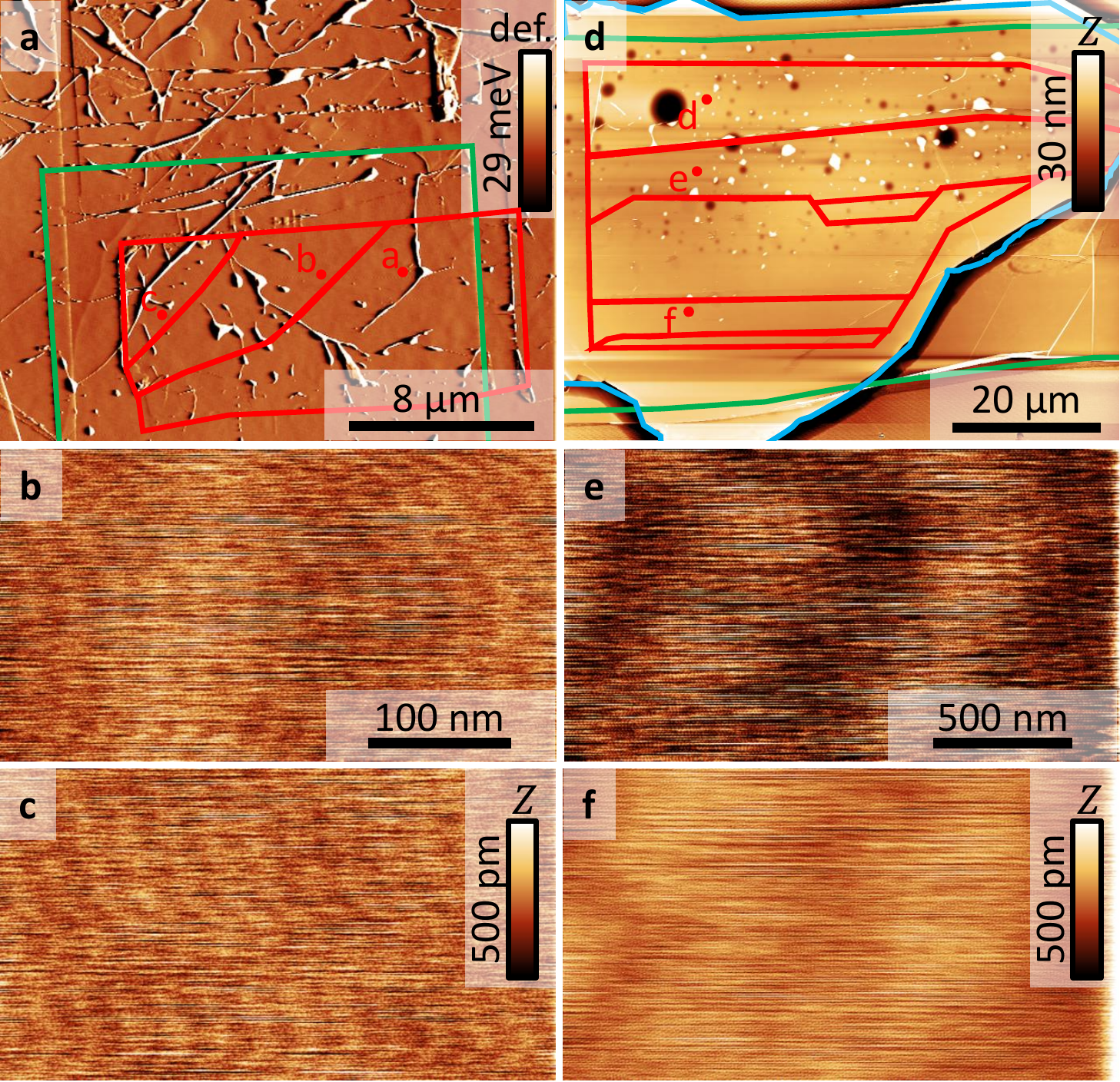}
    \caption{\captiontitle{Topography maps of the studied terraced graphene samples.} \captionlabel{a} Topography map of the sample measured in \cref{fig:experimental-data}\captionlabel{a-c}, as reflected by the deflection error channel in contact mode. The layer markings are as in \cref{fig:graphene-optical}\captionlabel{a}. \captionlabel{b-c} Topography maps in contact mode taken simultaneously with \cref{fig:experimental-data}\captionlabel{a-b} respectively (\captionlabel{b-c} share a color-scheme and a scale-bar).
    \captionlabel{d} Topography map of the sample measured in \cref{fig:experimental-data}\captionlabel{d-f}, as reflected by tapping mode scan. The layer markings are as in \cref{fig:graphene-optical}\captionlabel{b}. \captionlabel{e-f} Topography maps in contact mode taken simultaneously with \cref{fig:experimental-data}\captionlabel{d-e} respectively (\captionlabel{e-f} share a color-scheme and a scale-bar). The locations of the scans of \cref{fig:experimental-data}\captionlabel{a-f} are marked over \captionlabel{a-b}.}
    \label{fig:graphene-afm}
\end{figure}

Using an optical microscope, we identify exfoliated graphene flakes with tiered graphene thicknesses in close proximity to a bulk graphite flake from the same crystal. Samples are then assembled using a dry transfer technique using a slide with polycarbonate (PC) film on top of a polydimethyl siloxane (PDMS) dome~\cite{Wang2013b}. We use the slide to first pick up a BN flake of thickness $30-40\unit{nm}$, which is then placed into contact with the bulk graphite flake. After picking up the bulk graphite, we then rotationally misalign the transfer stage by a small angle of less than 1 degree before picking up the tiered graphene flake. The completed heterostructure remains on the polymer stamp slide during subsequent experimental measurements.
\Cref{fig:graphene-optical} presents optical images of the completed heterostructures studied in \cref{fig:experimental-data}. Topography data of these samples are shown in \cref{fig:graphene-afm}. While the large scale topography maps of \cref{fig:graphene-afm}\captionlabel{a,d} indicate the existence of bubbles and wrinkles in the structures, the samples have an abundance of flat regions that were meticulously studied. \cref{fig:graphene-afm}\captionlabel{a} was acquired with a Bruker Dimension Icon with a Nanoscope V Controller, and \cref{fig:graphene-afm}\captionlabel{d} was acquired by a Jupiter XR AFM by Oxford Instruments Asylum Research.
We use piezoresponse force microscopy (PFM) to image the moir\'e superlattice between the rotationally misaligned tiered graphene flake and the bulk graphite~\cite{McGilly2020c}. In PFM, we apply an AC bias between the conductive probe and the microscope stage
(on which the sample is directly mounted, giving an approximate distance of about $500\,\mathrm{\mu m}$ between biased tip and ground),
which induces a periodic deformation of the sample through the piezoelectric effect. 
Even though the sample is not contacted, and thus one may raise the concern of its potential being ill-defined, empirically the results of PFM studies do not seem to be sensitive to the sample being properly grounded. Apparently, due to the proximity to a grounded plane, the stage still provides a good dc-reference for the oscillating fast perturbations made by the metallic
tip (at hundreds of $\mathrm{kHz}$).
The phase and amplitude of the local deformation provide the strongest contrast to visualize moir\'e in our samples. AC bias magnitudes were $500-1000\unit{mV}$ with resonance frequencies in the range of $270-300\unit{kHz}$ for vertical and $750-850\unit{kHz}$ for lateral PFM. The experiments were performed in a Bruker Dimension Icon with a Nanoscope V Controller, using Oxford Instrument Asylum Research ASYELEC-01 Ti/Ir coated silicon probes.
While the PFM signal is commonly interpreted as a measure of local deformation resulting from the electric field exerted by the tip, i.e.\ a local piezoelectric or flexoelectric effect (see Ref.~\onlinecite{McGilly2020c}) the exact physical mechanism responsible for the contrast is not necessarily purely of such origin. When looking at the contact mode scans of \cref{fig:graphene-afm}\captionlabel{b-c,e-f}, taken simultaneously with those of \cref{fig:experimental-data}\captionlabel{a-b,d-e} respectively, one may notice faint periodic structure, which may reflect some topographic features of the moir\'e lattice. Regardless of the exact mechanism for the PFM contrast, we consider it to be correlated with the strength of the relaxation at the sample surface.

The PFM scans of twenty different locations in the $\unsim0.66^\circ$ sample revealed twist angles ranging from $0.63^\circ$ to $0.8^\circ$, based upon the moir\'e wavelength. For the $\unsim0.1^\circ$ sample, we took scans in forty-eight different locations, showing a rather uniform twist angle of $0.11^\circ \pm 0.02^\circ$ through the two to five top layer graphene and bulk top layer graphene regions. In both samples the thinner top-layers regions (MLG and to some extent the BLG as well) were more susceptible to strain, as indicated by the moir\'e superlattice.

One may wonder whether the observed moir\'e patterns may actually originate from hBN/graphene moir\'e interface. For the case of the single, bi- and trilayer graphene twisted at an angle of $0.66^\circ$ (cf.\ \cref{fig:experimental-data}\captionlabel{a-c}), we note that we can exclude a potential moir\'e lattice formed by aligned hBN and the bottom multi-layered graphene as a source of PFM contrast due to the different length scales involved. At maximum, hBN and graphene form a moir\'e lattice with $\lambda_\mathrm{hBN}^\mathrm{max}\approx14\unit{nm}$~\cite{moon2014electronic}, whereas the moir\'e wavelength observed experimentally is $\lambda_\mathrm{graphene}=a_\mathrm{graphene}/ (2\sin(\theta/2))\approx18-22\unit{nm}$, which is significantly larger than $14\unit{nm}$; and therefore corresponds to the mentioned graphene-graphene moir\'e lattice with twist angle of $\theta=0.66^\circ$.

\section{PdTe$_2$/Bi$_2$Se$_3$ experiment}
\label{App:PdTe2_exp}
The PdTe$_2$ thin films were grown on a Bi$_2$Se$_3$ substrate in an integrated MBE-STM ultrahigh vacuum (UHV) system with base pressure below $2\times10^{-10}\unit{mbar}$. The Bi$_2$Se$_3$ was prepared by in-situ cleaving the surface and subsequent annealing to $250\unit{^{\circ}C}$ for one hour to degas. Then, high-purity Pd (99.95\%) and Te (99.9999\%) were evaporated from an electron-beam evaporator and a standard Knudsen cell, respectively, with a flux ratio of 1:10. The deposition rate of Pd and Te atoms was monitored by a quartz oscillator. The temperature of substrate was kept at $210\unit{^{\circ}C}$ during the growth.

After the initial annealing of the Bi$_2$Se$_3$, the substrate was transferred in-situ to the Aarhus STM stage with a Tungsten tip. STM measurements were taken of the substrate to ensure the quality of the substrate. Once deposition of the Pd and Te was complete, the sample was transferred in-situ to the STM stage for surface topography mapping.

\section{Justifying assumptions of in-plane relaxation model: Estimating curvature contributions to the elastic energy}
\label{app:curvature}
\begin{figure}
    \centering
    \includegraphics[width=\columnwidth]{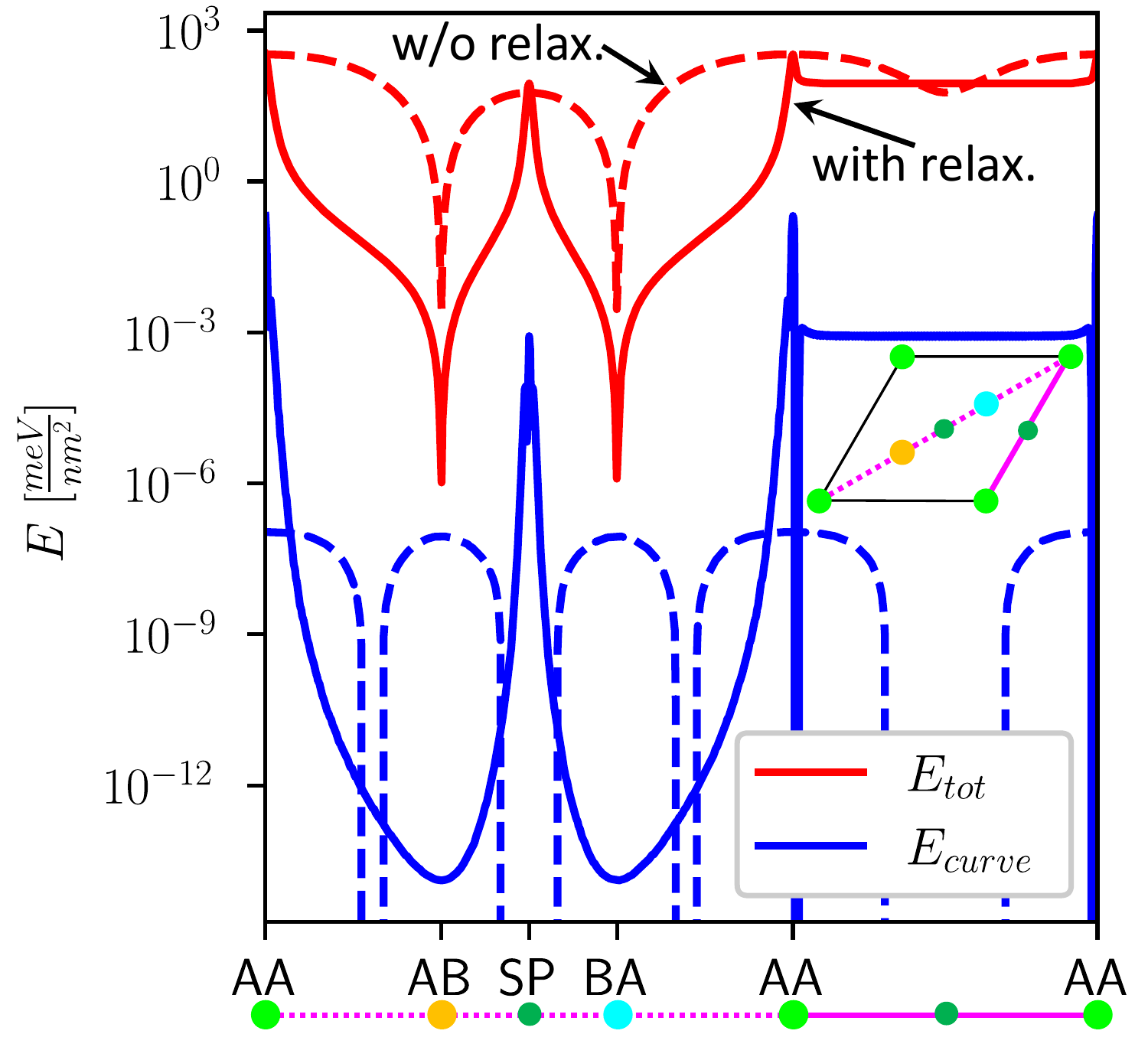}
    \caption{\captiontitle{Effects of curvature on atomic relaxation process in twisted bilayer graphene.} Comparing total energy density (red) to the elastic energy term resulting from membrane bending (blue - more details in the text) before and after relaxation (dashed and solid lines respectively) as a function of original (before relaxation) stacking configuration (on a path in configuration space indicated in the inset). The out-of-plane displacement was taken to be the layer spacing at the local configuration as presented in Table III of Ref.~\cite{Zhou2015e} (and independently corroborated).}
    \label{fig:curvature-effect}
\end{figure}

While our atomic relaxation model is restricted to in-plane relaxation, we argue that the most important contributions of out-of-plane strain are still captured. During the DFT calculations of the GSFE function, the inter-layer spacing is relaxed separately for each stacking configuration. The result is a map of the GSFE function $V_\mathrm{GSFE}\left(v, w\right)$, but also a stacking dependent inter-layer spacing function $d\left(v, w\right)$. The only effect that is omitted in this calculation is the contribution to the elastic energy emanating from the bending of the membrane. The bending energy can be written as $E_b=\int\mathrm{d}^2r\left(\frac{1}{2}\kappa H^2+\kappa_G K\right)$ \cite{Chen2011}, where $H$ and $K$ are the mean and Gaussian curvatures respectively and their respective rigidities $\kappa$ and $\kappa_G$. Similar to Ref.~\onlinecite{Chen2011} we approximate $H$ and $K$ as $H\approx \nabla^2f$ and $K\approx\det\left(\partial_i\partial_j f\right)$, where $f$ is the out of plane strain of the membrane. For the sake of the estimate we assume $f\approx d$, namely, that the shape of the membrane is dictated by the local stacking configuration. Only if the curvature related terms of the elastic energy density are significant relative to the total energy density at a given stacking configuration can one argue that this effect is of importance in the atomic relaxation process. \cref{fig:curvature-effect} compares the curvature terms using the above assumptions with the total energy omitting this effect, before and after relaxation. It is clear that the curvature contribution is secondary, therefore we omitted these terms, to greatly simply the model. However, there is no fundamental problem preventing from including this effect in the model. In the estimate of \cref{fig:curvature-effect} we took $\kappa=\kappa_G=10\unit{eV}$ following the calculations of Ref.~\onlinecite{Lai2016} and the material parameters for graphene detailed in \cref{tab:material_param}.

\section{Material parameters used for the relaxation calculations}
\label{app:material-parameters}

The material parameters used for the multilayer graphene relaxations, consisting of the GSFE coefficients $c_{0,\dots,5}$ and strain (bulk and shear) moduli $K$ and $G$, are taken from Ref.~\onlinecite{Carr2018}. For the PdTe$_2$/Bi$_2$Se$_3$ multilayer simulations, the three interfacial GSFE functionals, strain moduli, and monolayer lattice parameters were extracted from density functional calculations performed with the Vienna Ab initio Simulation Package (VASP)~\cite{Kresse1996}. A monolayer geometry was used to calculate the $\alpha$, $K$, and $G$ values for both materials, and a two-layer geometry was used for the three different GSFEs (PdTe$_2$/PdTe$_2$, PdTe$_2$/Bi$_2$Se$_3$, and Bi$_2$Se$_3$/Bi$_2$Se$_3$). The numerical values of all material parameters relevant for relaxing the atomic structures are given in \cref{tab:material_param}. In all cases, the out-of-plane $c$ axis (parallel to $z$) was set to $30\unit{\text{\AA}}$ to ensure no interaction between the periodic slabs. The van der Waals energy functional SCAN+rVV10~\cite{Peng2016} was used alongside PAW-PBE pseudo potentials for all atoms~\cite{Kresse1999}. A plane wave energy cutoff of $300\unit{eV}$ and electronic convergence of $10^{-5}\unit{eV}$ were used, with dipole corrections turned on for the out-of-plane ($c$, or $z$) axis. Optimization of the lattice constant for the monolayer, and out-of-plane relaxation of the atoms for the interfacial calculations, were performed via conjugate gradient descent with a $10^{-4}\unit{eV}$ convergence criterion.

\begin{table}[ht]
\centering
\caption{Material parameters for the PdTe$_2$/Bi$_2$Se$_3$ and graphene based heterostructures used for the relaxation calculations in this work. $c_{0,\dots,5}$ are the GSFE coefficients of \cref{eqn:Vgsfe} for the PdTe$_2$/PdTe$_2$, Bi$_2$Se$_3$/Bi$_2$Se$_3$, PdTe$_2$/Bi$_2$Se$_3$ and the graphene/graphene interfaces (u.c. stands for the area of the 2D unit cell). $\alpha$ are the lattice constants for PdTe$_2$, Bi$_2$Se$_3$ and graphene. $K$ and $G$ stand for the bulk and shear moduli. }
\label{tab:material_param}
\vspace{0.5em}
\begin{ruledtabular}
\def\arraystretch{1.5}
\newcommand{\mevuc}{\frac{\mathrm{meV}}{\mathrm{u.c.}}}
\newcommand{\sscr}[1]{\textsubscript{#1}}
\begin{tabular}{c|cccc}
    & \bfseries PdTe\sscr{2} & \bfseries Bi\sscr{2}Se\sscr{3} & \bfseries PdTe\sscr{2}/Bi\sscr{2}Se\sscr{3} & \bfseries Graphene \\\hline
    $c_0 \, [\mevuc]$ & \, 157.73 &\, 59.19 &\, 102.65 &\, 6.832 \\\hline
    $c_1 \, [\mevuc]$ & \, -41.10 &\,  22.80 &\, 28.21 &\, 4.064 \\\hline
    $c_2 \, [\mevuc]$ & \, -6.25 &\, -4.76 &\, -6.47 &\,  -0.374 \\\hline
    $c_3 \, [\mevuc]$ & \, -5.23 &\, -0.21 &\, -1.66 &\, -0.095 \\\hline
    $c_4 \, [\mevuc]$ & \, 5.63 &\, 1.71 &\, 11.17 &\, 0 \\\hline
    $c_5 \, [\mevuc]$ & \, 1.18 &\, 2.53 &\, 5.54 &\, 0 \\\hline
    $\alpha \, [\text{\AA}]$ & \, 4.014 &\, 4.129 &\, -- &\, 2.47 \\\hline
    $K \, [\mevuc]$ & \, 30902 &\, 47049 &\, -- &\, 69518 \\\hline
    $G \, [\mevuc]$ & \, 17759 &\, 25995 &\, -- &\, 47352
\end{tabular}
\end{ruledtabular}
\end{table}

\section{LDOS from continuum model}
\begin{figure*}
    \centering
    \includegraphics[width=0.95\textwidth]{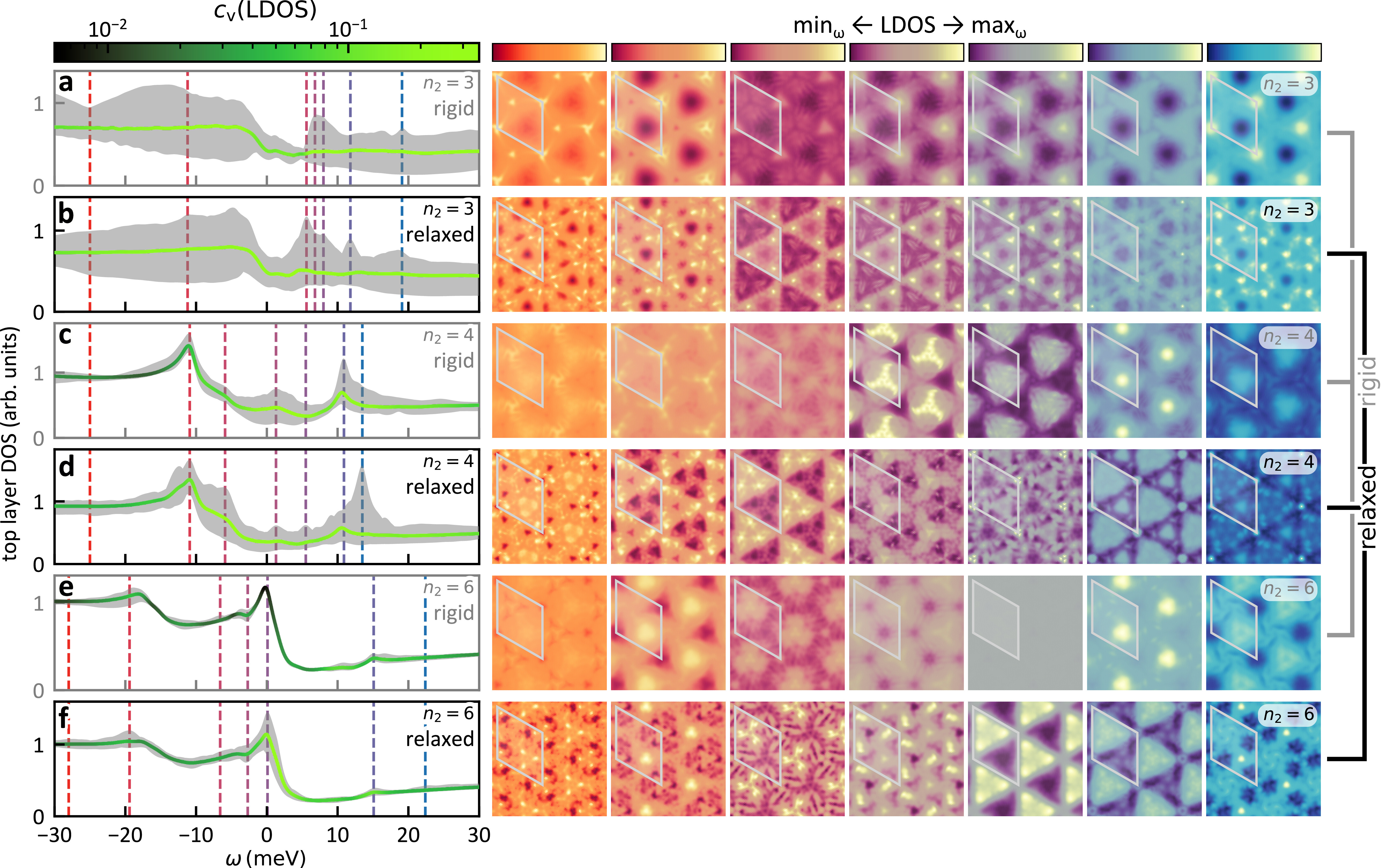}
    \caption{\captiontitle{Small angle ($\theta=0.1^\circ$) electronic structure of few-layer graphene on graphite.} The number of layers of the first graphite flake is set to $n_1=20$, whereas the number of layers of the second graphite flake is varied from $n_2=3$~\captionlabel{a,b} over $n_2=4$~\captionlabel{c,d} to $n_2=6$~\captionlabel{e,f}. In panels (rows)~\captionlabel{a,c,e} we used rigid atomic positions and in~\captionlabel{b,d,f} relaxed ones. The color of the lines in the left subpanels codes the coefficient of variation of the top layer LDOS, with green as non-uniform regions and black as uniform regions. The gray background is given by the minimum/maximum values of the LDOS. On the right, we show LDOS maps at selected frequencies indicated in the left subpanels. Each subsequent two rows [e.g.~\captionlabel{a} and~\captionlabel{b}] share a color-map for each frequency such that differences in relaxed and rigid structures are apparent.}
    \label{fig:continuum-modeling-small-add}
\end{figure*}
Spectral densities generally can be defined in an arbitrary basis of the Hamiltonian. The layer-resolved local density of states, i.e. $\mathrm{LDOS}(l,\omega,\bvec r)$, is nothing else but the spectral density in a basis consisting of real-space-, layer-, sublattice-, valley- and momentum-indices. The sublattice ($x$), valley and momentum degrees of freedom are summed over to obtain a function of $l$ and $\bvec r$, and frequency $\omega$. As our model \cref{eqn:full-continuum} is forumlated in reciprocal moir\'e lattice vector ($\bvec G$) space, we obtain the following Green's function:
\begin{align}
    \label{eqn:GF-spectral-basis}
    G^{\bvec G\bvec G',\xi,ll',xx'}(i\omega,\bvec k) &{}= \sum_b \frac{u_{\bvec G,\xi,l,x}^b(\bvec k) \big[u_{\bvec G',\xi,l',x'}^{b}(\bvec k)\big]^*}{i\omega-\epsilon_\xi^{b}(\bvec k)} \,,
\end{align}
with $b$ a band index and $\epsilon_\xi^b(\bvec k)$ the dispersion of band $b$ of valley $\xi$ at momentum $\bvec k$. Note that, since the Hamiltonian is diagonal in the valleys $\xi=\pm1$, the Green's function carries only one valley index as well. For a real-space description, we need to transform the $\bvec G$ indices to real-space:
\begin{multline}
    \label{eqn:GF-transform-basis}
    G^{\bvec r\bvec r',\xi,ll',xx'}(i\omega,\bvec k) = \frac1{N_{\bvec G}}\sum_{\bvec G\bvec G'} e^{-i(\bvec G\cdot\bvec r-\bvec G'\cdot\bvec r')} \\
    \times G^{\bvec G\bvec G',\xi,ll',xx'}(i\omega,\bvec k)  \,.
\end{multline}
Here, the number of reciprocal moir\'e lattice vectors $N_{\bvec G}$ is needed for proper normalization.
The representation in a diagonal basis, \cref{eqn:GF-spectral-basis}, numerically facilitates the Fourier transform: Instead of transforming both indices in a double sum [cf.~\cref{eqn:GF-transform-basis}], we can directly transform the ``Bloch"-functions $u_{\bvec G,\xi,l,x}^b(\bvec k)$ to real-space:
\begin{align}
    u_{\bvec r,\xi,l,x}^b(\bvec k) &{}= \frac1{\sqrt{N_{\bvec G}}}\sum_{\bvec G} e^{-i\bvec G\cdot\bvec r}\,u_{\bvec G,\xi,l,x}^b(\bvec k) \,.
\end{align}

Then we perform the analytic continuation $i\omega\rightarrow\omega+i\eta$ and trace out the sublattice-, valley- and momentum-indices to arrive at
\begin{align}
\mathrm{LDOS}(l,\omega,\bvec r) &{}= -\Im\sum_{\xi,x,\bvec k,b}\frac{|u_{\bvec r,\xi,l,x}^b(\bvec k)|^2}{\epsilon_\xi^b(\bvec k)-\omega+i\eta}\,.
\end{align}
The broadening parameter $\eta$ determines the energy resolution. We set $\eta = 5\,\mathrm{meV}$ for all simulations of the $\theta=0.8^\circ$ systems and $\eta=0.8\,\mathrm{meV}$ for the $\theta=0.1^\circ$ case. The momentum summation is carried out on an equi-spaced mesh with $14\times14$ points in a $C_3$ reduced wedge of the Brillouin zone.

For numerical calculations of the system's electronic properties we employ specialized code operating on a hybrid CPU/multi-GPU (OpenMP/CUDA) architecture. Convergence of the inverse moir\'e lattice vector expansion dictates large matrix sizes for the Hamiltonian. The expansion's cutoff $G_c$ is set by magnitude: $\|\bvec G\|<G_c$. In the case of $\theta=0.1^\circ$ we require $G_c=13\|\bvec B_{1/2}\|$ for convergence resulting in Hamilton matrix sizes of $\unsim 50,\!000$ (for $n_1=n_2=20$). In the large angle case ($\theta=0.8^\circ$) we set $G_c=5\|\bvec B_{1/2}\|$ to converge the expansion.

\section{Additional Electronic Structure Data}
\label{app:large-angle-electronic-structure}
\subsection{Small angle case with few-layered graphene}
\label{app:small-angle-add}
We here present the additional cases of $n_2=3$, $n_2=4$, and $n_2=6$ for the small angle ($\theta=0.1^\circ$) system with $n_1=20$. \Cref{fig:continuum-modeling-small-add} displays the results in the same manner as \cref{fig:continuum-modeling} in the main text.

\begin{figure*}
    \centering
    \includegraphics[width=0.95\textwidth]{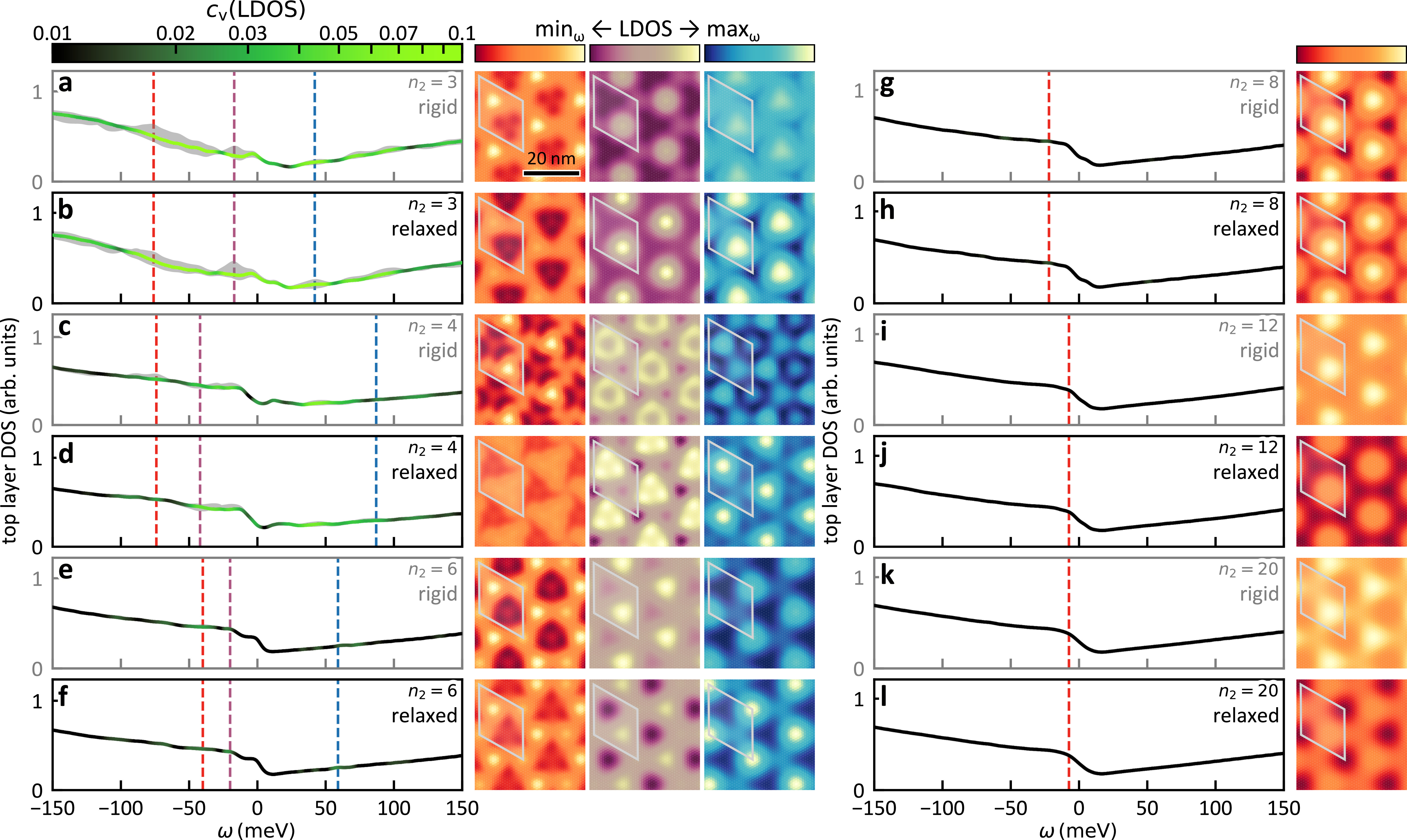}
    \caption{\captiontitle{Overview over electronic structure of many-layer graphene twisted at $\theta=0.8^\circ$ on graphite system} Similarly to \cref{fig:continuum-modeling,fig:continuum-modeling-small-add}, we display the integrated top layer DOS and additionally color-code the coefficient of variation $c_\mathrm{v}$. Furthermore, we mark the minimum/maximum values taken by the top layer LDOS as gray background. The system considered here also consists of $n_1=20$ layers of graphene as a bottom flake and $n_2=3,\dots,20$ layers of graphene as a top flake.
    In Panels~\captionlabel{a-f}, we vary $n_2$ from $3$ over $4$ to $6$, with subsequent rows corresponding to the relaxed and rigid cases of the same system. The subpanels on the right show the LDOS maps at selected frequencies indicated as vertical dashed lines in the left subpanels. The color-maps correspond to the vertical dashed lines' colors. Two subsequent rows [e.g.~\captionlabel{a} and~\captionlabel{b}] share a color-map for each frequency to make differences apparent.
    \captionlabel{g-l}~Same as panels~\captionlabel{a-f}, but for $n_2=8$, $n_2=12$, and $n_2=20$. As for these systems local variations in the electronic density of the top layer are very minuscule, we only show the LDOS maps at a single frequency. Apart from a small shift in absolute scale, these LDOS maps are very similar between each relaxed and rigid case in panels~\captionlabel{g-l}.}
    \label{fig:continuum-modeling-large}
\end{figure*}

The six panels \captionlabel{a-f} correspond to the LDOS data of rigid/relaxed systems at $n_2=3$, $n_2=4$, and $n_2=6$. The left subpanel shows the spatially integrated LDOS of the outermost layer with color coding as the coefficient of variation (over space). Here, black corresponds to small variation, i.e. a uniform LDOS and green corresponds to large variation, i.e. a spatially non-uniform LDOS. Additionally, we add the minimum/maximum values taken by the top layer LDOS as gray background. At selected frequencies, we show the explicit spacial distribution of the LDOS for each system on the right hand side in the same row. The color-maps of the LDOS maps are normalized for the rigid and relaxed cases of a single $n_2$, i.e. subsequent rows [e.g. panels~\captionlabel{a} and~\captionlabel{b}] share a color-map normalization. The color-scheme of the LDOS maps evolves as a function of frequency to match the color of the vertical dashed lines indicating the exact frequency values. Even in the few-layer (small angle) data in \cref{fig:continuum-modeling-small-add}, we observe the tendency that, when increasing $n_2$, the local variations in top layer LDOS are strongly enhanced for the relaxed case.

\subsection{Large angle case}
To highlight the importance of small twist angles for large penetration depths of relaxation effects into the stack, we here present the analysis of the electronic structure for a twist angle of $\theta=0.8^\circ$. As in \cref{sec:impact-electronic} and \cref{app:small-angle-add}, the number of graphene layers in the bottom flake is fixed at $n_1=20$ and we vary the number of graphene layers in the top flake from $n_2=3$ to $n_2=20$. The results are presented in \Cref{fig:continuum-modeling-large}. Overall, we observe that patterns in the LDOS of the top layer decay much faster as a function of increasing $n_2$. For example, even the $n_2=6$ and $n_2=8$ cases display no notable difference in $c_\mathrm{v}$ of the relaxed and rigid structures. For $n_2\geq8$, the value of $c_\mathrm{v}$ is too small to be visible on the selected scale (less than $1\%$). Furthermore, no such clear pattern of increased $c_\mathrm{v}$ in the relaxed case over the rigid case can be found as for the small angle ($\theta=0.1^\circ$) case. We interpret this behaviour as the relaxation decaying as fast -- or faster than -- the effects that the moir\'e interlayer coupling $U^{\bvec G,\bvec G'}$ has on the top layer electronic structure.

\subsection{Full frequency LDOS maps}
\label{app:full-freq-ldos-maps}
As supplemental material \cite{supplement}, we provide all LDOS maps for the systems studied within this work as videos where we change $\omega$ with each frame. The color-maps evolve as a function of $\omega$ similar to the color-maps of the LDOS maps in \cref{fig:continuum-modeling,fig:continuum-modeling-small-add,fig:continuum-modeling-large}. The upper panel corresponds to the rigid structure and the lower panel to the relaxed one. In each frame, we normalize the color-map separately with bright (yellow) colors indicating maxima and dark (red/blue) colors indicating minima. To display only significant LDOS variations in the movies, we set the minimum relative difference that needs to be covered by the color-map to $\pm2\%$ such that when the minimum and maximum values are closer than $2\%$ to their mean value, we artificially increase the color-mapped minimum and maximum values to be $2\%$ away from the mean value.

\bibliography{references}

\providecommand{\noopsort}[1]{}\providecommand{\singleletter}[1]{#1}%
\begin{thebibliography}{97}%
\makeatletter
\providecommand \@ifxundefined [1]{%
 \@ifx{#1\undefined}
}%
\providecommand \@ifnum [1]{%
 \ifnum #1\expandafter \@firstoftwo
 \else \expandafter \@secondoftwo
 \fi
}%
\providecommand \@ifx [1]{%
 \ifx #1\expandafter \@firstoftwo
 \else \expandafter \@secondoftwo
 \fi
}%
\providecommand \natexlab [1]{#1}%
\providecommand \enquote  [1]{``#1''}%
\providecommand \bibnamefont  [1]{#1}%
\providecommand \bibfnamefont [1]{#1}%
\providecommand \citenamefont [1]{#1}%
\providecommand \href@noop [0]{\@secondoftwo}%
\providecommand \href [0]{\begingroup \@sanitize@url \@href}%
\providecommand \@href[1]{\@@startlink{#1}\@@href}%
\providecommand \@@href[1]{\endgroup#1\@@endlink}%
\providecommand \@sanitize@url [0]{\catcode `\\12\catcode `\$12\catcode
  `\&12\catcode `\#12\catcode `\^12\catcode `\_12\catcode `\%12\relax}%
\providecommand \@@startlink[1]{}%
\providecommand \@@endlink[0]{}%
\providecommand \url  [0]{\begingroup\@sanitize@url \@url }%
\providecommand \@url [1]{\endgroup\@href {#1}{\urlprefix }}%
\providecommand \urlprefix  [0]{URL }%
\providecommand \Eprint [0]{\href }%
\providecommand \doibase [0]{https://doi.org/}%
\providecommand \selectlanguage [0]{\@gobble}%
\providecommand \bibinfo  [0]{\@secondoftwo}%
\providecommand \bibfield  [0]{\@secondoftwo}%
\providecommand \translation [1]{[#1]}%
\providecommand \BibitemOpen [0]{}%
\providecommand \bibitemStop [0]{}%
\providecommand \bibitemNoStop [0]{.\EOS\space}%
\providecommand \EOS [0]{\spacefactor3000\relax}%
\providecommand \BibitemShut  [1]{\csname bibitem#1\endcsname}%
\let\auto@bib@innerbib\@empty
\bibitem [{\citenamefont {Geim}\ and\ \citenamefont
  {Grigorieva}(2013)}]{geim2013van}%
  \BibitemOpen
  \bibfield  {author} {\bibinfo {author} {\bibfnamefont {A.~K.}\ \bibnamefont
  {Geim}}\ and\ \bibinfo {author} {\bibfnamefont {I.~V.}\ \bibnamefont
  {Grigorieva}},\ }\bibfield  {title} {\bibinfo {title} {Van der waals
  heterostructures},\ }\href@noop {} {\bibfield  {journal} {\bibinfo  {journal}
  {Nature}\ }\textbf {\bibinfo {volume} {499}},\ \bibinfo {pages} {419}
  (\bibinfo {year} {2013})}\BibitemShut {NoStop}%
\bibitem [{\citenamefont {Wang}\ \emph
  {et~al.}(2013{\natexlab{a}})\citenamefont {Wang}, \citenamefont {Jang},
  \citenamefont {Jang}, \citenamefont {Kim}, \citenamefont {Park},
  \citenamefont {Kim}, \citenamefont {Kahng}, \citenamefont {Choi},
  \citenamefont {Ruoff}, \citenamefont {Song},\ and\ \citenamefont
  {Lee}}]{wang2013platform}%
  \BibitemOpen
  \bibfield  {author} {\bibinfo {author} {\bibfnamefont {M.}~\bibnamefont
  {Wang}}, \bibinfo {author} {\bibfnamefont {S.~K.}\ \bibnamefont {Jang}},
  \bibinfo {author} {\bibfnamefont {W.-J.}\ \bibnamefont {Jang}}, \bibinfo
  {author} {\bibfnamefont {M.}~\bibnamefont {Kim}}, \bibinfo {author}
  {\bibfnamefont {S.-Y.}\ \bibnamefont {Park}}, \bibinfo {author}
  {\bibfnamefont {S.-W.}\ \bibnamefont {Kim}}, \bibinfo {author} {\bibfnamefont
  {S.-J.}\ \bibnamefont {Kahng}}, \bibinfo {author} {\bibfnamefont {J.-Y.}\
  \bibnamefont {Choi}}, \bibinfo {author} {\bibfnamefont {R.~S.}\ \bibnamefont
  {Ruoff}}, \bibinfo {author} {\bibfnamefont {Y.~J.}\ \bibnamefont {Song}},\
  and\ \bibinfo {author} {\bibfnamefont {S.}~\bibnamefont {Lee}},\ }\bibfield
  {title} {\bibinfo {title} {A platform for large-scale graphene electronics
  – cvd growth of single-layer graphene on cvd-grown hexagonal boron
  nitride},\ }\href {https://doi.org/https://doi.org/10.1002/adma.201204904}
  {\bibfield  {journal} {\bibinfo  {journal} {Advanced Materials}\ }\textbf
  {\bibinfo {volume} {25}},\ \bibinfo {pages} {2746} (\bibinfo {year}
  {2013}{\natexlab{a}})}\BibitemShut {NoStop}%
\bibitem [{\citenamefont {Wang}\ \emph {et~al.}(2014)\citenamefont {Wang},
  \citenamefont {Liu}, \citenamefont {Fu}, \citenamefont {Fang}, \citenamefont
  {Zhou},\ and\ \citenamefont {Liu}}]{wang2014two}%
  \BibitemOpen
  \bibfield  {author} {\bibinfo {author} {\bibfnamefont {H.}~\bibnamefont
  {Wang}}, \bibinfo {author} {\bibfnamefont {F.}~\bibnamefont {Liu}}, \bibinfo
  {author} {\bibfnamefont {W.}~\bibnamefont {Fu}}, \bibinfo {author}
  {\bibfnamefont {Z.}~\bibnamefont {Fang}}, \bibinfo {author} {\bibfnamefont
  {W.}~\bibnamefont {Zhou}},\ and\ \bibinfo {author} {\bibfnamefont
  {Z.}~\bibnamefont {Liu}},\ }\bibfield  {title} {\bibinfo {title}
  {Two-dimensional heterostructures: fabrication, characterization, and
  application},\ }\href@noop {} {\bibfield  {journal} {\bibinfo  {journal}
  {Nanoscale}\ }\textbf {\bibinfo {volume} {6}},\ \bibinfo {pages} {12250}
  (\bibinfo {year} {2014})}\BibitemShut {NoStop}%
\bibitem [{\citenamefont {Niu}\ and\ \citenamefont {Li}(2015)}]{niu2015two}%
  \BibitemOpen
  \bibfield  {author} {\bibinfo {author} {\bibfnamefont {T.}~\bibnamefont
  {Niu}}\ and\ \bibinfo {author} {\bibfnamefont {A.}~\bibnamefont {Li}},\
  }\bibfield  {title} {\bibinfo {title} {From two-dimensional materials to
  heterostructures},\ }\href@noop {} {\bibfield  {journal} {\bibinfo  {journal}
  {Progress in Surface Science}\ }\textbf {\bibinfo {volume} {90}},\ \bibinfo
  {pages} {21} (\bibinfo {year} {2015})}\BibitemShut {NoStop}%
\bibitem [{\citenamefont {Novoselov}\ \emph {et~al.}(2016)\citenamefont
  {Novoselov}, \citenamefont {Mishchenko}, \citenamefont {Carvalho},\ and\
  \citenamefont {Castro~Neto}}]{novoselov20162d}%
  \BibitemOpen
  \bibfield  {author} {\bibinfo {author} {\bibfnamefont {K.}~\bibnamefont
  {Novoselov}}, \bibinfo {author} {\bibfnamefont {o.~A.}\ \bibnamefont
  {Mishchenko}}, \bibinfo {author} {\bibfnamefont {o.~A.}\ \bibnamefont
  {Carvalho}},\ and\ \bibinfo {author} {\bibfnamefont {A.}~\bibnamefont
  {Castro~Neto}},\ }\bibfield  {title} {\bibinfo {title} {2d materials and van
  der waals heterostructures},\ }\href@noop {} {\bibfield  {journal} {\bibinfo
  {journal} {Science}\ }\textbf {\bibinfo {volume} {353}},\ \bibinfo {pages}
  {aac9439} (\bibinfo {year} {2016})}\BibitemShut {NoStop}%
\bibitem [{\citenamefont {Cao}\ \emph {et~al.}(2018{\natexlab{a}})\citenamefont
  {Cao}, \citenamefont {Fatemi}, \citenamefont {Fang}, \citenamefont
  {Watanabe}, \citenamefont {Taniguchi}, \citenamefont {Kaxiras},\ and\
  \citenamefont {Jarillo-Herrero}}]{cao2018unconventional}%
  \BibitemOpen
  \bibfield  {author} {\bibinfo {author} {\bibfnamefont {Y.}~\bibnamefont
  {Cao}}, \bibinfo {author} {\bibfnamefont {V.}~\bibnamefont {Fatemi}},
  \bibinfo {author} {\bibfnamefont {S.}~\bibnamefont {Fang}}, \bibinfo {author}
  {\bibfnamefont {K.}~\bibnamefont {Watanabe}}, \bibinfo {author}
  {\bibfnamefont {T.}~\bibnamefont {Taniguchi}}, \bibinfo {author}
  {\bibfnamefont {E.}~\bibnamefont {Kaxiras}},\ and\ \bibinfo {author}
  {\bibfnamefont {P.}~\bibnamefont {Jarillo-Herrero}},\ }\bibfield  {title}
  {\bibinfo {title} {Unconventional superconductivity in magic-angle graphene
  superlattices},\ }\href@noop {} {\bibfield  {journal} {\bibinfo  {journal}
  {Nature}\ }\textbf {\bibinfo {volume} {556}},\ \bibinfo {pages} {43}
  (\bibinfo {year} {2018}{\natexlab{a}})}\BibitemShut {NoStop}%
\bibitem [{\citenamefont {Lu}\ \emph {et~al.}(2019)\citenamefont {Lu},
  \citenamefont {Stepanov}, \citenamefont {Yang}, \citenamefont {Xie},
  \citenamefont {Aamir}, \citenamefont {Das}, \citenamefont {Urgell},
  \citenamefont {Watanabe}, \citenamefont {Taniguchi}, \citenamefont {Zhang},
  \citenamefont {Bachtold}, \citenamefont {MacDonald},\ and\ \citenamefont
  {Efetov}}]{lu2019superconductors}%
  \BibitemOpen
  \bibfield  {author} {\bibinfo {author} {\bibfnamefont {X.}~\bibnamefont
  {Lu}}, \bibinfo {author} {\bibfnamefont {P.}~\bibnamefont {Stepanov}},
  \bibinfo {author} {\bibfnamefont {W.}~\bibnamefont {Yang}}, \bibinfo {author}
  {\bibfnamefont {M.}~\bibnamefont {Xie}}, \bibinfo {author} {\bibfnamefont
  {M.~A.}\ \bibnamefont {Aamir}}, \bibinfo {author} {\bibfnamefont
  {I.}~\bibnamefont {Das}}, \bibinfo {author} {\bibfnamefont {C.}~\bibnamefont
  {Urgell}}, \bibinfo {author} {\bibfnamefont {K.}~\bibnamefont {Watanabe}},
  \bibinfo {author} {\bibfnamefont {T.}~\bibnamefont {Taniguchi}}, \bibinfo
  {author} {\bibfnamefont {G.}~\bibnamefont {Zhang}}, \bibinfo {author}
  {\bibfnamefont {A.}~\bibnamefont {Bachtold}}, \bibinfo {author}
  {\bibfnamefont {A.~H.}\ \bibnamefont {MacDonald}},\ and\ \bibinfo {author}
  {\bibfnamefont {D.~K.}\ \bibnamefont {Efetov}},\ }\bibfield  {title}
  {\bibinfo {title} {Superconductors, orbital magnets and correlated states in
  magic-angle bilayer graphene},\ }\href@noop {} {\bibfield  {journal}
  {\bibinfo  {journal} {Nature}\ }\textbf {\bibinfo {volume} {574}},\ \bibinfo
  {pages} {653} (\bibinfo {year} {2019})}\BibitemShut {NoStop}%
\bibitem [{\citenamefont {Yankowitz}\ \emph {et~al.}(2019)\citenamefont
  {Yankowitz}, \citenamefont {Chen}, \citenamefont {Polshyn}, \citenamefont
  {Zhang}, \citenamefont {Watanabe}, \citenamefont {Taniguchi}, \citenamefont
  {Graf}, \citenamefont {Young},\ and\ \citenamefont
  {Dean}}]{yankowitz2019tuning}%
  \BibitemOpen
  \bibfield  {author} {\bibinfo {author} {\bibfnamefont {M.}~\bibnamefont
  {Yankowitz}}, \bibinfo {author} {\bibfnamefont {S.}~\bibnamefont {Chen}},
  \bibinfo {author} {\bibfnamefont {H.}~\bibnamefont {Polshyn}}, \bibinfo
  {author} {\bibfnamefont {Y.}~\bibnamefont {Zhang}}, \bibinfo {author}
  {\bibfnamefont {K.}~\bibnamefont {Watanabe}}, \bibinfo {author}
  {\bibfnamefont {T.}~\bibnamefont {Taniguchi}}, \bibinfo {author}
  {\bibfnamefont {D.}~\bibnamefont {Graf}}, \bibinfo {author} {\bibfnamefont
  {A.~F.}\ \bibnamefont {Young}},\ and\ \bibinfo {author} {\bibfnamefont
  {C.~R.}\ \bibnamefont {Dean}},\ }\bibfield  {title} {\bibinfo {title} {Tuning
  superconductivity in twisted bilayer graphene},\ }\href@noop {} {\bibfield
  {journal} {\bibinfo  {journal} {Science}\ }\textbf {\bibinfo {volume}
  {363}},\ \bibinfo {pages} {1059} (\bibinfo {year} {2019})}\BibitemShut
  {NoStop}%
\bibitem [{\citenamefont {Cao}\ \emph {et~al.}(2020{\natexlab{a}})\citenamefont
  {Cao}, \citenamefont {Rodan-Legrain}, \citenamefont {Park}, \citenamefont
  {Yuan}, \citenamefont {Watanabe}, \citenamefont {Taniguchi}, \citenamefont
  {Fernandes}, \citenamefont {Fu},\ and\ \citenamefont
  {Jarillo-Herrero}}]{cao2020nematicity}%
  \BibitemOpen
  \bibfield  {author} {\bibinfo {author} {\bibfnamefont {Y.}~\bibnamefont
  {Cao}}, \bibinfo {author} {\bibfnamefont {D.}~\bibnamefont {Rodan-Legrain}},
  \bibinfo {author} {\bibfnamefont {J.~M.}\ \bibnamefont {Park}}, \bibinfo
  {author} {\bibfnamefont {F.~N.}\ \bibnamefont {Yuan}}, \bibinfo {author}
  {\bibfnamefont {K.}~\bibnamefont {Watanabe}}, \bibinfo {author}
  {\bibfnamefont {T.}~\bibnamefont {Taniguchi}}, \bibinfo {author}
  {\bibfnamefont {R.~M.}\ \bibnamefont {Fernandes}}, \bibinfo {author}
  {\bibfnamefont {L.}~\bibnamefont {Fu}},\ and\ \bibinfo {author}
  {\bibfnamefont {P.}~\bibnamefont {Jarillo-Herrero}},\ }\bibfield  {title}
  {\bibinfo {title} {Nematicity and competing orders in superconducting
  magic-angle graphene},\ }\href@noop {} {\bibfield  {journal} {\bibinfo
  {journal} {arXiv preprint arXiv:2004.04148}\ } (\bibinfo {year}
  {2020}{\natexlab{a}})}\BibitemShut {NoStop}%
\bibitem [{\citenamefont {Liu}\ \emph {et~al.}(2021)\citenamefont {Liu},
  \citenamefont {Wang}, \citenamefont {Watanabe}, \citenamefont {Taniguchi},
  \citenamefont {Vafek},\ and\ \citenamefont {Li}}]{liu2021tuning}%
  \BibitemOpen
  \bibfield  {author} {\bibinfo {author} {\bibfnamefont {X.}~\bibnamefont
  {Liu}}, \bibinfo {author} {\bibfnamefont {Z.}~\bibnamefont {Wang}}, \bibinfo
  {author} {\bibfnamefont {K.}~\bibnamefont {Watanabe}}, \bibinfo {author}
  {\bibfnamefont {T.}~\bibnamefont {Taniguchi}}, \bibinfo {author}
  {\bibfnamefont {O.}~\bibnamefont {Vafek}},\ and\ \bibinfo {author}
  {\bibfnamefont {J.}~\bibnamefont {Li}},\ }\bibfield  {title} {\bibinfo
  {title} {Tuning electron correlation in magic-angle twisted bilayer graphene
  using coulomb screening},\ }\href@noop {} {\bibfield  {journal} {\bibinfo
  {journal} {Science}\ }\textbf {\bibinfo {volume} {371}},\ \bibinfo {pages}
  {1261} (\bibinfo {year} {2021})}\BibitemShut {NoStop}%
\bibitem [{\citenamefont {Stepanov}\ \emph {et~al.}(2020)\citenamefont
  {Stepanov}, \citenamefont {Das}, \citenamefont {Lu}, \citenamefont
  {Fahimniya}, \citenamefont {Watanabe}, \citenamefont {Taniguchi},
  \citenamefont {Koppens}, \citenamefont {Lischner}, \citenamefont {Levitov},\
  and\ \citenamefont {Efetov}}]{stepanov2020untying}%
  \BibitemOpen
  \bibfield  {author} {\bibinfo {author} {\bibfnamefont {P.}~\bibnamefont
  {Stepanov}}, \bibinfo {author} {\bibfnamefont {I.}~\bibnamefont {Das}},
  \bibinfo {author} {\bibfnamefont {X.}~\bibnamefont {Lu}}, \bibinfo {author}
  {\bibfnamefont {A.}~\bibnamefont {Fahimniya}}, \bibinfo {author}
  {\bibfnamefont {K.}~\bibnamefont {Watanabe}}, \bibinfo {author}
  {\bibfnamefont {T.}~\bibnamefont {Taniguchi}}, \bibinfo {author}
  {\bibfnamefont {F.~H.}\ \bibnamefont {Koppens}}, \bibinfo {author}
  {\bibfnamefont {J.}~\bibnamefont {Lischner}}, \bibinfo {author}
  {\bibfnamefont {L.}~\bibnamefont {Levitov}},\ and\ \bibinfo {author}
  {\bibfnamefont {D.~K.}\ \bibnamefont {Efetov}},\ }\bibfield  {title}
  {\bibinfo {title} {Untying the insulating and superconducting orders in
  magic-angle graphene},\ }\href@noop {} {\bibfield  {journal} {\bibinfo
  {journal} {Nature}\ }\textbf {\bibinfo {volume} {583}},\ \bibinfo {pages}
  {375} (\bibinfo {year} {2020})}\BibitemShut {NoStop}%
\bibitem [{\citenamefont {Arora}\ \emph {et~al.}(2020)\citenamefont {Arora},
  \citenamefont {Polski}, \citenamefont {Zhang}, \citenamefont {Thomson},
  \citenamefont {Choi}, \citenamefont {Kim}, \citenamefont {Lin}, \citenamefont
  {Wilson}, \citenamefont {Xu}, \citenamefont {Chu} \emph
  {et~al.}}]{arora2020superconductivity}%
  \BibitemOpen
  \bibfield  {author} {\bibinfo {author} {\bibfnamefont {H.~S.}\ \bibnamefont
  {Arora}}, \bibinfo {author} {\bibfnamefont {R.}~\bibnamefont {Polski}},
  \bibinfo {author} {\bibfnamefont {Y.}~\bibnamefont {Zhang}}, \bibinfo
  {author} {\bibfnamefont {A.}~\bibnamefont {Thomson}}, \bibinfo {author}
  {\bibfnamefont {Y.}~\bibnamefont {Choi}}, \bibinfo {author} {\bibfnamefont
  {H.}~\bibnamefont {Kim}}, \bibinfo {author} {\bibfnamefont {Z.}~\bibnamefont
  {Lin}}, \bibinfo {author} {\bibfnamefont {I.~Z.}\ \bibnamefont {Wilson}},
  \bibinfo {author} {\bibfnamefont {X.}~\bibnamefont {Xu}}, \bibinfo {author}
  {\bibfnamefont {J.-H.}\ \bibnamefont {Chu}}, \emph {et~al.},\ }\bibfield
  {title} {\bibinfo {title} {Superconductivity in metallic twisted bilayer
  graphene stabilized by wse 2},\ }\href@noop {} {\bibfield  {journal}
  {\bibinfo  {journal} {Nature}\ }\textbf {\bibinfo {volume} {583}},\ \bibinfo
  {pages} {379} (\bibinfo {year} {2020})}\BibitemShut {NoStop}%
\bibitem [{\citenamefont {Oh}\ \emph {et~al.}(2021)\citenamefont {Oh},
  \citenamefont {Nuckolls}, \citenamefont {Wong}, \citenamefont {Lee},
  \citenamefont {Liu}, \citenamefont {Watanabe}, \citenamefont {Taniguchi},\
  and\ \citenamefont {Yazdani}}]{oh2021evidence}%
  \BibitemOpen
  \bibfield  {author} {\bibinfo {author} {\bibfnamefont {M.}~\bibnamefont
  {Oh}}, \bibinfo {author} {\bibfnamefont {K.~P.}\ \bibnamefont {Nuckolls}},
  \bibinfo {author} {\bibfnamefont {D.}~\bibnamefont {Wong}}, \bibinfo {author}
  {\bibfnamefont {R.~L.}\ \bibnamefont {Lee}}, \bibinfo {author} {\bibfnamefont
  {X.}~\bibnamefont {Liu}}, \bibinfo {author} {\bibfnamefont {K.}~\bibnamefont
  {Watanabe}}, \bibinfo {author} {\bibfnamefont {T.}~\bibnamefont
  {Taniguchi}},\ and\ \bibinfo {author} {\bibfnamefont {A.}~\bibnamefont
  {Yazdani}},\ }\bibfield  {title} {\bibinfo {title} {Evidence for
  unconventional superconductivity in twisted bilayer graphene},\ }\href
  {https://doi.org/10.1038/s41586-021-04121-x} {\bibfield  {journal} {\bibinfo
  {journal} {Nature}\ }\textbf {\bibinfo {volume} {600}},\ \bibinfo {pages}
  {240–245} (\bibinfo {year} {2021})}\BibitemShut {NoStop}%
\bibitem [{\citenamefont {Park}\ \emph
  {et~al.}(2021{\natexlab{a}})\citenamefont {Park}, \citenamefont {Cao},
  \citenamefont {Watanabe}, \citenamefont {Taniguchi},\ and\ \citenamefont
  {Jarillo-Herrero}}]{park2021tunable}%
  \BibitemOpen
  \bibfield  {author} {\bibinfo {author} {\bibfnamefont {J.~M.}\ \bibnamefont
  {Park}}, \bibinfo {author} {\bibfnamefont {Y.}~\bibnamefont {Cao}}, \bibinfo
  {author} {\bibfnamefont {K.}~\bibnamefont {Watanabe}}, \bibinfo {author}
  {\bibfnamefont {T.}~\bibnamefont {Taniguchi}},\ and\ \bibinfo {author}
  {\bibfnamefont {P.}~\bibnamefont {Jarillo-Herrero}},\ }\bibfield  {title}
  {\bibinfo {title} {Tunable strongly coupled superconductivity in magic-angle
  twisted trilayer graphene},\ }\href
  {https://doi.org/10.1038/s41586-021-03192-0} {\bibfield  {journal} {\bibinfo
  {journal} {Nature}\ }\textbf {\bibinfo {volume} {590}},\ \bibinfo {pages}
  {249} (\bibinfo {year} {2021}{\natexlab{a}})}\BibitemShut {NoStop}%
\bibitem [{\citenamefont {Cao}\ \emph {et~al.}(2021)\citenamefont {Cao},
  \citenamefont {Park}, \citenamefont {Watanabe}, \citenamefont {Taniguchi},\
  and\ \citenamefont {Jarillo-Herrero}}]{cao2021large}%
  \BibitemOpen
  \bibfield  {author} {\bibinfo {author} {\bibfnamefont {Y.}~\bibnamefont
  {Cao}}, \bibinfo {author} {\bibfnamefont {J.~M.}\ \bibnamefont {Park}},
  \bibinfo {author} {\bibfnamefont {K.}~\bibnamefont {Watanabe}}, \bibinfo
  {author} {\bibfnamefont {T.}~\bibnamefont {Taniguchi}},\ and\ \bibinfo
  {author} {\bibfnamefont {P.}~\bibnamefont {Jarillo-Herrero}},\ }\bibfield
  {title} {\bibinfo {title} {Pauli-limit violation and re-entrant
  superconductivity in moiré graphene},\ }\href
  {https://doi.org/10.1038/s41586-021-03685-y} {\bibfield  {journal} {\bibinfo
  {journal} {Nature}\ }\textbf {\bibinfo {volume} {595}},\ \bibinfo {pages}
  {526–531} (\bibinfo {year} {2021})}\BibitemShut {NoStop}%
\bibitem [{\citenamefont {Hao}\ \emph {et~al.}(2021)\citenamefont {Hao},
  \citenamefont {Zimmerman}, \citenamefont {Ledwith}, \citenamefont {Khalaf},
  \citenamefont {Najafabadi}, \citenamefont {Watanabe}, \citenamefont
  {Taniguchi}, \citenamefont {Vishwanath},\ and\ \citenamefont
  {Kim}}]{hao2021electric}%
  \BibitemOpen
  \bibfield  {author} {\bibinfo {author} {\bibfnamefont {Z.}~\bibnamefont
  {Hao}}, \bibinfo {author} {\bibfnamefont {A.}~\bibnamefont {Zimmerman}},
  \bibinfo {author} {\bibfnamefont {P.}~\bibnamefont {Ledwith}}, \bibinfo
  {author} {\bibfnamefont {E.}~\bibnamefont {Khalaf}}, \bibinfo {author}
  {\bibfnamefont {D.~H.}\ \bibnamefont {Najafabadi}}, \bibinfo {author}
  {\bibfnamefont {K.}~\bibnamefont {Watanabe}}, \bibinfo {author}
  {\bibfnamefont {T.}~\bibnamefont {Taniguchi}}, \bibinfo {author}
  {\bibfnamefont {A.}~\bibnamefont {Vishwanath}},\ and\ \bibinfo {author}
  {\bibfnamefont {P.}~\bibnamefont {Kim}},\ }\bibfield  {title} {\bibinfo
  {title} {Electric field--tunable superconductivity in alternating-twist
  magic-angle trilayer graphene},\ }\href@noop {} {\bibfield  {journal}
  {\bibinfo  {journal} {Science}\ }\textbf {\bibinfo {volume} {371}},\ \bibinfo
  {pages} {1133} (\bibinfo {year} {2021})}\BibitemShut {NoStop}%
\bibitem [{\citenamefont {Kim}\ \emph {et~al.}(2021)\citenamefont {Kim},
  \citenamefont {Choi}, \citenamefont {Lewandowski}, \citenamefont {Thomson},
  \citenamefont {Zhang}, \citenamefont {Polski}, \citenamefont {Watanabe},
  \citenamefont {Taniguchi}, \citenamefont {Alicea},\ and\ \citenamefont
  {Nadj-Perge}}]{kim2021spectroscopic}%
  \BibitemOpen
  \bibfield  {author} {\bibinfo {author} {\bibfnamefont {H.}~\bibnamefont
  {Kim}}, \bibinfo {author} {\bibfnamefont {Y.}~\bibnamefont {Choi}}, \bibinfo
  {author} {\bibfnamefont {C.}~\bibnamefont {Lewandowski}}, \bibinfo {author}
  {\bibfnamefont {A.}~\bibnamefont {Thomson}}, \bibinfo {author} {\bibfnamefont
  {Y.}~\bibnamefont {Zhang}}, \bibinfo {author} {\bibfnamefont
  {R.}~\bibnamefont {Polski}}, \bibinfo {author} {\bibfnamefont
  {K.}~\bibnamefont {Watanabe}}, \bibinfo {author} {\bibfnamefont
  {T.}~\bibnamefont {Taniguchi}}, \bibinfo {author} {\bibfnamefont
  {J.}~\bibnamefont {Alicea}},\ and\ \bibinfo {author} {\bibfnamefont
  {S.}~\bibnamefont {Nadj-Perge}},\ }\href@noop {} {\bibinfo {title}
  {Spectroscopic signatures of strong correlations and unconventional
  superconductivity in twisted trilayer graphene}} (\bibinfo {year} {2021}),\
  \Eprint {https://arxiv.org/abs/2109.12127} {arXiv:2109.12127
  [cond-mat.mes-hall]} \BibitemShut {NoStop}%
\bibitem [{\citenamefont {Cao}\ \emph {et~al.}(2018{\natexlab{b}})\citenamefont
  {Cao}, \citenamefont {Fatemi}, \citenamefont {Demir}, \citenamefont {Fang},
  \citenamefont {Tomarken}, \citenamefont {Luo}, \citenamefont
  {Sanchez-Yamagishi}, \citenamefont {Watanabe}, \citenamefont {Taniguchi},
  \citenamefont {Kaxiras}, \citenamefont {Ashoori},\ and\ \citenamefont
  {Jarillo-Herrero}}]{cao2018mott}%
  \BibitemOpen
  \bibfield  {author} {\bibinfo {author} {\bibfnamefont {Y.}~\bibnamefont
  {Cao}}, \bibinfo {author} {\bibfnamefont {V.}~\bibnamefont {Fatemi}},
  \bibinfo {author} {\bibfnamefont {A.}~\bibnamefont {Demir}}, \bibinfo
  {author} {\bibfnamefont {S.}~\bibnamefont {Fang}}, \bibinfo {author}
  {\bibfnamefont {S.~L.}\ \bibnamefont {Tomarken}}, \bibinfo {author}
  {\bibfnamefont {J.~Y.}\ \bibnamefont {Luo}}, \bibinfo {author} {\bibfnamefont
  {J.~D.}\ \bibnamefont {Sanchez-Yamagishi}}, \bibinfo {author} {\bibfnamefont
  {K.}~\bibnamefont {Watanabe}}, \bibinfo {author} {\bibfnamefont
  {T.}~\bibnamefont {Taniguchi}}, \bibinfo {author} {\bibfnamefont
  {E.}~\bibnamefont {Kaxiras}}, \bibinfo {author} {\bibfnamefont {R.~C.}\
  \bibnamefont {Ashoori}},\ and\ \bibinfo {author} {\bibfnamefont
  {P.}~\bibnamefont {Jarillo-Herrero}},\ }\bibfield  {title} {\bibinfo {title}
  {Correlated insulator behaviour at half-filling in magic-angle graphene
  superlattices},\ }\href@noop {} {\bibfield  {journal} {\bibinfo  {journal}
  {Nature}\ }\textbf {\bibinfo {volume} {556}},\ \bibinfo {pages} {80}
  (\bibinfo {year} {2018}{\natexlab{b}})}\BibitemShut {NoStop}%
\bibitem [{\citenamefont {Cao}\ \emph {et~al.}(2020{\natexlab{b}})\citenamefont
  {Cao}, \citenamefont {Chowdhury}, \citenamefont {Rodan-Legrain},
  \citenamefont {Rubies-Bigord\`a}, \citenamefont {Watanabe}, \citenamefont
  {Taniguchi}, \citenamefont {Senthil},\ and\ \citenamefont
  {Jarillo-Herrero}}]{cao2020strange}%
  \BibitemOpen
  \bibfield  {author} {\bibinfo {author} {\bibfnamefont {Y.}~\bibnamefont
  {Cao}}, \bibinfo {author} {\bibfnamefont {D.}~\bibnamefont {Chowdhury}},
  \bibinfo {author} {\bibfnamefont {D.}~\bibnamefont {Rodan-Legrain}}, \bibinfo
  {author} {\bibfnamefont {O.}~\bibnamefont {Rubies-Bigord\`a}}, \bibinfo
  {author} {\bibfnamefont {K.}~\bibnamefont {Watanabe}}, \bibinfo {author}
  {\bibfnamefont {T.}~\bibnamefont {Taniguchi}}, \bibinfo {author}
  {\bibfnamefont {T.}~\bibnamefont {Senthil}},\ and\ \bibinfo {author}
  {\bibfnamefont {P.}~\bibnamefont {Jarillo-Herrero}},\ }\bibfield  {title}
  {\bibinfo {title} {Strange metal in magic-angle graphene with near planckian
  dissipation},\ }\href@noop {} {\bibfield  {journal} {\bibinfo  {journal}
  {Phys. Rev. Lett.}\ }\textbf {\bibinfo {volume} {124}},\ \bibinfo {pages}
  {076801} (\bibinfo {year} {2020}{\natexlab{b}})}\BibitemShut {NoStop}%
\bibitem [{\citenamefont {Polshyn}\ \emph {et~al.}(2019)\citenamefont
  {Polshyn}, \citenamefont {Yankowitz}, \citenamefont {Chen}, \citenamefont
  {Zhang}, \citenamefont {Watanabe}, \citenamefont {Taniguchi}, \citenamefont
  {Dean},\ and\ \citenamefont {Young}}]{polshyn2019large}%
  \BibitemOpen
  \bibfield  {author} {\bibinfo {author} {\bibfnamefont {H.}~\bibnamefont
  {Polshyn}}, \bibinfo {author} {\bibfnamefont {M.}~\bibnamefont {Yankowitz}},
  \bibinfo {author} {\bibfnamefont {S.}~\bibnamefont {Chen}}, \bibinfo {author}
  {\bibfnamefont {Y.}~\bibnamefont {Zhang}}, \bibinfo {author} {\bibfnamefont
  {K.}~\bibnamefont {Watanabe}}, \bibinfo {author} {\bibfnamefont
  {T.}~\bibnamefont {Taniguchi}}, \bibinfo {author} {\bibfnamefont {C.~R.}\
  \bibnamefont {Dean}},\ and\ \bibinfo {author} {\bibfnamefont {A.~F.}\
  \bibnamefont {Young}},\ }\bibfield  {title} {\bibinfo {title} {Large
  linear-in-temperature resistivity in twisted bilayer graphene},\ }\href@noop
  {} {\bibfield  {journal} {\bibinfo  {journal} {Nat. Phys.}\ }\textbf
  {\bibinfo {volume} {15}},\ \bibinfo {pages} {1011} (\bibinfo {year}
  {2019})}\BibitemShut {NoStop}%
\bibitem [{\citenamefont {Zondiner}\ \emph {et~al.}(2020)\citenamefont
  {Zondiner}, \citenamefont {Rozen}, \citenamefont {Rodan-Legrain},
  \citenamefont {Cao}, \citenamefont {Queiroz}, \citenamefont {Taniguchi},
  \citenamefont {Watanabe}, \citenamefont {Oreg}, \citenamefont {von Oppen},
  \citenamefont {Stern}, \citenamefont {Berg}, \citenamefont
  {Jarillo-Herrero},\ and\ \citenamefont {Ilani}}]{zondiner2020cascade}%
  \BibitemOpen
  \bibfield  {author} {\bibinfo {author} {\bibfnamefont {U.}~\bibnamefont
  {Zondiner}}, \bibinfo {author} {\bibfnamefont {A.}~\bibnamefont {Rozen}},
  \bibinfo {author} {\bibfnamefont {D.}~\bibnamefont {Rodan-Legrain}}, \bibinfo
  {author} {\bibfnamefont {Y.}~\bibnamefont {Cao}}, \bibinfo {author}
  {\bibfnamefont {R.}~\bibnamefont {Queiroz}}, \bibinfo {author} {\bibfnamefont
  {T.}~\bibnamefont {Taniguchi}}, \bibinfo {author} {\bibfnamefont
  {K.}~\bibnamefont {Watanabe}}, \bibinfo {author} {\bibfnamefont
  {Y.}~\bibnamefont {Oreg}}, \bibinfo {author} {\bibfnamefont {F.}~\bibnamefont
  {von Oppen}}, \bibinfo {author} {\bibfnamefont {A.}~\bibnamefont {Stern}},
  \bibinfo {author} {\bibfnamefont {E.}~\bibnamefont {Berg}}, \bibinfo {author}
  {\bibfnamefont {P.}~\bibnamefont {Jarillo-Herrero}},\ and\ \bibinfo {author}
  {\bibfnamefont {S.}~\bibnamefont {Ilani}},\ }\bibfield  {title} {\bibinfo
  {title} {Cascade of phase transitions and dirac revivals in magic-angle
  graphene},\ }\href@noop {} {\bibfield  {journal} {\bibinfo  {journal}
  {Nature}\ }\textbf {\bibinfo {volume} {582}},\ \bibinfo {pages} {203}
  (\bibinfo {year} {2020})}\BibitemShut {NoStop}%
\bibitem [{\citenamefont {Wong}\ \emph {et~al.}(2020)\citenamefont {Wong},
  \citenamefont {Nuckolls}, \citenamefont {Oh}, \citenamefont {Lian},
  \citenamefont {Yonglong~Xie}, \citenamefont {Watanabe}, \citenamefont
  {Taniguchi}, \citenamefont {Bernevig},\ and\ \citenamefont
  {Yazdani}}]{wong2020cascade}%
  \BibitemOpen
  \bibfield  {author} {\bibinfo {author} {\bibfnamefont {D.}~\bibnamefont
  {Wong}}, \bibinfo {author} {\bibfnamefont {K.~P.}\ \bibnamefont {Nuckolls}},
  \bibinfo {author} {\bibfnamefont {M.}~\bibnamefont {Oh}}, \bibinfo {author}
  {\bibfnamefont {B.}~\bibnamefont {Lian}}, \bibinfo {author} {\bibfnamefont
  {S.~J.}\ \bibnamefont {Yonglong~Xie}}, \bibinfo {author} {\bibfnamefont
  {K.}~\bibnamefont {Watanabe}}, \bibinfo {author} {\bibfnamefont
  {T.}~\bibnamefont {Taniguchi}}, \bibinfo {author} {\bibfnamefont {B.~A.}\
  \bibnamefont {Bernevig}},\ and\ \bibinfo {author} {\bibfnamefont
  {A.}~\bibnamefont {Yazdani}},\ }\bibfield  {title} {\bibinfo {title} {Cascade
  of electronic transitions in magic-angle twisted bilayer graphene},\
  }\href@noop {} {\bibfield  {journal} {\bibinfo  {journal} {Nature}\ }\textbf
  {\bibinfo {volume} {582}},\ \bibinfo {pages} {198–202} (\bibinfo {year}
  {2020})}\BibitemShut {NoStop}%
\bibitem [{\citenamefont {Xie}\ \emph {et~al.}(2019)\citenamefont {Xie},
  \citenamefont {Lian}, \citenamefont {J\"{a}ck}, \citenamefont {Liu},
  \citenamefont {Chiu}, \citenamefont {Watanabe}, \citenamefont {Taniguchi},
  \citenamefont {Bernevig},\ and\ \citenamefont
  {Yazdani}}]{xie2019spectrosopic}%
  \BibitemOpen
  \bibfield  {author} {\bibinfo {author} {\bibfnamefont {Y.}~\bibnamefont
  {Xie}}, \bibinfo {author} {\bibfnamefont {B.}~\bibnamefont {Lian}}, \bibinfo
  {author} {\bibfnamefont {B.}~\bibnamefont {J\"{a}ck}}, \bibinfo {author}
  {\bibfnamefont {X.}~\bibnamefont {Liu}}, \bibinfo {author} {\bibfnamefont
  {C.-L.}\ \bibnamefont {Chiu}}, \bibinfo {author} {\bibfnamefont
  {K.}~\bibnamefont {Watanabe}}, \bibinfo {author} {\bibfnamefont
  {T.}~\bibnamefont {Taniguchi}}, \bibinfo {author} {\bibfnamefont {B.~A.}\
  \bibnamefont {Bernevig}},\ and\ \bibinfo {author} {\bibfnamefont
  {A.}~\bibnamefont {Yazdani}},\ }\bibfield  {title} {\bibinfo {title}
  {Spectroscopic signatures of many-body correlations in magic-angle twisted
  bilayer graphene},\ }\href@noop {} {\bibfield  {journal} {\bibinfo  {journal}
  {Nature}\ }\textbf {\bibinfo {volume} {572}},\ \bibinfo {pages} {101}
  (\bibinfo {year} {2019})}\BibitemShut {NoStop}%
\bibitem [{\citenamefont {Kerelsky}\ \emph {et~al.}(2019)\citenamefont
  {Kerelsky}, \citenamefont {McGilly}, \citenamefont {Kennes}, \citenamefont
  {Xian}, \citenamefont {Yankowitz}, \citenamefont {Chen}, \citenamefont
  {Watanabe}, \citenamefont {Taniguchi}, \citenamefont {Hone}, \citenamefont
  {Dean}, \citenamefont {Rubio},\ and\ \citenamefont
  {Pasupathy}}]{kerelsky2019maximized}%
  \BibitemOpen
  \bibfield  {author} {\bibinfo {author} {\bibfnamefont {A.}~\bibnamefont
  {Kerelsky}}, \bibinfo {author} {\bibfnamefont {L.~J.}\ \bibnamefont
  {McGilly}}, \bibinfo {author} {\bibfnamefont {D.~M.}\ \bibnamefont {Kennes}},
  \bibinfo {author} {\bibfnamefont {L.}~\bibnamefont {Xian}}, \bibinfo {author}
  {\bibfnamefont {M.}~\bibnamefont {Yankowitz}}, \bibinfo {author}
  {\bibfnamefont {S.}~\bibnamefont {Chen}}, \bibinfo {author} {\bibfnamefont
  {K.}~\bibnamefont {Watanabe}}, \bibinfo {author} {\bibfnamefont
  {T.}~\bibnamefont {Taniguchi}}, \bibinfo {author} {\bibfnamefont
  {J.}~\bibnamefont {Hone}}, \bibinfo {author} {\bibfnamefont {C.}~\bibnamefont
  {Dean}}, \bibinfo {author} {\bibfnamefont {A.}~\bibnamefont {Rubio}},\ and\
  \bibinfo {author} {\bibfnamefont {A.~N.}\ \bibnamefont {Pasupathy}},\
  }\bibfield  {title} {\bibinfo {title} {Maximized electron interactions at the
  magic angle in twisted bilayer graphene},\ }\href@noop {} {\bibfield
  {journal} {\bibinfo  {journal} {Nature}\ }\textbf {\bibinfo {volume} {572}},\
  \bibinfo {pages} {95} (\bibinfo {year} {2019})}\BibitemShut {NoStop}%
\bibitem [{\citenamefont {Jiang}\ \emph {et~al.}(2019)\citenamefont {Jiang},
  \citenamefont {Lai}, \citenamefont {Watanabe}, \citenamefont {Taniguchi},
  \citenamefont {Haule}, \citenamefont {Mao},\ and\ \citenamefont
  {Andrei}}]{jiang2019charge}%
  \BibitemOpen
  \bibfield  {author} {\bibinfo {author} {\bibfnamefont {Y.}~\bibnamefont
  {Jiang}}, \bibinfo {author} {\bibfnamefont {X.}~\bibnamefont {Lai}}, \bibinfo
  {author} {\bibfnamefont {K.}~\bibnamefont {Watanabe}}, \bibinfo {author}
  {\bibfnamefont {T.}~\bibnamefont {Taniguchi}}, \bibinfo {author}
  {\bibfnamefont {K.}~\bibnamefont {Haule}}, \bibinfo {author} {\bibfnamefont
  {J.}~\bibnamefont {Mao}},\ and\ \bibinfo {author} {\bibfnamefont {E.~Y.}\
  \bibnamefont {Andrei}},\ }\bibfield  {title} {\bibinfo {title} {Charge order
  and broken rotational symmetry in magic-angle twisted bilayer graphene},\
  }\href@noop {} {\bibfield  {journal} {\bibinfo  {journal} {Nature}\ }\textbf
  {\bibinfo {volume} {573}},\ \bibinfo {pages} {91} (\bibinfo {year}
  {2019})}\BibitemShut {NoStop}%
\bibitem [{\citenamefont {Choi}\ \emph {et~al.}(2019)\citenamefont {Choi},
  \citenamefont {Kemmer}, \citenamefont {Peng}, \citenamefont {Thomson},
  \citenamefont {Arora}, \citenamefont {Polski}, \citenamefont {Zhang},
  \citenamefont {Ren}, \citenamefont {Alicea}, \citenamefont {Refael},
  \citenamefont {von Oppen}, \citenamefont {Watanabe}, \citenamefont
  {Taniguchi},\ and\ \citenamefont {Nadj-Perge}}]{choi2019correlations}%
  \BibitemOpen
  \bibfield  {author} {\bibinfo {author} {\bibfnamefont {Y.}~\bibnamefont
  {Choi}}, \bibinfo {author} {\bibfnamefont {J.}~\bibnamefont {Kemmer}},
  \bibinfo {author} {\bibfnamefont {Y.}~\bibnamefont {Peng}}, \bibinfo {author}
  {\bibfnamefont {A.}~\bibnamefont {Thomson}}, \bibinfo {author} {\bibfnamefont
  {H.}~\bibnamefont {Arora}}, \bibinfo {author} {\bibfnamefont
  {R.}~\bibnamefont {Polski}}, \bibinfo {author} {\bibfnamefont
  {Y.}~\bibnamefont {Zhang}}, \bibinfo {author} {\bibfnamefont
  {H.}~\bibnamefont {Ren}}, \bibinfo {author} {\bibfnamefont {J.}~\bibnamefont
  {Alicea}}, \bibinfo {author} {\bibfnamefont {G.}~\bibnamefont {Refael}},
  \bibinfo {author} {\bibfnamefont {F.}~\bibnamefont {von Oppen}}, \bibinfo
  {author} {\bibfnamefont {K.}~\bibnamefont {Watanabe}}, \bibinfo {author}
  {\bibfnamefont {T.}~\bibnamefont {Taniguchi}},\ and\ \bibinfo {author}
  {\bibfnamefont {S.}~\bibnamefont {Nadj-Perge}},\ }\bibfield  {title}
  {\bibinfo {title} {Electronic correlations in twisted bilayer graphene near
  the magic angle},\ }\href@noop {} {\bibfield  {journal} {\bibinfo  {journal}
  {Nat. Phys.}\ }\textbf {\bibinfo {volume} {15}},\ \bibinfo {pages} {1174}
  (\bibinfo {year} {2019})}\BibitemShut {NoStop}%
\bibitem [{\citenamefont {Chen}\ \emph {et~al.}(2021)\citenamefont {Chen},
  \citenamefont {He}, \citenamefont {Zhang}, \citenamefont {Hsieh},
  \citenamefont {Fei}, \citenamefont {Watanabe}, \citenamefont {Taniguchi},
  \citenamefont {Cobden}, \citenamefont {Xu}, \citenamefont {Dean} \emph
  {et~al.}}]{chen2021electrically}%
  \BibitemOpen
  \bibfield  {author} {\bibinfo {author} {\bibfnamefont {S.}~\bibnamefont
  {Chen}}, \bibinfo {author} {\bibfnamefont {M.}~\bibnamefont {He}}, \bibinfo
  {author} {\bibfnamefont {Y.-H.}\ \bibnamefont {Zhang}}, \bibinfo {author}
  {\bibfnamefont {V.}~\bibnamefont {Hsieh}}, \bibinfo {author} {\bibfnamefont
  {Z.}~\bibnamefont {Fei}}, \bibinfo {author} {\bibfnamefont {K.}~\bibnamefont
  {Watanabe}}, \bibinfo {author} {\bibfnamefont {T.}~\bibnamefont {Taniguchi}},
  \bibinfo {author} {\bibfnamefont {D.~H.}\ \bibnamefont {Cobden}}, \bibinfo
  {author} {\bibfnamefont {X.}~\bibnamefont {Xu}}, \bibinfo {author}
  {\bibfnamefont {C.~R.}\ \bibnamefont {Dean}}, \emph {et~al.},\ }\bibfield
  {title} {\bibinfo {title} {Electrically tunable correlated and topological
  states in twisted monolayer--bilayer graphene},\ }\href@noop {} {\bibfield
  {journal} {\bibinfo  {journal} {Nature Physics}\ }\textbf {\bibinfo {volume}
  {17}},\ \bibinfo {pages} {374} (\bibinfo {year} {2021})}\BibitemShut
  {NoStop}%
\bibitem [{\citenamefont {Shi}\ \emph {et~al.}(2020)\citenamefont {Shi},
  \citenamefont {Xu}, \citenamefont {Ezzi}, \citenamefont {Balakrishnan},
  \citenamefont {Garcia-Ruiz}, \citenamefont {Tsim}, \citenamefont {Mullan},
  \citenamefont {Barrier}, \citenamefont {Xin}, \citenamefont {Piot} \emph
  {et~al.}}]{shi2020tunable}%
  \BibitemOpen
  \bibfield  {author} {\bibinfo {author} {\bibfnamefont {Y.}~\bibnamefont
  {Shi}}, \bibinfo {author} {\bibfnamefont {S.}~\bibnamefont {Xu}}, \bibinfo
  {author} {\bibfnamefont {M.~M.~A.}\ \bibnamefont {Ezzi}}, \bibinfo {author}
  {\bibfnamefont {N.}~\bibnamefont {Balakrishnan}}, \bibinfo {author}
  {\bibfnamefont {A.}~\bibnamefont {Garcia-Ruiz}}, \bibinfo {author}
  {\bibfnamefont {B.}~\bibnamefont {Tsim}}, \bibinfo {author} {\bibfnamefont
  {C.}~\bibnamefont {Mullan}}, \bibinfo {author} {\bibfnamefont
  {J.}~\bibnamefont {Barrier}}, \bibinfo {author} {\bibfnamefont
  {N.}~\bibnamefont {Xin}}, \bibinfo {author} {\bibfnamefont {B.~A.}\
  \bibnamefont {Piot}}, \emph {et~al.},\ }\bibfield  {title} {\bibinfo {title}
  {Tunable van hove singularities and correlated states in twisted trilayer
  graphene},\ }\href@noop {} {\bibfield  {journal} {\bibinfo  {journal} {arXiv
  preprint arXiv:2004.12414}\ } (\bibinfo {year} {2020})}\BibitemShut {NoStop}%
\bibitem [{\citenamefont {Liu}\ \emph {et~al.}(2020)\citenamefont {Liu},
  \citenamefont {Hao}, \citenamefont {Khalaf}, \citenamefont {Lee},
  \citenamefont {Ronen}, \citenamefont {Yoo}, \citenamefont {Haei~Najafabadi},
  \citenamefont {Watanabe}, \citenamefont {Taniguchi}, \citenamefont
  {Vishwanath} \emph {et~al.}}]{liu2020tunable}%
  \BibitemOpen
  \bibfield  {author} {\bibinfo {author} {\bibfnamefont {X.}~\bibnamefont
  {Liu}}, \bibinfo {author} {\bibfnamefont {Z.}~\bibnamefont {Hao}}, \bibinfo
  {author} {\bibfnamefont {E.}~\bibnamefont {Khalaf}}, \bibinfo {author}
  {\bibfnamefont {J.~Y.}\ \bibnamefont {Lee}}, \bibinfo {author} {\bibfnamefont
  {Y.}~\bibnamefont {Ronen}}, \bibinfo {author} {\bibfnamefont
  {H.}~\bibnamefont {Yoo}}, \bibinfo {author} {\bibfnamefont {D.}~\bibnamefont
  {Haei~Najafabadi}}, \bibinfo {author} {\bibfnamefont {K.}~\bibnamefont
  {Watanabe}}, \bibinfo {author} {\bibfnamefont {T.}~\bibnamefont {Taniguchi}},
  \bibinfo {author} {\bibfnamefont {A.}~\bibnamefont {Vishwanath}}, \emph
  {et~al.},\ }\bibfield  {title} {\bibinfo {title} {Tunable spin-polarized
  correlated states in twisted double bilayer graphene},\ }\href@noop {}
  {\bibfield  {journal} {\bibinfo  {journal} {Nature}\ }\textbf {\bibinfo
  {volume} {583}},\ \bibinfo {pages} {221} (\bibinfo {year}
  {2020})}\BibitemShut {NoStop}%
\bibitem [{\citenamefont {Shen}\ \emph {et~al.}(2020)\citenamefont {Shen},
  \citenamefont {Chu}, \citenamefont {Wu}, \citenamefont {Li}, \citenamefont
  {Wang}, \citenamefont {Zhao}, \citenamefont {Tang}, \citenamefont {Liu},
  \citenamefont {Tian}, \citenamefont {Watanabe} \emph
  {et~al.}}]{shen2020correlated}%
  \BibitemOpen
  \bibfield  {author} {\bibinfo {author} {\bibfnamefont {C.}~\bibnamefont
  {Shen}}, \bibinfo {author} {\bibfnamefont {Y.}~\bibnamefont {Chu}}, \bibinfo
  {author} {\bibfnamefont {Q.}~\bibnamefont {Wu}}, \bibinfo {author}
  {\bibfnamefont {N.}~\bibnamefont {Li}}, \bibinfo {author} {\bibfnamefont
  {S.}~\bibnamefont {Wang}}, \bibinfo {author} {\bibfnamefont {Y.}~\bibnamefont
  {Zhao}}, \bibinfo {author} {\bibfnamefont {J.}~\bibnamefont {Tang}}, \bibinfo
  {author} {\bibfnamefont {J.}~\bibnamefont {Liu}}, \bibinfo {author}
  {\bibfnamefont {J.}~\bibnamefont {Tian}}, \bibinfo {author} {\bibfnamefont
  {K.}~\bibnamefont {Watanabe}}, \emph {et~al.},\ }\bibfield  {title} {\bibinfo
  {title} {Correlated states in twisted double bilayer graphene},\ }\href@noop
  {} {\bibfield  {journal} {\bibinfo  {journal} {Nature Physics}\ }\textbf
  {\bibinfo {volume} {16}},\ \bibinfo {pages} {520} (\bibinfo {year}
  {2020})}\BibitemShut {NoStop}%
\bibitem [{\citenamefont {Cao}\ \emph {et~al.}(2020{\natexlab{c}})\citenamefont
  {Cao}, \citenamefont {Rodan-Legrain}, \citenamefont {Rubies-Bigorda},
  \citenamefont {Park}, \citenamefont {Watanabe}, \citenamefont {Taniguchi},\
  and\ \citenamefont {Jarillo-Herrero}}]{cao2020tunable}%
  \BibitemOpen
  \bibfield  {author} {\bibinfo {author} {\bibfnamefont {Y.}~\bibnamefont
  {Cao}}, \bibinfo {author} {\bibfnamefont {D.}~\bibnamefont {Rodan-Legrain}},
  \bibinfo {author} {\bibfnamefont {O.}~\bibnamefont {Rubies-Bigorda}},
  \bibinfo {author} {\bibfnamefont {J.~M.}\ \bibnamefont {Park}}, \bibinfo
  {author} {\bibfnamefont {K.}~\bibnamefont {Watanabe}}, \bibinfo {author}
  {\bibfnamefont {T.}~\bibnamefont {Taniguchi}},\ and\ \bibinfo {author}
  {\bibfnamefont {P.}~\bibnamefont {Jarillo-Herrero}},\ }\bibfield  {title}
  {\bibinfo {title} {Tunable correlated states and spin-polarized phases in
  twisted bilayer--bilayer graphene},\ }\href@noop {} {\bibfield  {journal}
  {\bibinfo  {journal} {Nature}\ }\textbf {\bibinfo {volume} {583}},\ \bibinfo
  {pages} {215} (\bibinfo {year} {2020}{\natexlab{c}})}\BibitemShut {NoStop}%
\bibitem [{\citenamefont {Wang}\ \emph {et~al.}(2020)\citenamefont {Wang},
  \citenamefont {Shih}, \citenamefont {Ghiotto}, \citenamefont {Xian},
  \citenamefont {Rhodes}, \citenamefont {Tan}, \citenamefont {Claassen},
  \citenamefont {Kennes}, \citenamefont {Bai}, \citenamefont {Kim} \emph
  {et~al.}}]{wang2020correlated}%
  \BibitemOpen
  \bibfield  {author} {\bibinfo {author} {\bibfnamefont {L.}~\bibnamefont
  {Wang}}, \bibinfo {author} {\bibfnamefont {E.-M.}\ \bibnamefont {Shih}},
  \bibinfo {author} {\bibfnamefont {A.}~\bibnamefont {Ghiotto}}, \bibinfo
  {author} {\bibfnamefont {L.}~\bibnamefont {Xian}}, \bibinfo {author}
  {\bibfnamefont {D.~A.}\ \bibnamefont {Rhodes}}, \bibinfo {author}
  {\bibfnamefont {C.}~\bibnamefont {Tan}}, \bibinfo {author} {\bibfnamefont
  {M.}~\bibnamefont {Claassen}}, \bibinfo {author} {\bibfnamefont {D.~M.}\
  \bibnamefont {Kennes}}, \bibinfo {author} {\bibfnamefont {Y.}~\bibnamefont
  {Bai}}, \bibinfo {author} {\bibfnamefont {B.}~\bibnamefont {Kim}}, \emph
  {et~al.},\ }\bibfield  {title} {\bibinfo {title} {Correlated electronic
  phases in twisted bilayer transition metal dichalcogenides},\ }\href@noop {}
  {\bibfield  {journal} {\bibinfo  {journal} {Nature materials}\ }\textbf
  {\bibinfo {volume} {19}},\ \bibinfo {pages} {861} (\bibinfo {year}
  {2020})}\BibitemShut {NoStop}%
\bibitem [{\citenamefont {Tang}\ \emph {et~al.}(2020)\citenamefont {Tang},
  \citenamefont {Li}, \citenamefont {Li}, \citenamefont {Xu}, \citenamefont
  {Liu}, \citenamefont {Barmak}, \citenamefont {Watanabe}, \citenamefont
  {Taniguchi}, \citenamefont {MacDonald}, \citenamefont {Shan} \emph
  {et~al.}}]{tang2020simulation}%
  \BibitemOpen
  \bibfield  {author} {\bibinfo {author} {\bibfnamefont {Y.}~\bibnamefont
  {Tang}}, \bibinfo {author} {\bibfnamefont {L.}~\bibnamefont {Li}}, \bibinfo
  {author} {\bibfnamefont {T.}~\bibnamefont {Li}}, \bibinfo {author}
  {\bibfnamefont {Y.}~\bibnamefont {Xu}}, \bibinfo {author} {\bibfnamefont
  {S.}~\bibnamefont {Liu}}, \bibinfo {author} {\bibfnamefont {K.}~\bibnamefont
  {Barmak}}, \bibinfo {author} {\bibfnamefont {K.}~\bibnamefont {Watanabe}},
  \bibinfo {author} {\bibfnamefont {T.}~\bibnamefont {Taniguchi}}, \bibinfo
  {author} {\bibfnamefont {A.~H.}\ \bibnamefont {MacDonald}}, \bibinfo {author}
  {\bibfnamefont {J.}~\bibnamefont {Shan}}, \emph {et~al.},\ }\bibfield
  {title} {\bibinfo {title} {Simulation of hubbard model physics in wse2/ws2
  moir{\'e} superlattices},\ }\href@noop {} {\bibfield  {journal} {\bibinfo
  {journal} {Nature}\ }\textbf {\bibinfo {volume} {579}},\ \bibinfo {pages}
  {353} (\bibinfo {year} {2020})}\BibitemShut {NoStop}%
\bibitem [{\citenamefont {Regan}\ \emph {et~al.}(2020)\citenamefont {Regan},
  \citenamefont {Wang}, \citenamefont {Jin}, \citenamefont {Bakti~Utama},
  \citenamefont {Gao}, \citenamefont {Wei}, \citenamefont {Zhao}, \citenamefont
  {Zhao}, \citenamefont {Zhang}, \citenamefont {Yumigeta} \emph
  {et~al.}}]{regan2020mott}%
  \BibitemOpen
  \bibfield  {author} {\bibinfo {author} {\bibfnamefont {E.~C.}\ \bibnamefont
  {Regan}}, \bibinfo {author} {\bibfnamefont {D.}~\bibnamefont {Wang}},
  \bibinfo {author} {\bibfnamefont {C.}~\bibnamefont {Jin}}, \bibinfo {author}
  {\bibfnamefont {M.~I.}\ \bibnamefont {Bakti~Utama}}, \bibinfo {author}
  {\bibfnamefont {B.}~\bibnamefont {Gao}}, \bibinfo {author} {\bibfnamefont
  {X.}~\bibnamefont {Wei}}, \bibinfo {author} {\bibfnamefont {S.}~\bibnamefont
  {Zhao}}, \bibinfo {author} {\bibfnamefont {W.}~\bibnamefont {Zhao}}, \bibinfo
  {author} {\bibfnamefont {Z.}~\bibnamefont {Zhang}}, \bibinfo {author}
  {\bibfnamefont {K.}~\bibnamefont {Yumigeta}}, \emph {et~al.},\ }\bibfield
  {title} {\bibinfo {title} {Mott and generalized wigner crystal states in
  wse2/ws2 moir{\'e} superlattices},\ }\href@noop {} {\bibfield  {journal}
  {\bibinfo  {journal} {Nature}\ }\textbf {\bibinfo {volume} {579}},\ \bibinfo
  {pages} {359} (\bibinfo {year} {2020})}\BibitemShut {NoStop}%
\bibitem [{\citenamefont {Nayak}\ \emph {et~al.}(2017)\citenamefont {Nayak},
  \citenamefont {Horbatenko}, \citenamefont {Ahn}, \citenamefont {Kim},
  \citenamefont {Lee}, \citenamefont {Ma}, \citenamefont {Jang}, \citenamefont
  {Lim}, \citenamefont {Kim}, \citenamefont {Ryu} \emph
  {et~al.}}]{nayak2017probing}%
  \BibitemOpen
  \bibfield  {author} {\bibinfo {author} {\bibfnamefont {P.~K.}\ \bibnamefont
  {Nayak}}, \bibinfo {author} {\bibfnamefont {Y.}~\bibnamefont {Horbatenko}},
  \bibinfo {author} {\bibfnamefont {S.}~\bibnamefont {Ahn}}, \bibinfo {author}
  {\bibfnamefont {G.}~\bibnamefont {Kim}}, \bibinfo {author} {\bibfnamefont
  {J.-U.}\ \bibnamefont {Lee}}, \bibinfo {author} {\bibfnamefont {K.~Y.}\
  \bibnamefont {Ma}}, \bibinfo {author} {\bibfnamefont {A.-R.}\ \bibnamefont
  {Jang}}, \bibinfo {author} {\bibfnamefont {H.}~\bibnamefont {Lim}}, \bibinfo
  {author} {\bibfnamefont {D.}~\bibnamefont {Kim}}, \bibinfo {author}
  {\bibfnamefont {S.}~\bibnamefont {Ryu}}, \emph {et~al.},\ }\bibfield  {title}
  {\bibinfo {title} {Probing evolution of twist-angle-dependent interlayer
  excitons in mose2/wse2 van der waals heterostructures},\ }\href@noop {}
  {\bibfield  {journal} {\bibinfo  {journal} {ACS nano}\ }\textbf {\bibinfo
  {volume} {11}},\ \bibinfo {pages} {4041} (\bibinfo {year}
  {2017})}\BibitemShut {NoStop}%
\bibitem [{\citenamefont {Rivera}\ \emph {et~al.}(2018)\citenamefont {Rivera},
  \citenamefont {Yu}, \citenamefont {Seyler}, \citenamefont {Wilson},
  \citenamefont {Yao},\ and\ \citenamefont {Xu}}]{rivera2018interlayer}%
  \BibitemOpen
  \bibfield  {author} {\bibinfo {author} {\bibfnamefont {P.}~\bibnamefont
  {Rivera}}, \bibinfo {author} {\bibfnamefont {H.}~\bibnamefont {Yu}}, \bibinfo
  {author} {\bibfnamefont {K.~L.}\ \bibnamefont {Seyler}}, \bibinfo {author}
  {\bibfnamefont {N.~P.}\ \bibnamefont {Wilson}}, \bibinfo {author}
  {\bibfnamefont {W.}~\bibnamefont {Yao}},\ and\ \bibinfo {author}
  {\bibfnamefont {X.}~\bibnamefont {Xu}},\ }\bibfield  {title} {\bibinfo
  {title} {Interlayer valley excitons in heterobilayers of transition metal
  dichalcogenides},\ }\href@noop {} {\bibfield  {journal} {\bibinfo  {journal}
  {Nature nanotechnology}\ }\textbf {\bibinfo {volume} {13}},\ \bibinfo {pages}
  {1004} (\bibinfo {year} {2018})}\BibitemShut {NoStop}%
\bibitem [{\citenamefont {Alexeev}\ \emph {et~al.}(2019)\citenamefont
  {Alexeev}, \citenamefont {Ruiz-Tijerina}, \citenamefont {Danovich},
  \citenamefont {Hamer}, \citenamefont {Terry}, \citenamefont {Nayak},
  \citenamefont {Ahn}, \citenamefont {Pak}, \citenamefont {Lee}, \citenamefont
  {Sohn} \emph {et~al.}}]{alexeev2019resonantly}%
  \BibitemOpen
  \bibfield  {author} {\bibinfo {author} {\bibfnamefont {E.~M.}\ \bibnamefont
  {Alexeev}}, \bibinfo {author} {\bibfnamefont {D.~A.}\ \bibnamefont
  {Ruiz-Tijerina}}, \bibinfo {author} {\bibfnamefont {M.}~\bibnamefont
  {Danovich}}, \bibinfo {author} {\bibfnamefont {M.~J.}\ \bibnamefont {Hamer}},
  \bibinfo {author} {\bibfnamefont {D.~J.}\ \bibnamefont {Terry}}, \bibinfo
  {author} {\bibfnamefont {P.~K.}\ \bibnamefont {Nayak}}, \bibinfo {author}
  {\bibfnamefont {S.}~\bibnamefont {Ahn}}, \bibinfo {author} {\bibfnamefont
  {S.}~\bibnamefont {Pak}}, \bibinfo {author} {\bibfnamefont {J.}~\bibnamefont
  {Lee}}, \bibinfo {author} {\bibfnamefont {J.~I.}\ \bibnamefont {Sohn}}, \emph
  {et~al.},\ }\bibfield  {title} {\bibinfo {title} {Resonantly hybridized
  excitons in moir{\'e} superlattices in van der waals heterostructures},\
  }\href@noop {} {\bibfield  {journal} {\bibinfo  {journal} {Nature}\ }\textbf
  {\bibinfo {volume} {567}},\ \bibinfo {pages} {81} (\bibinfo {year}
  {2019})}\BibitemShut {NoStop}%
\bibitem [{\citenamefont {Andersen}\ \emph {et~al.}(2021)\citenamefont
  {Andersen}, \citenamefont {Scuri}, \citenamefont {Sushko}, \citenamefont
  {De~Greve}, \citenamefont {Sung}, \citenamefont {Zhou}, \citenamefont {Wild},
  \citenamefont {Gelly}, \citenamefont {Heo}, \citenamefont {B{\'e}rub{\'e}}
  \emph {et~al.}}]{andersen2021excitons}%
  \BibitemOpen
  \bibfield  {author} {\bibinfo {author} {\bibfnamefont {T.~I.}\ \bibnamefont
  {Andersen}}, \bibinfo {author} {\bibfnamefont {G.}~\bibnamefont {Scuri}},
  \bibinfo {author} {\bibfnamefont {A.}~\bibnamefont {Sushko}}, \bibinfo
  {author} {\bibfnamefont {K.}~\bibnamefont {De~Greve}}, \bibinfo {author}
  {\bibfnamefont {J.}~\bibnamefont {Sung}}, \bibinfo {author} {\bibfnamefont
  {Y.}~\bibnamefont {Zhou}}, \bibinfo {author} {\bibfnamefont {D.~S.}\
  \bibnamefont {Wild}}, \bibinfo {author} {\bibfnamefont {R.~J.}\ \bibnamefont
  {Gelly}}, \bibinfo {author} {\bibfnamefont {H.}~\bibnamefont {Heo}}, \bibinfo
  {author} {\bibfnamefont {D.}~\bibnamefont {B{\'e}rub{\'e}}}, \emph {et~al.},\
  }\bibfield  {title} {\bibinfo {title} {Excitons in a reconstructed moir{\'e}
  potential in twisted wse2/wse2 homobilayers},\ }\href@noop {} {\bibfield
  {journal} {\bibinfo  {journal} {Nature Materials}\ }\textbf {\bibinfo
  {volume} {20}},\ \bibinfo {pages} {480} (\bibinfo {year} {2021})}\BibitemShut
  {NoStop}%
\bibitem [{\citenamefont {Jin}\ \emph {et~al.}(2019)\citenamefont {Jin},
  \citenamefont {Regan}, \citenamefont {Yan}, \citenamefont {Iqbal
  Bakti~Utama}, \citenamefont {Wang}, \citenamefont {Zhao}, \citenamefont
  {Qin}, \citenamefont {Yang}, \citenamefont {Zheng}, \citenamefont {Shi} \emph
  {et~al.}}]{jin2019observation}%
  \BibitemOpen
  \bibfield  {author} {\bibinfo {author} {\bibfnamefont {C.}~\bibnamefont
  {Jin}}, \bibinfo {author} {\bibfnamefont {E.~C.}\ \bibnamefont {Regan}},
  \bibinfo {author} {\bibfnamefont {A.}~\bibnamefont {Yan}}, \bibinfo {author}
  {\bibfnamefont {M.}~\bibnamefont {Iqbal Bakti~Utama}}, \bibinfo {author}
  {\bibfnamefont {D.}~\bibnamefont {Wang}}, \bibinfo {author} {\bibfnamefont
  {S.}~\bibnamefont {Zhao}}, \bibinfo {author} {\bibfnamefont {Y.}~\bibnamefont
  {Qin}}, \bibinfo {author} {\bibfnamefont {S.}~\bibnamefont {Yang}}, \bibinfo
  {author} {\bibfnamefont {Z.}~\bibnamefont {Zheng}}, \bibinfo {author}
  {\bibfnamefont {S.}~\bibnamefont {Shi}}, \emph {et~al.},\ }\bibfield  {title}
  {\bibinfo {title} {Observation of moir{\'e} excitons in wse2/ws2
  heterostructure superlattices},\ }\href@noop {} {\bibfield  {journal}
  {\bibinfo  {journal} {Nature}\ }\textbf {\bibinfo {volume} {567}},\ \bibinfo
  {pages} {76} (\bibinfo {year} {2019})}\BibitemShut {NoStop}%
\bibitem [{\citenamefont {Burg}\ \emph {et~al.}(2019)\citenamefont {Burg},
  \citenamefont {Zhu}, \citenamefont {Taniguchi}, \citenamefont {Watanabe},
  \citenamefont {MacDonald},\ and\ \citenamefont {Tutuc}}]{burg2019correlated}%
  \BibitemOpen
  \bibfield  {author} {\bibinfo {author} {\bibfnamefont {G.~W.}\ \bibnamefont
  {Burg}}, \bibinfo {author} {\bibfnamefont {J.}~\bibnamefont {Zhu}}, \bibinfo
  {author} {\bibfnamefont {T.}~\bibnamefont {Taniguchi}}, \bibinfo {author}
  {\bibfnamefont {K.}~\bibnamefont {Watanabe}}, \bibinfo {author}
  {\bibfnamefont {A.~H.}\ \bibnamefont {MacDonald}},\ and\ \bibinfo {author}
  {\bibfnamefont {E.}~\bibnamefont {Tutuc}},\ }\bibfield  {title} {\bibinfo
  {title} {Correlated insulating states in twisted double bilayer graphene},\
  }\href@noop {} {\bibfield  {journal} {\bibinfo  {journal} {Physical review
  letters}\ }\textbf {\bibinfo {volume} {123}},\ \bibinfo {pages} {197702}
  (\bibinfo {year} {2019})}\BibitemShut {NoStop}%
\bibitem [{\citenamefont {Rubio-Verd{\'u}}\ \emph {et~al.}(2021)\citenamefont
  {Rubio-Verd{\'u}}, \citenamefont {Turkel}, \citenamefont {Song},
  \citenamefont {Klebl}, \citenamefont {Samajdar}, \citenamefont {Scheurer},
  \citenamefont {Venderbos}, \citenamefont {Watanabe}, \citenamefont
  {Taniguchi}, \citenamefont {Ochoa} \emph {et~al.}}]{rubio2021moire}%
  \BibitemOpen
  \bibfield  {author} {\bibinfo {author} {\bibfnamefont {C.}~\bibnamefont
  {Rubio-Verd{\'u}}}, \bibinfo {author} {\bibfnamefont {S.}~\bibnamefont
  {Turkel}}, \bibinfo {author} {\bibfnamefont {Y.}~\bibnamefont {Song}},
  \bibinfo {author} {\bibfnamefont {L.}~\bibnamefont {Klebl}}, \bibinfo
  {author} {\bibfnamefont {R.}~\bibnamefont {Samajdar}}, \bibinfo {author}
  {\bibfnamefont {M.~S.}\ \bibnamefont {Scheurer}}, \bibinfo {author}
  {\bibfnamefont {J.~W.}\ \bibnamefont {Venderbos}}, \bibinfo {author}
  {\bibfnamefont {K.}~\bibnamefont {Watanabe}}, \bibinfo {author}
  {\bibfnamefont {T.}~\bibnamefont {Taniguchi}}, \bibinfo {author}
  {\bibfnamefont {H.}~\bibnamefont {Ochoa}}, \emph {et~al.},\ }\bibfield
  {title} {\bibinfo {title} {Moir{\'e} nematic phase in twisted double bilayer
  graphene},\ }\href@noop {} {\bibfield  {journal} {\bibinfo  {journal} {Nature
  Physics}\ ,\ \bibinfo {pages} {1}} (\bibinfo {year} {2021})}\BibitemShut
  {NoStop}%
\bibitem [{\citenamefont {Chen}\ \emph
  {et~al.}(2019{\natexlab{a}})\citenamefont {Chen}, \citenamefont {Sharpe},
  \citenamefont {Gallagher}, \citenamefont {Rosen}, \citenamefont {Fox},
  \citenamefont {Jiang}, \citenamefont {Lyu}, \citenamefont {Li}, \citenamefont
  {Watanabe}, \citenamefont {Taniguchi} \emph {et~al.}}]{chen2019signatures}%
  \BibitemOpen
  \bibfield  {author} {\bibinfo {author} {\bibfnamefont {G.}~\bibnamefont
  {Chen}}, \bibinfo {author} {\bibfnamefont {A.~L.}\ \bibnamefont {Sharpe}},
  \bibinfo {author} {\bibfnamefont {P.}~\bibnamefont {Gallagher}}, \bibinfo
  {author} {\bibfnamefont {I.~T.}\ \bibnamefont {Rosen}}, \bibinfo {author}
  {\bibfnamefont {E.~J.}\ \bibnamefont {Fox}}, \bibinfo {author} {\bibfnamefont
  {L.}~\bibnamefont {Jiang}}, \bibinfo {author} {\bibfnamefont
  {B.}~\bibnamefont {Lyu}}, \bibinfo {author} {\bibfnamefont {H.}~\bibnamefont
  {Li}}, \bibinfo {author} {\bibfnamefont {K.}~\bibnamefont {Watanabe}},
  \bibinfo {author} {\bibfnamefont {T.}~\bibnamefont {Taniguchi}}, \emph
  {et~al.},\ }\bibfield  {title} {\bibinfo {title} {Signatures of tunable
  superconductivity in a trilayer graphene moir{\'e} superlattice},\
  }\href@noop {} {\bibfield  {journal} {\bibinfo  {journal} {Nature}\ }\textbf
  {\bibinfo {volume} {572}},\ \bibinfo {pages} {215} (\bibinfo {year}
  {2019}{\natexlab{a}})}\BibitemShut {NoStop}%
\bibitem [{\citenamefont {Chen}\ \emph
  {et~al.}(2019{\natexlab{b}})\citenamefont {Chen}, \citenamefont {Jiang},
  \citenamefont {Wu}, \citenamefont {Lyu}, \citenamefont {Li}, \citenamefont
  {Chittari}, \citenamefont {Watanabe}, \citenamefont {Taniguchi},
  \citenamefont {Shi}, \citenamefont {Jung} \emph {et~al.}}]{chen2019evidence}%
  \BibitemOpen
  \bibfield  {author} {\bibinfo {author} {\bibfnamefont {G.}~\bibnamefont
  {Chen}}, \bibinfo {author} {\bibfnamefont {L.}~\bibnamefont {Jiang}},
  \bibinfo {author} {\bibfnamefont {S.}~\bibnamefont {Wu}}, \bibinfo {author}
  {\bibfnamefont {B.}~\bibnamefont {Lyu}}, \bibinfo {author} {\bibfnamefont
  {H.}~\bibnamefont {Li}}, \bibinfo {author} {\bibfnamefont {B.~L.}\
  \bibnamefont {Chittari}}, \bibinfo {author} {\bibfnamefont {K.}~\bibnamefont
  {Watanabe}}, \bibinfo {author} {\bibfnamefont {T.}~\bibnamefont {Taniguchi}},
  \bibinfo {author} {\bibfnamefont {Z.}~\bibnamefont {Shi}}, \bibinfo {author}
  {\bibfnamefont {J.}~\bibnamefont {Jung}}, \emph {et~al.},\ }\bibfield
  {title} {\bibinfo {title} {Evidence of a gate-tunable mott insulator in a
  trilayer graphene moir{\'e} superlattice},\ }\href@noop {} {\bibfield
  {journal} {\bibinfo  {journal} {Nature Physics}\ }\textbf {\bibinfo {volume}
  {15}},\ \bibinfo {pages} {237} (\bibinfo {year}
  {2019}{\natexlab{b}})}\BibitemShut {NoStop}%
\bibitem [{\citenamefont {Chen}\ \emph {et~al.}(2020)\citenamefont {Chen},
  \citenamefont {Sharpe}, \citenamefont {Fox}, \citenamefont {Zhang},
  \citenamefont {Wang}, \citenamefont {Jiang}, \citenamefont {Lyu},
  \citenamefont {Li}, \citenamefont {Watanabe}, \citenamefont {Taniguchi} \emph
  {et~al.}}]{chen2020tunable}%
  \BibitemOpen
  \bibfield  {author} {\bibinfo {author} {\bibfnamefont {G.}~\bibnamefont
  {Chen}}, \bibinfo {author} {\bibfnamefont {A.~L.}\ \bibnamefont {Sharpe}},
  \bibinfo {author} {\bibfnamefont {E.~J.}\ \bibnamefont {Fox}}, \bibinfo
  {author} {\bibfnamefont {Y.-H.}\ \bibnamefont {Zhang}}, \bibinfo {author}
  {\bibfnamefont {S.}~\bibnamefont {Wang}}, \bibinfo {author} {\bibfnamefont
  {L.}~\bibnamefont {Jiang}}, \bibinfo {author} {\bibfnamefont
  {B.}~\bibnamefont {Lyu}}, \bibinfo {author} {\bibfnamefont {H.}~\bibnamefont
  {Li}}, \bibinfo {author} {\bibfnamefont {K.}~\bibnamefont {Watanabe}},
  \bibinfo {author} {\bibfnamefont {T.}~\bibnamefont {Taniguchi}}, \emph
  {et~al.},\ }\bibfield  {title} {\bibinfo {title} {Tunable correlated chern
  insulator and ferromagnetism in a moir{\'e} superlattice},\ }\href@noop {}
  {\bibfield  {journal} {\bibinfo  {journal} {Nature}\ }\textbf {\bibinfo
  {volume} {579}},\ \bibinfo {pages} {56} (\bibinfo {year} {2020})}\BibitemShut
  {NoStop}%
\bibitem [{\citenamefont {Nuckolls}\ \emph {et~al.}(2020)\citenamefont
  {Nuckolls}, \citenamefont {Oh}, \citenamefont {Wong}, \citenamefont {Lian},
  \citenamefont {Watanabe}, \citenamefont {Taniguchi}, \citenamefont
  {Bernevig},\ and\ \citenamefont {Yazdani}}]{nuckolls2020strongly}%
  \BibitemOpen
  \bibfield  {author} {\bibinfo {author} {\bibfnamefont {K.~P.}\ \bibnamefont
  {Nuckolls}}, \bibinfo {author} {\bibfnamefont {M.}~\bibnamefont {Oh}},
  \bibinfo {author} {\bibfnamefont {D.}~\bibnamefont {Wong}}, \bibinfo {author}
  {\bibfnamefont {B.}~\bibnamefont {Lian}}, \bibinfo {author} {\bibfnamefont
  {K.}~\bibnamefont {Watanabe}}, \bibinfo {author} {\bibfnamefont
  {T.}~\bibnamefont {Taniguchi}}, \bibinfo {author} {\bibfnamefont {B.~A.}\
  \bibnamefont {Bernevig}},\ and\ \bibinfo {author} {\bibfnamefont
  {A.}~\bibnamefont {Yazdani}},\ }\bibfield  {title} {\bibinfo {title}
  {Strongly correlated chern insulators in magic-angle twisted bilayer
  graphene},\ }\href@noop {} {\bibfield  {journal} {\bibinfo  {journal}
  {Nature}\ }\textbf {\bibinfo {volume} {588}},\ \bibinfo {pages} {610}
  (\bibinfo {year} {2020})}\BibitemShut {NoStop}%
\bibitem [{\citenamefont {Xie}\ \emph {et~al.}(2021)\citenamefont {Xie},
  \citenamefont {Pierce}, \citenamefont {Park}, \citenamefont {Parker},
  \citenamefont {Khalaf}, \citenamefont {Ledwith}, \citenamefont {Cao},
  \citenamefont {Lee}, \citenamefont {Chen}, \citenamefont {Forrester} \emph
  {et~al.}}]{xie2021fractional}%
  \BibitemOpen
  \bibfield  {author} {\bibinfo {author} {\bibfnamefont {Y.}~\bibnamefont
  {Xie}}, \bibinfo {author} {\bibfnamefont {A.~T.}\ \bibnamefont {Pierce}},
  \bibinfo {author} {\bibfnamefont {J.~M.}\ \bibnamefont {Park}}, \bibinfo
  {author} {\bibfnamefont {D.~E.}\ \bibnamefont {Parker}}, \bibinfo {author}
  {\bibfnamefont {E.}~\bibnamefont {Khalaf}}, \bibinfo {author} {\bibfnamefont
  {P.}~\bibnamefont {Ledwith}}, \bibinfo {author} {\bibfnamefont
  {Y.}~\bibnamefont {Cao}}, \bibinfo {author} {\bibfnamefont {S.~H.}\
  \bibnamefont {Lee}}, \bibinfo {author} {\bibfnamefont {S.}~\bibnamefont
  {Chen}}, \bibinfo {author} {\bibfnamefont {P.~R.}\ \bibnamefont {Forrester}},
  \emph {et~al.},\ }\bibfield  {title} {\bibinfo {title} {Fractional chern
  insulators in magic-angle twisted bilayer graphene},\ }\href@noop {}
  {\bibfield  {journal} {\bibinfo  {journal} {Nature}\ }\textbf {\bibinfo
  {volume} {600}},\ \bibinfo {pages} {439} (\bibinfo {year}
  {2021})}\BibitemShut {NoStop}%
\bibitem [{\citenamefont {Pierce}\ \emph {et~al.}(2021)\citenamefont {Pierce},
  \citenamefont {Xie}, \citenamefont {Park}, \citenamefont {Khalaf},
  \citenamefont {Lee}, \citenamefont {Cao}, \citenamefont {Parker},
  \citenamefont {Forrester}, \citenamefont {Chen}, \citenamefont {Watanabe}
  \emph {et~al.}}]{pierce2021unconventional}%
  \BibitemOpen
  \bibfield  {author} {\bibinfo {author} {\bibfnamefont {A.~T.}\ \bibnamefont
  {Pierce}}, \bibinfo {author} {\bibfnamefont {Y.}~\bibnamefont {Xie}},
  \bibinfo {author} {\bibfnamefont {J.~M.}\ \bibnamefont {Park}}, \bibinfo
  {author} {\bibfnamefont {E.}~\bibnamefont {Khalaf}}, \bibinfo {author}
  {\bibfnamefont {S.~H.}\ \bibnamefont {Lee}}, \bibinfo {author} {\bibfnamefont
  {Y.}~\bibnamefont {Cao}}, \bibinfo {author} {\bibfnamefont {D.~E.}\
  \bibnamefont {Parker}}, \bibinfo {author} {\bibfnamefont {P.~R.}\
  \bibnamefont {Forrester}}, \bibinfo {author} {\bibfnamefont {S.}~\bibnamefont
  {Chen}}, \bibinfo {author} {\bibfnamefont {K.}~\bibnamefont {Watanabe}},
  \emph {et~al.},\ }\bibfield  {title} {\bibinfo {title} {Unconventional
  sequence of correlated chern insulators in magic-angle twisted bilayer
  graphene},\ }\href@noop {} {\bibfield  {journal} {\bibinfo  {journal} {Nature
  Physics}\ }\textbf {\bibinfo {volume} {17}},\ \bibinfo {pages} {1210}
  (\bibinfo {year} {2021})}\BibitemShut {NoStop}%
\bibitem [{\citenamefont {Stepanov}\ \emph {et~al.}(2021)\citenamefont
  {Stepanov}, \citenamefont {Xie}, \citenamefont {Taniguchi}, \citenamefont
  {Watanabe}, \citenamefont {Lu}, \citenamefont {MacDonald}, \citenamefont
  {Bernevig},\ and\ \citenamefont {Efetov}}]{stepanov2021competing}%
  \BibitemOpen
  \bibfield  {author} {\bibinfo {author} {\bibfnamefont {P.}~\bibnamefont
  {Stepanov}}, \bibinfo {author} {\bibfnamefont {M.}~\bibnamefont {Xie}},
  \bibinfo {author} {\bibfnamefont {T.}~\bibnamefont {Taniguchi}}, \bibinfo
  {author} {\bibfnamefont {K.}~\bibnamefont {Watanabe}}, \bibinfo {author}
  {\bibfnamefont {X.}~\bibnamefont {Lu}}, \bibinfo {author} {\bibfnamefont
  {A.~H.}\ \bibnamefont {MacDonald}}, \bibinfo {author} {\bibfnamefont {B.~A.}\
  \bibnamefont {Bernevig}},\ and\ \bibinfo {author} {\bibfnamefont {D.~K.}\
  \bibnamefont {Efetov}},\ }\bibfield  {title} {\bibinfo {title} {Competing
  zero-field chern insulators in superconducting twisted bilayer graphene},\
  }\href@noop {} {\bibfield  {journal} {\bibinfo  {journal} {Physical review
  letters}\ }\textbf {\bibinfo {volume} {127}},\ \bibinfo {pages} {197701}
  (\bibinfo {year} {2021})}\BibitemShut {NoStop}%
\bibitem [{\citenamefont {Choi}\ \emph {et~al.}(2020)\citenamefont {Choi},
  \citenamefont {Kim}, \citenamefont {Peng}, \citenamefont {Thomson},
  \citenamefont {Lewandowski}, \citenamefont {Polski}, \citenamefont {Zhang},
  \citenamefont {Arora}, \citenamefont {Watanabe}, \citenamefont {Taniguchi}
  \emph {et~al.}}]{choi2020tracing}%
  \BibitemOpen
  \bibfield  {author} {\bibinfo {author} {\bibfnamefont {Y.}~\bibnamefont
  {Choi}}, \bibinfo {author} {\bibfnamefont {H.}~\bibnamefont {Kim}}, \bibinfo
  {author} {\bibfnamefont {Y.}~\bibnamefont {Peng}}, \bibinfo {author}
  {\bibfnamefont {A.}~\bibnamefont {Thomson}}, \bibinfo {author} {\bibfnamefont
  {C.}~\bibnamefont {Lewandowski}}, \bibinfo {author} {\bibfnamefont
  {R.}~\bibnamefont {Polski}}, \bibinfo {author} {\bibfnamefont
  {Y.}~\bibnamefont {Zhang}}, \bibinfo {author} {\bibfnamefont {H.~S.}\
  \bibnamefont {Arora}}, \bibinfo {author} {\bibfnamefont {K.}~\bibnamefont
  {Watanabe}}, \bibinfo {author} {\bibfnamefont {T.}~\bibnamefont {Taniguchi}},
  \emph {et~al.},\ }\bibfield  {title} {\bibinfo {title} {Tracing out
  correlated chern insulators in magic angle twisted bilayer graphene},\
  }\href@noop {} {\bibfield  {journal} {\bibinfo  {journal} {arXiv preprint
  arXiv:2008.11746}\ } (\bibinfo {year} {2020})}\BibitemShut {NoStop}%
\bibitem [{\citenamefont {Sharpe}\ \emph {et~al.}(2019)\citenamefont {Sharpe},
  \citenamefont {Fox}, \citenamefont {Barnard}, \citenamefont {Finney},
  \citenamefont {Watanabe}, \citenamefont {Taniguchi}, \citenamefont
  {Kastner},\ and\ \citenamefont {Goldhaber-Gordon}}]{sharpe2019emergent}%
  \BibitemOpen
  \bibfield  {author} {\bibinfo {author} {\bibfnamefont {A.~L.}\ \bibnamefont
  {Sharpe}}, \bibinfo {author} {\bibfnamefont {E.~J.}\ \bibnamefont {Fox}},
  \bibinfo {author} {\bibfnamefont {A.~W.}\ \bibnamefont {Barnard}}, \bibinfo
  {author} {\bibfnamefont {J.}~\bibnamefont {Finney}}, \bibinfo {author}
  {\bibfnamefont {K.}~\bibnamefont {Watanabe}}, \bibinfo {author}
  {\bibfnamefont {T.}~\bibnamefont {Taniguchi}}, \bibinfo {author}
  {\bibfnamefont {M.}~\bibnamefont {Kastner}},\ and\ \bibinfo {author}
  {\bibfnamefont {D.}~\bibnamefont {Goldhaber-Gordon}},\ }\bibfield  {title}
  {\bibinfo {title} {Emergent ferromagnetism near three-quarters filling in
  twisted bilayer graphene},\ }\href@noop {} {\bibfield  {journal} {\bibinfo
  {journal} {Science}\ }\textbf {\bibinfo {volume} {365}},\ \bibinfo {pages}
  {605} (\bibinfo {year} {2019})}\BibitemShut {NoStop}%
\bibitem [{\citenamefont {Wu}\ \emph {et~al.}(2021)\citenamefont {Wu},
  \citenamefont {Zhang}, \citenamefont {Watanabe}, \citenamefont {Taniguchi},\
  and\ \citenamefont {Andrei}}]{wu2021chern}%
  \BibitemOpen
  \bibfield  {author} {\bibinfo {author} {\bibfnamefont {S.}~\bibnamefont
  {Wu}}, \bibinfo {author} {\bibfnamefont {Z.}~\bibnamefont {Zhang}}, \bibinfo
  {author} {\bibfnamefont {K.}~\bibnamefont {Watanabe}}, \bibinfo {author}
  {\bibfnamefont {T.}~\bibnamefont {Taniguchi}},\ and\ \bibinfo {author}
  {\bibfnamefont {E.~Y.}\ \bibnamefont {Andrei}},\ }\bibfield  {title}
  {\bibinfo {title} {Chern insulators, van hove singularities and topological
  flat bands in magic-angle twisted bilayer graphene},\ }\href@noop {}
  {\bibfield  {journal} {\bibinfo  {journal} {Nature materials}\ }\textbf
  {\bibinfo {volume} {20}},\ \bibinfo {pages} {488} (\bibinfo {year}
  {2021})}\BibitemShut {NoStop}%
\bibitem [{\citenamefont {Das}\ \emph {et~al.}(2021)\citenamefont {Das},
  \citenamefont {Lu}, \citenamefont {Herzog-Arbeitman}, \citenamefont {Song},
  \citenamefont {Watanabe}, \citenamefont {Taniguchi}, \citenamefont
  {Bernevig},\ and\ \citenamefont {Efetov}}]{das2021symmetry}%
  \BibitemOpen
  \bibfield  {author} {\bibinfo {author} {\bibfnamefont {I.}~\bibnamefont
  {Das}}, \bibinfo {author} {\bibfnamefont {X.}~\bibnamefont {Lu}}, \bibinfo
  {author} {\bibfnamefont {J.}~\bibnamefont {Herzog-Arbeitman}}, \bibinfo
  {author} {\bibfnamefont {Z.-D.}\ \bibnamefont {Song}}, \bibinfo {author}
  {\bibfnamefont {K.}~\bibnamefont {Watanabe}}, \bibinfo {author}
  {\bibfnamefont {T.}~\bibnamefont {Taniguchi}}, \bibinfo {author}
  {\bibfnamefont {B.~A.}\ \bibnamefont {Bernevig}},\ and\ \bibinfo {author}
  {\bibfnamefont {D.~K.}\ \bibnamefont {Efetov}},\ }\bibfield  {title}
  {\bibinfo {title} {Symmetry-broken chern insulators and rashba-like
  landau-level crossings in magic-angle bilayer graphene},\ }\href@noop {}
  {\bibfield  {journal} {\bibinfo  {journal} {Nature Physics}\ }\textbf
  {\bibinfo {volume} {17}},\ \bibinfo {pages} {710} (\bibinfo {year}
  {2021})}\BibitemShut {NoStop}%
\bibitem [{\citenamefont {Park}\ \emph
  {et~al.}(2021{\natexlab{b}})\citenamefont {Park}, \citenamefont {Cao},
  \citenamefont {Watanabe}, \citenamefont {Taniguchi},\ and\ \citenamefont
  {Jarillo-Herrero}}]{park2021flavour}%
  \BibitemOpen
  \bibfield  {author} {\bibinfo {author} {\bibfnamefont {J.~M.}\ \bibnamefont
  {Park}}, \bibinfo {author} {\bibfnamefont {Y.}~\bibnamefont {Cao}}, \bibinfo
  {author} {\bibfnamefont {K.}~\bibnamefont {Watanabe}}, \bibinfo {author}
  {\bibfnamefont {T.}~\bibnamefont {Taniguchi}},\ and\ \bibinfo {author}
  {\bibfnamefont {P.}~\bibnamefont {Jarillo-Herrero}},\ }\bibfield  {title}
  {\bibinfo {title} {Flavour hund’s coupling, chern gaps and charge
  diffusivity in moir{\'e} graphene},\ }\href@noop {} {\bibfield  {journal}
  {\bibinfo  {journal} {Nature}\ }\textbf {\bibinfo {volume} {592}},\ \bibinfo
  {pages} {43} (\bibinfo {year} {2021}{\natexlab{b}})}\BibitemShut {NoStop}%
\bibitem [{\citenamefont {Saito}\ \emph {et~al.}(2020)\citenamefont {Saito},
  \citenamefont {Ge}, \citenamefont {Watanabe}, \citenamefont {Taniguchi},\
  and\ \citenamefont {Young}}]{saito2020independent}%
  \BibitemOpen
  \bibfield  {author} {\bibinfo {author} {\bibfnamefont {Y.}~\bibnamefont
  {Saito}}, \bibinfo {author} {\bibfnamefont {J.}~\bibnamefont {Ge}}, \bibinfo
  {author} {\bibfnamefont {K.}~\bibnamefont {Watanabe}}, \bibinfo {author}
  {\bibfnamefont {T.}~\bibnamefont {Taniguchi}},\ and\ \bibinfo {author}
  {\bibfnamefont {A.~F.}\ \bibnamefont {Young}},\ }\bibfield  {title} {\bibinfo
  {title} {Independent superconductors and correlated insulators in twisted
  bilayer graphene},\ }\href@noop {} {\bibfield  {journal} {\bibinfo  {journal}
  {Nature Physics}\ }\textbf {\bibinfo {volume} {16}},\ \bibinfo {pages} {926}
  (\bibinfo {year} {2020})}\BibitemShut {NoStop}%
\bibitem [{\citenamefont {Kennes}\ \emph {et~al.}(2021)\citenamefont {Kennes},
  \citenamefont {Claassen}, \citenamefont {Xian}, \citenamefont {Georges},
  \citenamefont {Millis}, \citenamefont {Hone}, \citenamefont {Dean},
  \citenamefont {Basov}, \citenamefont {Pasupathy},\ and\ \citenamefont
  {Rubio}}]{kennes2021moire}%
  \BibitemOpen
  \bibfield  {author} {\bibinfo {author} {\bibfnamefont {D.~M.}\ \bibnamefont
  {Kennes}}, \bibinfo {author} {\bibfnamefont {M.}~\bibnamefont {Claassen}},
  \bibinfo {author} {\bibfnamefont {L.}~\bibnamefont {Xian}}, \bibinfo {author}
  {\bibfnamefont {A.}~\bibnamefont {Georges}}, \bibinfo {author} {\bibfnamefont
  {A.~J.}\ \bibnamefont {Millis}}, \bibinfo {author} {\bibfnamefont
  {J.}~\bibnamefont {Hone}}, \bibinfo {author} {\bibfnamefont {C.~R.}\
  \bibnamefont {Dean}}, \bibinfo {author} {\bibfnamefont {D.~N.}\ \bibnamefont
  {Basov}}, \bibinfo {author} {\bibfnamefont {A.~N.}\ \bibnamefont
  {Pasupathy}},\ and\ \bibinfo {author} {\bibfnamefont {A.}~\bibnamefont
  {Rubio}},\ }\bibfield  {title} {\bibinfo {title} {Moir{\'e} heterostructures
  as a condensed-matter quantum simulator},\ }\href
  {https://doi.org/10.1038/s41567-020-01154-3} {\bibfield  {journal} {\bibinfo
  {journal} {Nature Physics}\ }\textbf {\bibinfo {volume} {17}},\ \bibinfo
  {pages} {155} (\bibinfo {year} {2021})}\BibitemShut {NoStop}%
\bibitem [{\citenamefont {Carr}\ \emph {et~al.}(2020)\citenamefont {Carr},
  \citenamefont {Fang},\ and\ \citenamefont {Kaxiras}}]{carr2020electronic}%
  \BibitemOpen
  \bibfield  {author} {\bibinfo {author} {\bibfnamefont {S.}~\bibnamefont
  {Carr}}, \bibinfo {author} {\bibfnamefont {S.}~\bibnamefont {Fang}},\ and\
  \bibinfo {author} {\bibfnamefont {E.}~\bibnamefont {Kaxiras}},\ }\bibfield
  {title} {\bibinfo {title} {Electronic-structure methods for twisted moiré
  layers},\ }\href {https://doi.org/10.1038/s41578-020-0214-0} {\bibfield
  {journal} {\bibinfo  {journal} {Nature Reviews Materials}\ }\textbf {\bibinfo
  {volume} {5}},\ \bibinfo {pages} {748–763} (\bibinfo {year}
  {2020})}\BibitemShut {NoStop}%
\bibitem [{\citenamefont {Uchida}\ \emph {et~al.}(2014)\citenamefont {Uchida},
  \citenamefont {Furuya}, \citenamefont {Iwata},\ and\ \citenamefont
  {Oshiyama}}]{uchida2014atomic}%
  \BibitemOpen
  \bibfield  {author} {\bibinfo {author} {\bibfnamefont {K.}~\bibnamefont
  {Uchida}}, \bibinfo {author} {\bibfnamefont {S.}~\bibnamefont {Furuya}},
  \bibinfo {author} {\bibfnamefont {J.-I.}\ \bibnamefont {Iwata}},\ and\
  \bibinfo {author} {\bibfnamefont {A.}~\bibnamefont {Oshiyama}},\ }\bibfield
  {title} {\bibinfo {title} {Atomic corrugation and electron localization due
  to moir\'e patterns in twisted bilayer graphenes},\ }\href
  {https://doi.org/10.1103/PhysRevB.90.155451} {\bibfield  {journal} {\bibinfo
  {journal} {Phys. Rev. B}\ }\textbf {\bibinfo {volume} {90}},\ \bibinfo
  {pages} {155451} (\bibinfo {year} {2014})}\BibitemShut {NoStop}%
\bibitem [{\citenamefont {Lucignano}\ \emph {et~al.}(2019)\citenamefont
  {Lucignano}, \citenamefont {Alf\`e}, \citenamefont {Cataudella},
  \citenamefont {Ninno},\ and\ \citenamefont {Cantele}}]{lucignano2019crucial}%
  \BibitemOpen
  \bibfield  {author} {\bibinfo {author} {\bibfnamefont {P.}~\bibnamefont
  {Lucignano}}, \bibinfo {author} {\bibfnamefont {D.}~\bibnamefont {Alf\`e}},
  \bibinfo {author} {\bibfnamefont {V.}~\bibnamefont {Cataudella}}, \bibinfo
  {author} {\bibfnamefont {D.}~\bibnamefont {Ninno}},\ and\ \bibinfo {author}
  {\bibfnamefont {G.}~\bibnamefont {Cantele}},\ }\bibfield  {title} {\bibinfo
  {title} {Crucial role of atomic corrugation on the flat bands and energy gaps
  of twisted bilayer graphene at the magic angle
  $\ensuremath{\theta}\ensuremath{\sim}1.{08}^{\ensuremath{\circ}}$},\ }\href
  {https://doi.org/10.1103/PhysRevB.99.195419} {\bibfield  {journal} {\bibinfo
  {journal} {Phys. Rev. B}\ }\textbf {\bibinfo {volume} {99}},\ \bibinfo
  {pages} {195419} (\bibinfo {year} {2019})}\BibitemShut {NoStop}%
\bibitem [{\citenamefont {Guinea}\ and\ \citenamefont
  {Walet}(2019)}]{guinea2019continuum}%
  \BibitemOpen
  \bibfield  {author} {\bibinfo {author} {\bibfnamefont {F.}~\bibnamefont
  {Guinea}}\ and\ \bibinfo {author} {\bibfnamefont {N.~R.}\ \bibnamefont
  {Walet}},\ }\bibfield  {title} {\bibinfo {title} {Continuum models for
  twisted bilayer graphene: the effects of lattice deformation and hopping
  parameters},\ }\href@noop {} {\bibfield  {journal} {\bibinfo  {journal}
  {Phys. Rev. B}\ }\textbf {\bibinfo {volume} {99}},\ \bibinfo {pages} {205134}
  (\bibinfo {year} {2019})}\BibitemShut {NoStop}%
\bibitem [{\citenamefont {Cantele}\ \emph {et~al.}(2020)\citenamefont
  {Cantele}, \citenamefont {Alf\`e}, \citenamefont {Conte}, \citenamefont
  {Cataudella}, \citenamefont {Ninno},\ and\ \citenamefont
  {Lucignano}}]{cantele2020structural}%
  \BibitemOpen
  \bibfield  {author} {\bibinfo {author} {\bibfnamefont {G.}~\bibnamefont
  {Cantele}}, \bibinfo {author} {\bibfnamefont {D.}~\bibnamefont {Alf\`e}},
  \bibinfo {author} {\bibfnamefont {F.}~\bibnamefont {Conte}}, \bibinfo
  {author} {\bibfnamefont {V.}~\bibnamefont {Cataudella}}, \bibinfo {author}
  {\bibfnamefont {D.}~\bibnamefont {Ninno}},\ and\ \bibinfo {author}
  {\bibfnamefont {P.}~\bibnamefont {Lucignano}},\ }\bibfield  {title} {\bibinfo
  {title} {Structural relaxation and low energy properties of twisted bilayer
  graphene},\ }\href@noop {} {\bibfield  {journal} {\bibinfo  {journal}
  {arXiv:2004.14323v1}\ } (\bibinfo {year} {2020})}\BibitemShut {NoStop}%
\bibitem [{\citenamefont {McGilly}\ \emph {et~al.}(2020)\citenamefont
  {McGilly}, \citenamefont {Kerelsky}, \citenamefont {Finney}, \citenamefont
  {Shapovalov}, \citenamefont {Shih}, \citenamefont {Ghiotto}, \citenamefont
  {Zeng}, \citenamefont {Moore}, \citenamefont {Wu}, \citenamefont {Bai},
  \citenamefont {Watanabe}, \citenamefont {Taniguchi}, \citenamefont {Stengel},
  \citenamefont {Zhou}, \citenamefont {Hone}, \citenamefont {Zhu},
  \citenamefont {Basov}, \citenamefont {Dean}, \citenamefont {Dreyer},\ and\
  \citenamefont {Pasupathy}}]{McGilly2020c}%
  \BibitemOpen
  \bibfield  {author} {\bibinfo {author} {\bibfnamefont {L.~J.}\ \bibnamefont
  {McGilly}}, \bibinfo {author} {\bibfnamefont {A.}~\bibnamefont {Kerelsky}},
  \bibinfo {author} {\bibfnamefont {N.~R.}\ \bibnamefont {Finney}}, \bibinfo
  {author} {\bibfnamefont {K.}~\bibnamefont {Shapovalov}}, \bibinfo {author}
  {\bibfnamefont {E.~M.}\ \bibnamefont {Shih}}, \bibinfo {author}
  {\bibfnamefont {A.}~\bibnamefont {Ghiotto}}, \bibinfo {author} {\bibfnamefont
  {Y.}~\bibnamefont {Zeng}}, \bibinfo {author} {\bibfnamefont {S.~L.}\
  \bibnamefont {Moore}}, \bibinfo {author} {\bibfnamefont {W.}~\bibnamefont
  {Wu}}, \bibinfo {author} {\bibfnamefont {Y.}~\bibnamefont {Bai}}, \bibinfo
  {author} {\bibfnamefont {K.}~\bibnamefont {Watanabe}}, \bibinfo {author}
  {\bibfnamefont {T.}~\bibnamefont {Taniguchi}}, \bibinfo {author}
  {\bibfnamefont {M.}~\bibnamefont {Stengel}}, \bibinfo {author} {\bibfnamefont
  {L.}~\bibnamefont {Zhou}}, \bibinfo {author} {\bibfnamefont {J.}~\bibnamefont
  {Hone}}, \bibinfo {author} {\bibfnamefont {X.}~\bibnamefont {Zhu}}, \bibinfo
  {author} {\bibfnamefont {D.~N.}\ \bibnamefont {Basov}}, \bibinfo {author}
  {\bibfnamefont {C.}~\bibnamefont {Dean}}, \bibinfo {author} {\bibfnamefont
  {C.~E.}\ \bibnamefont {Dreyer}},\ and\ \bibinfo {author} {\bibfnamefont
  {A.~N.}\ \bibnamefont {Pasupathy}},\ }\bibfield  {title} {\bibinfo {title}
  {{Visualization of moir{\'{e}} superlattices}},\ }\href
  {https://doi.org/10.1038/s41565-020-0708-3} {\bibfield  {journal} {\bibinfo
  {journal} {Nature Nanotechnology}\ }\textbf {\bibinfo {volume} {15}},\
  \bibinfo {pages} {580} (\bibinfo {year} {2020})}\BibitemShut {NoStop}%
\bibitem [{\citenamefont {Cook}\ \emph {et~al.}(2022)\citenamefont {Cook},
  \citenamefont {Halbertal}, \citenamefont {Lu}, \citenamefont {Snyder},
  \citenamefont {Hor}, \citenamefont {Basov},\ and\ \citenamefont
  {Bian}}]{cook2022moire}%
  \BibitemOpen
  \bibfield  {author} {\bibinfo {author} {\bibfnamefont {J.}~\bibnamefont
  {Cook}}, \bibinfo {author} {\bibfnamefont {D.}~\bibnamefont {Halbertal}},
  \bibinfo {author} {\bibfnamefont {Q.}~\bibnamefont {Lu}}, \bibinfo {author}
  {\bibfnamefont {M.}~\bibnamefont {Snyder}}, \bibinfo {author} {\bibfnamefont
  {Y.~S.}\ \bibnamefont {Hor}}, \bibinfo {author} {\bibfnamefont {D.~N.}\
  \bibnamefont {Basov}},\ and\ \bibinfo {author} {\bibfnamefont
  {G.}~\bibnamefont {Bian}},\ }\href
  {https://doi.org/10.48550/ARXIV.2206.07035} {\bibinfo {title} {Moiré
  modulated lattice strain and thickness-dependent lattice expansion in
  epitaxial ultrathin films of pdte$_2$}} (\bibinfo {year} {2022})\BibitemShut
  {NoStop}%
\bibitem [{\citenamefont {Naik}\ and\ \citenamefont {Jain}(2018)}]{Naik2018}%
  \BibitemOpen
  \bibfield  {author} {\bibinfo {author} {\bibfnamefont {M.~H.}\ \bibnamefont
  {Naik}}\ and\ \bibinfo {author} {\bibfnamefont {M.}~\bibnamefont {Jain}},\
  }\bibfield  {title} {\bibinfo {title} {{Ultraflatbands and Shear Solitons in
  Moir{\'{e}} Patterns of Twisted Bilayer Transition Metal Dichalcogenides}},\
  }\href {https://doi.org/10.1103/PhysRevLett.121.266401} {\bibfield  {journal}
  {\bibinfo  {journal} {Physical Review Letters}\ }\textbf {\bibinfo {volume}
  {121}},\ \bibinfo {pages} {266401} (\bibinfo {year} {2018})}\BibitemShut
  {NoStop}%
\bibitem [{\citenamefont {Gargiulo}\ and\ \citenamefont
  {Yazyev}(2018)}]{Gargiulo2018}%
  \BibitemOpen
  \bibfield  {author} {\bibinfo {author} {\bibfnamefont {F.}~\bibnamefont
  {Gargiulo}}\ and\ \bibinfo {author} {\bibfnamefont {O.~V.}\ \bibnamefont
  {Yazyev}},\ }\bibfield  {title} {\bibinfo {title} {{Structural and electronic
  transformation in low-angle twisted bilayer graphene}},\ }\bibfield
  {journal} {\bibinfo  {journal} {2D Materials}\ }\textbf {\bibinfo {volume}
  {5}},\ \href {https://doi.org/10.1088/2053-1583/aa9640}
  {10.1088/2053-1583/aa9640} (\bibinfo {year} {2018})\BibitemShut {NoStop}%
\bibitem [{\citenamefont {Haddadi}\ \emph {et~al.}(2020)\citenamefont
  {Haddadi}, \citenamefont {Wu}, \citenamefont {Kruchkov},\ and\ \citenamefont
  {Yazyev}}]{Haddadi2020}%
  \BibitemOpen
  \bibfield  {author} {\bibinfo {author} {\bibfnamefont {F.}~\bibnamefont
  {Haddadi}}, \bibinfo {author} {\bibfnamefont {Q.~S.}\ \bibnamefont {Wu}},
  \bibinfo {author} {\bibfnamefont {A.~J.}\ \bibnamefont {Kruchkov}},\ and\
  \bibinfo {author} {\bibfnamefont {O.~V.}\ \bibnamefont {Yazyev}},\ }\bibfield
   {title} {\bibinfo {title} {{Moir{\'{e}} Flat Bands in Twisted Double Bilayer
  Graphene}},\ }\href {https://doi.org/10.1021/acs.nanolett.9b05117} {\bibfield
   {journal} {\bibinfo  {journal} {Nano Letters}\ }\textbf {\bibinfo {volume}
  {20}},\ \bibinfo {pages} {2410} (\bibinfo {year} {2020})}\BibitemShut
  {NoStop}%
\bibitem [{\citenamefont {Van~Wijk}\ \emph {et~al.}(2015)\citenamefont
  {Van~Wijk}, \citenamefont {Schuring}, \citenamefont {Katsnelson},\ and\
  \citenamefont {Fasolino}}]{van2015relaxation}%
  \BibitemOpen
  \bibfield  {author} {\bibinfo {author} {\bibfnamefont {M.}~\bibnamefont
  {Van~Wijk}}, \bibinfo {author} {\bibfnamefont {A.}~\bibnamefont {Schuring}},
  \bibinfo {author} {\bibfnamefont {M.}~\bibnamefont {Katsnelson}},\ and\
  \bibinfo {author} {\bibfnamefont {A.}~\bibnamefont {Fasolino}},\ }\bibfield
  {title} {\bibinfo {title} {Relaxation of moir{\'e} patterns for slightly
  misaligned identical lattices: graphene on graphite},\ }\href@noop {}
  {\bibfield  {journal} {\bibinfo  {journal} {2D Materials}\ }\textbf {\bibinfo
  {volume} {2}},\ \bibinfo {pages} {034010} (\bibinfo {year}
  {2015})}\BibitemShut {NoStop}%
\bibitem [{\citenamefont {Maity}\ \emph {et~al.}(2021)\citenamefont {Maity},
  \citenamefont {Maiti}, \citenamefont {Krishnamurthy},\ and\ \citenamefont
  {Jain}}]{Maity2021}%
  \BibitemOpen
  \bibfield  {author} {\bibinfo {author} {\bibfnamefont {I.}~\bibnamefont
  {Maity}}, \bibinfo {author} {\bibfnamefont {P.~K.}\ \bibnamefont {Maiti}},
  \bibinfo {author} {\bibfnamefont {H.~R.}\ \bibnamefont {Krishnamurthy}},\
  and\ \bibinfo {author} {\bibfnamefont {M.}~\bibnamefont {Jain}},\ }\bibfield
  {title} {\bibinfo {title} {{Reconstruction of moir{\'{e}} lattices in twisted
  transition metal dichalcogenide bilayers}},\ }\href
  {https://doi.org/10.1103/PhysRevB.103.L121102} {\bibfield  {journal}
  {\bibinfo  {journal} {Physical Review B}\ }\textbf {\bibinfo {volume}
  {103}},\ \bibinfo {pages} {L121102} (\bibinfo {year} {2021})}\BibitemShut
  {NoStop}%
\bibitem [{\citenamefont {Jain}\ \emph {et~al.}(2016)\citenamefont {Jain},
  \citenamefont {Juri{\v{c}}i{\'c}},\ and\ \citenamefont
  {Barkema}}]{jain2016structure}%
  \BibitemOpen
  \bibfield  {author} {\bibinfo {author} {\bibfnamefont {S.~K.}\ \bibnamefont
  {Jain}}, \bibinfo {author} {\bibfnamefont {V.}~\bibnamefont
  {Juri{\v{c}}i{\'c}}},\ and\ \bibinfo {author} {\bibfnamefont {G.~T.}\
  \bibnamefont {Barkema}},\ }\bibfield  {title} {\bibinfo {title} {Structure of
  twisted and buckled bilayer graphene},\ }\href@noop {} {\bibfield  {journal}
  {\bibinfo  {journal} {2D Materials}\ }\textbf {\bibinfo {volume} {4}},\
  \bibinfo {pages} {015018} (\bibinfo {year} {2016})}\BibitemShut {NoStop}%
\bibitem [{\citenamefont {Carr}\ \emph {et~al.}(2018)\citenamefont {Carr},
  \citenamefont {Massatt}, \citenamefont {Torrisi}, \citenamefont {Cazeaux},
  \citenamefont {Luskin},\ and\ \citenamefont {Kaxiras}}]{Carr2018}%
  \BibitemOpen
  \bibfield  {author} {\bibinfo {author} {\bibfnamefont {S.}~\bibnamefont
  {Carr}}, \bibinfo {author} {\bibfnamefont {D.}~\bibnamefont {Massatt}},
  \bibinfo {author} {\bibfnamefont {S.~B.}\ \bibnamefont {Torrisi}}, \bibinfo
  {author} {\bibfnamefont {P.}~\bibnamefont {Cazeaux}}, \bibinfo {author}
  {\bibfnamefont {M.}~\bibnamefont {Luskin}},\ and\ \bibinfo {author}
  {\bibfnamefont {E.}~\bibnamefont {Kaxiras}},\ }\bibfield  {title} {\bibinfo
  {title} {{Relaxation and domain formation in incommensurate two-dimensional
  heterostructures}},\ }\href {https://doi.org/10.1103/PhysRevB.98.224102}
  {\bibfield  {journal} {\bibinfo  {journal} {Physical Review B}\ }\textbf
  {\bibinfo {volume} {98}},\ \bibinfo {pages} {224102} (\bibinfo {year}
  {2018})}\BibitemShut {NoStop}%
\bibitem [{\citenamefont {Zhu}\ \emph {et~al.}(2020)\citenamefont {Zhu},
  \citenamefont {Carr}, \citenamefont {Massatt}, \citenamefont {Luskin},\ and\
  \citenamefont {Kaxiras}}]{Zhu2020}%
  \BibitemOpen
  \bibfield  {author} {\bibinfo {author} {\bibfnamefont {Z.}~\bibnamefont
  {Zhu}}, \bibinfo {author} {\bibfnamefont {S.}~\bibnamefont {Carr}}, \bibinfo
  {author} {\bibfnamefont {D.}~\bibnamefont {Massatt}}, \bibinfo {author}
  {\bibfnamefont {M.}~\bibnamefont {Luskin}},\ and\ \bibinfo {author}
  {\bibfnamefont {E.}~\bibnamefont {Kaxiras}},\ }\bibfield  {title} {\bibinfo
  {title} {{Twisted Trilayer Graphene: A Precisely Tunable Platform for
  Correlated Electrons}},\ }\href
  {https://doi.org/10.1103/PHYSREVLETT.125.116404} {\bibfield  {journal}
  {\bibinfo  {journal} {Physical Review Letters}\ }\textbf {\bibinfo {volume}
  {125}},\ \bibinfo {pages} {1} (\bibinfo {year} {2020})}\BibitemShut {NoStop}%
\bibitem [{\citenamefont {Cazeaux}\ \emph {et~al.}(2020)\citenamefont
  {Cazeaux}, \citenamefont {Luskin},\ and\ \citenamefont
  {Massatt}}]{Cazeaux2020}%
  \BibitemOpen
  \bibfield  {author} {\bibinfo {author} {\bibfnamefont {P.}~\bibnamefont
  {Cazeaux}}, \bibinfo {author} {\bibfnamefont {M.}~\bibnamefont {Luskin}},\
  and\ \bibinfo {author} {\bibfnamefont {D.}~\bibnamefont {Massatt}},\
  }\bibfield  {title} {\bibinfo {title} {{Energy Minimization of Two
  Dimensional Incommensurate Heterostructures}},\ }\href
  {https://doi.org/10.1007/s00205-019-01444-y} {\bibfield  {journal} {\bibinfo
  {journal} {Archive for Rational Mechanics and Analysis}\ }\textbf {\bibinfo
  {volume} {235}},\ \bibinfo {pages} {1289} (\bibinfo {year}
  {2020})}\BibitemShut {NoStop}%
\bibitem [{\citenamefont {Halbertal}\ \emph {et~al.}(2021)\citenamefont
  {Halbertal}, \citenamefont {Finney}, \citenamefont {Sunku}, \citenamefont
  {Kerelsky}, \citenamefont {Rubio-Verd{\'{u}}}, \citenamefont {Shabani},
  \citenamefont {Xian}, \citenamefont {Carr}, \citenamefont {Chen},
  \citenamefont {Zhang}, \citenamefont {Wang}, \citenamefont
  {Gonzalez-Acevedo}, \citenamefont {McLeod}, \citenamefont {Rhodes},
  \citenamefont {Watanabe}, \citenamefont {Taniguchi}, \citenamefont {Kaxiras},
  \citenamefont {Dean}, \citenamefont {Hone}, \citenamefont {Pasupathy},
  \citenamefont {Kennes}, \citenamefont {Rubio},\ and\ \citenamefont
  {Basov}}]{Halbertal2021}%
  \BibitemOpen
  \bibfield  {author} {\bibinfo {author} {\bibfnamefont {D.}~\bibnamefont
  {Halbertal}}, \bibinfo {author} {\bibfnamefont {N.~R.}\ \bibnamefont
  {Finney}}, \bibinfo {author} {\bibfnamefont {S.~S.}\ \bibnamefont {Sunku}},
  \bibinfo {author} {\bibfnamefont {A.}~\bibnamefont {Kerelsky}}, \bibinfo
  {author} {\bibfnamefont {C.}~\bibnamefont {Rubio-Verd{\'{u}}}}, \bibinfo
  {author} {\bibfnamefont {S.}~\bibnamefont {Shabani}}, \bibinfo {author}
  {\bibfnamefont {L.}~\bibnamefont {Xian}}, \bibinfo {author} {\bibfnamefont
  {S.}~\bibnamefont {Carr}}, \bibinfo {author} {\bibfnamefont {S.}~\bibnamefont
  {Chen}}, \bibinfo {author} {\bibfnamefont {C.}~\bibnamefont {Zhang}},
  \bibinfo {author} {\bibfnamefont {L.}~\bibnamefont {Wang}}, \bibinfo {author}
  {\bibfnamefont {D.}~\bibnamefont {Gonzalez-Acevedo}}, \bibinfo {author}
  {\bibfnamefont {A.~S.}\ \bibnamefont {McLeod}}, \bibinfo {author}
  {\bibfnamefont {D.}~\bibnamefont {Rhodes}}, \bibinfo {author} {\bibfnamefont
  {K.}~\bibnamefont {Watanabe}}, \bibinfo {author} {\bibfnamefont
  {T.}~\bibnamefont {Taniguchi}}, \bibinfo {author} {\bibfnamefont
  {E.}~\bibnamefont {Kaxiras}}, \bibinfo {author} {\bibfnamefont {C.~R.}\
  \bibnamefont {Dean}}, \bibinfo {author} {\bibfnamefont {J.~C.}\ \bibnamefont
  {Hone}}, \bibinfo {author} {\bibfnamefont {A.~N.}\ \bibnamefont {Pasupathy}},
  \bibinfo {author} {\bibfnamefont {D.~M.}\ \bibnamefont {Kennes}}, \bibinfo
  {author} {\bibfnamefont {A.}~\bibnamefont {Rubio}},\ and\ \bibinfo {author}
  {\bibfnamefont {D.~N.}\ \bibnamefont {Basov}},\ }\bibfield  {title} {\bibinfo
  {title} {{Moir{\'{e}} metrology of energy landscapes in van der Waals
  heterostructures}},\ }\href {https://doi.org/10.1038/s41467-020-20428-1}
  {\bibfield  {journal} {\bibinfo  {journal} {Nature Communications}\ }\textbf
  {\bibinfo {volume} {12}},\ \bibinfo {pages} {1} (\bibinfo {year}
  {2021})}\BibitemShut {NoStop}%
\bibitem [{\citenamefont {Halbertal}\ \emph {et~al.}(2022)\citenamefont
  {Halbertal}, \citenamefont {Klebl}, \citenamefont {Hsieh}, \citenamefont
  {Cook}, \citenamefont {Carr}, \citenamefont {Bian}, \citenamefont {Dean},
  \citenamefont {Kennes},\ and\ \citenamefont {Basov}}]{supplement}%
  \BibitemOpen
  \bibfield  {author} {\bibinfo {author} {\bibfnamefont {D.}~\bibnamefont
  {Halbertal}}, \bibinfo {author} {\bibfnamefont {L.}~\bibnamefont {Klebl}},
  \bibinfo {author} {\bibfnamefont {V.}~\bibnamefont {Hsieh}}, \bibinfo
  {author} {\bibfnamefont {J.}~\bibnamefont {Cook}}, \bibinfo {author}
  {\bibfnamefont {S.}~\bibnamefont {Carr}}, \bibinfo {author} {\bibfnamefont
  {G.}~\bibnamefont {Bian}}, \bibinfo {author} {\bibfnamefont {C.}~\bibnamefont
  {Dean}}, \bibinfo {author} {\bibfnamefont {D.~M.}\ \bibnamefont {Kennes}},\
  and\ \bibinfo {author} {\bibfnamefont {D.~N.}\ \bibnamefont {Basov}},\
  }\href@noop {} {\bibinfo {title} {Supplementary material}} (\bibinfo {year}
  {2022})\BibitemShut {NoStop}%
\bibitem [{\citenamefont {Morozov}\ \emph {et~al.}(2006)\citenamefont
  {Morozov}, \citenamefont {Novoselov}, \citenamefont {Katsnelson},
  \citenamefont {Schedin}, \citenamefont {Ponomarenko}, \citenamefont {Jiang},\
  and\ \citenamefont {Geim}}]{morozov2006strong}%
  \BibitemOpen
  \bibfield  {author} {\bibinfo {author} {\bibfnamefont {S.~V.}\ \bibnamefont
  {Morozov}}, \bibinfo {author} {\bibfnamefont {K.~S.}\ \bibnamefont
  {Novoselov}}, \bibinfo {author} {\bibfnamefont {M.~I.}\ \bibnamefont
  {Katsnelson}}, \bibinfo {author} {\bibfnamefont {F.}~\bibnamefont {Schedin}},
  \bibinfo {author} {\bibfnamefont {L.~A.}\ \bibnamefont {Ponomarenko}},
  \bibinfo {author} {\bibfnamefont {D.}~\bibnamefont {Jiang}},\ and\ \bibinfo
  {author} {\bibfnamefont {A.~K.}\ \bibnamefont {Geim}},\ }\bibfield  {title}
  {\bibinfo {title} {Strong suppression of weak localization in graphene},\
  }\href {https://doi.org/10.1103/PhysRevLett.97.016801} {\bibfield  {journal}
  {\bibinfo  {journal} {Phys. Rev. Lett.}\ }\textbf {\bibinfo {volume} {97}},\
  \bibinfo {pages} {016801} (\bibinfo {year} {2006})}\BibitemShut {NoStop}%
\bibitem [{\citenamefont {Vozmediano}\ \emph {et~al.}(2010)\citenamefont
  {Vozmediano}, \citenamefont {Katsnelson},\ and\ \citenamefont
  {Guinea}}]{vozmediano2010gauge}%
  \BibitemOpen
  \bibfield  {author} {\bibinfo {author} {\bibfnamefont {M.}~\bibnamefont
  {Vozmediano}}, \bibinfo {author} {\bibfnamefont {M.}~\bibnamefont
  {Katsnelson}},\ and\ \bibinfo {author} {\bibfnamefont {F.}~\bibnamefont
  {Guinea}},\ }\bibfield  {title} {\bibinfo {title} {Gauge fields in
  graphene},\ }\href
  {https://doi.org/https://doi.org/10.1016/j.physrep.2010.07.003} {\bibfield
  {journal} {\bibinfo  {journal} {Physics Reports}\ }\textbf {\bibinfo {volume}
  {496}},\ \bibinfo {pages} {109} (\bibinfo {year} {2010})}\BibitemShut
  {NoStop}%
\bibitem [{\citenamefont {Katsnelson}(2012)}]{katsnelson2012graphene}%
  \BibitemOpen
  \bibfield  {author} {\bibinfo {author} {\bibfnamefont {M.~I.}\ \bibnamefont
  {Katsnelson}},\ }\href {https://doi.org/10.1017/CBO9781139031080} {\emph
  {\bibinfo {title} {Graphene: Carbon in Two Dimensions}}}\ (\bibinfo
  {publisher} {Cambridge University Press},\ \bibinfo {year}
  {2012})\BibitemShut {NoStop}%
\bibitem [{\citenamefont {Amorim}\ \emph {et~al.}(2016)\citenamefont {Amorim},
  \citenamefont {Cortijo}, \citenamefont {{de Juan}}, \citenamefont {Grushin},
  \citenamefont {Guinea}, \citenamefont {Gutiérrez-Rubio}, \citenamefont
  {Ochoa}, \citenamefont {Parente}, \citenamefont {Roldán}, \citenamefont
  {San-Jose}, \citenamefont {Schiefele}, \citenamefont {Sturla},\ and\
  \citenamefont {Vozmediano}}]{amorim2016novel}%
  \BibitemOpen
  \bibfield  {author} {\bibinfo {author} {\bibfnamefont {B.}~\bibnamefont
  {Amorim}}, \bibinfo {author} {\bibfnamefont {A.}~\bibnamefont {Cortijo}},
  \bibinfo {author} {\bibfnamefont {F.}~\bibnamefont {{de Juan}}}, \bibinfo
  {author} {\bibfnamefont {A.}~\bibnamefont {Grushin}}, \bibinfo {author}
  {\bibfnamefont {F.}~\bibnamefont {Guinea}}, \bibinfo {author} {\bibfnamefont
  {A.}~\bibnamefont {Gutiérrez-Rubio}}, \bibinfo {author} {\bibfnamefont
  {H.}~\bibnamefont {Ochoa}}, \bibinfo {author} {\bibfnamefont
  {V.}~\bibnamefont {Parente}}, \bibinfo {author} {\bibfnamefont
  {R.}~\bibnamefont {Roldán}}, \bibinfo {author} {\bibfnamefont
  {P.}~\bibnamefont {San-Jose}}, \bibinfo {author} {\bibfnamefont
  {J.}~\bibnamefont {Schiefele}}, \bibinfo {author} {\bibfnamefont
  {M.}~\bibnamefont {Sturla}},\ and\ \bibinfo {author} {\bibfnamefont
  {M.}~\bibnamefont {Vozmediano}},\ }\bibfield  {title} {\bibinfo {title}
  {Novel effects of strains in graphene and other two dimensional materials},\
  }\href {https://doi.org/https://doi.org/10.1016/j.physrep.2015.12.006}
  {\bibfield  {journal} {\bibinfo  {journal} {Physics Reports}\ }\textbf
  {\bibinfo {volume} {617}},\ \bibinfo {pages} {1} (\bibinfo {year} {2016})},\
  \bibinfo {note} {novel effects of strains in graphene and other two
  dimensional materials}\BibitemShut {NoStop}%
\bibitem [{\citenamefont {Nam}\ and\ \citenamefont {Koshino}(2017)}]{Nam2017}%
  \BibitemOpen
  \bibfield  {author} {\bibinfo {author} {\bibfnamefont {N.~N.}\ \bibnamefont
  {Nam}}\ and\ \bibinfo {author} {\bibfnamefont {M.}~\bibnamefont {Koshino}},\
  }\bibfield  {title} {\bibinfo {title} {{Lattice relaxation and energy band
  modulation in twisted bilayer graphene}},\ }\href
  {https://doi.org/10.1103/PhysRevB.96.075311} {\bibfield  {journal} {\bibinfo
  {journal} {Physical Review B}\ }\textbf {\bibinfo {volume} {96}},\ \bibinfo
  {pages} {1} (\bibinfo {year} {2017})}\BibitemShut {NoStop}%
\bibitem [{\citenamefont {Liang}\ \emph {et~al.}(2020)\citenamefont {Liang},
  \citenamefont {Goodwin}, \citenamefont {Vitale}, \citenamefont {Corsetti},
  \citenamefont {Mostofi},\ and\ \citenamefont {Lischner}}]{liang2020effect}%
  \BibitemOpen
  \bibfield  {author} {\bibinfo {author} {\bibfnamefont {X.}~\bibnamefont
  {Liang}}, \bibinfo {author} {\bibfnamefont {Z.~A.~H.}\ \bibnamefont
  {Goodwin}}, \bibinfo {author} {\bibfnamefont {V.}~\bibnamefont {Vitale}},
  \bibinfo {author} {\bibfnamefont {F.}~\bibnamefont {Corsetti}}, \bibinfo
  {author} {\bibfnamefont {A.~A.}\ \bibnamefont {Mostofi}},\ and\ \bibinfo
  {author} {\bibfnamefont {J.}~\bibnamefont {Lischner}},\ }\bibfield  {title}
  {\bibinfo {title} {Effect of bilayer stacking on the atomic and electronic
  structure of twisted double bilayer graphene},\ }\href@noop {} {\bibfield
  {journal} {\bibinfo  {journal} {Phys. Rev. B}\ }\textbf {\bibinfo {volume}
  {102}},\ \bibinfo {pages} {155146} (\bibinfo {year} {2020})}\BibitemShut
  {NoStop}%
\bibitem [{\citenamefont {Bistritzer}\ and\ \citenamefont
  {MacDonald}(2011)}]{bistritzer2011moire}%
  \BibitemOpen
  \bibfield  {author} {\bibinfo {author} {\bibfnamefont {R.}~\bibnamefont
  {Bistritzer}}\ and\ \bibinfo {author} {\bibfnamefont {A.~H.}\ \bibnamefont
  {MacDonald}},\ }\bibfield  {title} {\bibinfo {title} {Moir{\'e} bands in
  twisted double-layer graphene},\ }\href
  {https://doi.org/10.1073/pnas.1108174108} {\bibfield  {journal} {\bibinfo
  {journal} {Proceedings of the National Academy of Sciences}\ }\textbf
  {\bibinfo {volume} {108}},\ \bibinfo {pages} {12233} (\bibinfo {year}
  {2011})},\ \Eprint
  {https://arxiv.org/abs/https://www.pnas.org/content/108/30/12233.full.pdf}
  {https://www.pnas.org/content/108/30/12233.full.pdf} \BibitemShut {NoStop}%
\bibitem [{\citenamefont {Lopes~dos Santos}\ \emph {et~al.}(2012)\citenamefont
  {Lopes~dos Santos}, \citenamefont {Peres},\ and\ \citenamefont
  {Castro~Neto}}]{lopes2012continuum}%
  \BibitemOpen
  \bibfield  {author} {\bibinfo {author} {\bibfnamefont {J.~M.~B.}\
  \bibnamefont {Lopes~dos Santos}}, \bibinfo {author} {\bibfnamefont
  {N.~M.~R.}\ \bibnamefont {Peres}},\ and\ \bibinfo {author} {\bibfnamefont
  {A.~H.}\ \bibnamefont {Castro~Neto}},\ }\bibfield  {title} {\bibinfo {title}
  {Continuum model of the twisted graphene bilayer},\ }\href
  {https://doi.org/10.1103/PhysRevB.86.155449} {\bibfield  {journal} {\bibinfo
  {journal} {Phys. Rev. B}\ }\textbf {\bibinfo {volume} {86}},\ \bibinfo
  {pages} {155449} (\bibinfo {year} {2012})}\BibitemShut {NoStop}%
\bibitem [{\citenamefont {Lopes~dos Santos}\ \emph {et~al.}(2007)\citenamefont
  {Lopes~dos Santos}, \citenamefont {Peres},\ and\ \citenamefont
  {Castro~Neto}}]{lopes2007graphene}%
  \BibitemOpen
  \bibfield  {author} {\bibinfo {author} {\bibfnamefont {J.~M.~B.}\
  \bibnamefont {Lopes~dos Santos}}, \bibinfo {author} {\bibfnamefont
  {N.~M.~R.}\ \bibnamefont {Peres}},\ and\ \bibinfo {author} {\bibfnamefont
  {A.~H.}\ \bibnamefont {Castro~Neto}},\ }\bibfield  {title} {\bibinfo {title}
  {Graphene bilayer with a twist: Electronic structure},\ }\href
  {https://doi.org/10.1103/PhysRevLett.99.256802} {\bibfield  {journal}
  {\bibinfo  {journal} {Phys. Rev. Lett.}\ }\textbf {\bibinfo {volume} {99}},\
  \bibinfo {pages} {256802} (\bibinfo {year} {2007})}\BibitemShut {NoStop}%
\bibitem [{\citenamefont {Koshino}\ \emph {et~al.}(2018)\citenamefont
  {Koshino}, \citenamefont {Yuan}, \citenamefont {Koretsune}, \citenamefont
  {Ochi}, \citenamefont {Kuroki},\ and\ \citenamefont
  {Fu}}]{koshino2018maximally}%
  \BibitemOpen
  \bibfield  {author} {\bibinfo {author} {\bibfnamefont {M.}~\bibnamefont
  {Koshino}}, \bibinfo {author} {\bibfnamefont {N.~F.~Q.}\ \bibnamefont
  {Yuan}}, \bibinfo {author} {\bibfnamefont {T.}~\bibnamefont {Koretsune}},
  \bibinfo {author} {\bibfnamefont {M.}~\bibnamefont {Ochi}}, \bibinfo {author}
  {\bibfnamefont {K.}~\bibnamefont {Kuroki}},\ and\ \bibinfo {author}
  {\bibfnamefont {L.}~\bibnamefont {Fu}},\ }\bibfield  {title} {\bibinfo
  {title} {Maximally localized wannier orbitals and the extended hubbard model
  for twisted bilayer graphene},\ }\href
  {https://doi.org/10.1103/PhysRevX.8.031087} {\bibfield  {journal} {\bibinfo
  {journal} {Phys. Rev. X}\ }\textbf {\bibinfo {volume} {8}},\ \bibinfo {pages}
  {031087} (\bibinfo {year} {2018})}\BibitemShut {NoStop}%
\bibitem [{\citenamefont {Cea}\ \emph {et~al.}(2019)\citenamefont {Cea},
  \citenamefont {Walet},\ and\ \citenamefont {Guinea}}]{cea2019twists}%
  \BibitemOpen
  \bibfield  {author} {\bibinfo {author} {\bibfnamefont {T.}~\bibnamefont
  {Cea}}, \bibinfo {author} {\bibfnamefont {N.~R.}\ \bibnamefont {Walet}},\
  and\ \bibinfo {author} {\bibfnamefont {F.}~\bibnamefont {Guinea}},\
  }\bibfield  {title} {\bibinfo {title} {Twists and the electronic structure of
  graphitic materials},\ }\href {https://doi.org/10.1021/acs.nanolett.9b03335}
  {\bibfield  {journal} {\bibinfo  {journal} {Nano Letters}\ }\textbf {\bibinfo
  {volume} {19}},\ \bibinfo {pages} {8683} (\bibinfo {year} {2019})},\ \bibinfo
  {note} {pMID: 31743649},\ \Eprint
  {https://arxiv.org/abs/https://doi.org/10.1021/acs.nanolett.9b03335}
  {https://doi.org/10.1021/acs.nanolett.9b03335} \BibitemShut {NoStop}%
\bibitem [{\citenamefont {Balents}(2019)}]{balents2019general}%
  \BibitemOpen
  \bibfield  {author} {\bibinfo {author} {\bibfnamefont {L.}~\bibnamefont
  {Balents}},\ }\bibfield  {title} {\bibinfo {title} {{General continuum model
  for twisted bilayer graphene and arbitrary smooth deformations}},\ }\href
  {https://doi.org/10.21468/SciPostPhys.7.4.048} {\bibfield  {journal}
  {\bibinfo  {journal} {SciPost Phys.}\ }\textbf {\bibinfo {volume} {7}},\
  \bibinfo {pages} {48} (\bibinfo {year} {2019})}\BibitemShut {NoStop}%
\bibitem [{\citenamefont {Koshino}\ and\ \citenamefont
  {Nam}(2020)}]{koshino2020effective}%
  \BibitemOpen
  \bibfield  {author} {\bibinfo {author} {\bibfnamefont {M.}~\bibnamefont
  {Koshino}}\ and\ \bibinfo {author} {\bibfnamefont {N.~N.~T.}\ \bibnamefont
  {Nam}},\ }\bibfield  {title} {\bibinfo {title} {Effective continuum model for
  relaxed twisted bilayer graphene and moir\'e electron-phonon interaction},\
  }\href {https://doi.org/10.1103/PhysRevB.101.195425} {\bibfield  {journal}
  {\bibinfo  {journal} {Phys. Rev. B}\ }\textbf {\bibinfo {volume} {101}},\
  \bibinfo {pages} {195425} (\bibinfo {year} {2020})}\BibitemShut {NoStop}%
\bibitem [{\citenamefont {Gr\"uneis}\ \emph {et~al.}(2008)\citenamefont
  {Gr\"uneis}, \citenamefont {Attaccalite}, \citenamefont {Wirtz},
  \citenamefont {Shiozawa}, \citenamefont {Saito}, \citenamefont {Pichler},\
  and\ \citenamefont {Rubio}}]{gruneis2008tightbinding}%
  \BibitemOpen
  \bibfield  {author} {\bibinfo {author} {\bibfnamefont {A.}~\bibnamefont
  {Gr\"uneis}}, \bibinfo {author} {\bibfnamefont {C.}~\bibnamefont
  {Attaccalite}}, \bibinfo {author} {\bibfnamefont {L.}~\bibnamefont {Wirtz}},
  \bibinfo {author} {\bibfnamefont {H.}~\bibnamefont {Shiozawa}}, \bibinfo
  {author} {\bibfnamefont {R.}~\bibnamefont {Saito}}, \bibinfo {author}
  {\bibfnamefont {T.}~\bibnamefont {Pichler}},\ and\ \bibinfo {author}
  {\bibfnamefont {A.}~\bibnamefont {Rubio}},\ }\bibfield  {title} {\bibinfo
  {title} {Tight-binding description of the quasiparticle dispersion of
  graphite and few-layer graphene},\ }\href
  {https://doi.org/10.1103/PhysRevB.78.205425} {\bibfield  {journal} {\bibinfo
  {journal} {Phys. Rev. B}\ }\textbf {\bibinfo {volume} {78}},\ \bibinfo
  {pages} {205425} (\bibinfo {year} {2008})}\BibitemShut {NoStop}%
\bibitem [{\citenamefont {Koshino}(2019)}]{koshino2019band}%
  \BibitemOpen
  \bibfield  {author} {\bibinfo {author} {\bibfnamefont {M.}~\bibnamefont
  {Koshino}},\ }\bibfield  {title} {\bibinfo {title} {Band structure and
  topological properties of twisted double bilayer graphene},\ }\href
  {https://doi.org/10.1103/PhysRevB.99.235406} {\bibfield  {journal} {\bibinfo
  {journal} {Phys. Rev. B}\ }\textbf {\bibinfo {volume} {99}},\ \bibinfo
  {pages} {235406} (\bibinfo {year} {2019})}\BibitemShut {NoStop}%
\bibitem [{\citenamefont {Samajdar}\ and\ \citenamefont
  {Scheurer}(2020)}]{samajdar2020microscopic}%
  \BibitemOpen
  \bibfield  {author} {\bibinfo {author} {\bibfnamefont {R.}~\bibnamefont
  {Samajdar}}\ and\ \bibinfo {author} {\bibfnamefont {M.~S.}\ \bibnamefont
  {Scheurer}},\ }\bibfield  {title} {\bibinfo {title} {Microscopic pairing
  mechanism, order parameter, and disorder sensitivity in moir\'e
  superlattices: Applications to twisted double-bilayer graphene},\ }\href
  {https://doi.org/10.1103/PhysRevB.102.064501} {\bibfield  {journal} {\bibinfo
   {journal} {Phys. Rev. B}\ }\textbf {\bibinfo {volume} {102}},\ \bibinfo
  {pages} {064501} (\bibinfo {year} {2020})}\BibitemShut {NoStop}%
\bibitem [{\citenamefont {Wang}\ \emph
  {et~al.}(2013{\natexlab{b}})\citenamefont {Wang}, \citenamefont {Meric},
  \citenamefont {Huang}, \citenamefont {Gao}, \citenamefont {Gao},
  \citenamefont {Tran}, \citenamefont {Taniguchi}, \citenamefont {Watanabe},
  \citenamefont {Campos}, \citenamefont {Muller}, \citenamefont {Guo},
  \citenamefont {Kim}, \citenamefont {Hone}, \citenamefont {Shepard},\ and\
  \citenamefont {Dean}}]{Wang2013b}%
  \BibitemOpen
  \bibfield  {author} {\bibinfo {author} {\bibfnamefont {L.}~\bibnamefont
  {Wang}}, \bibinfo {author} {\bibfnamefont {I.}~\bibnamefont {Meric}},
  \bibinfo {author} {\bibfnamefont {P.~Y.}\ \bibnamefont {Huang}}, \bibinfo
  {author} {\bibfnamefont {Q.}~\bibnamefont {Gao}}, \bibinfo {author}
  {\bibfnamefont {Y.}~\bibnamefont {Gao}}, \bibinfo {author} {\bibfnamefont
  {H.}~\bibnamefont {Tran}}, \bibinfo {author} {\bibfnamefont {T.}~\bibnamefont
  {Taniguchi}}, \bibinfo {author} {\bibfnamefont {K.}~\bibnamefont {Watanabe}},
  \bibinfo {author} {\bibfnamefont {L.~M.}\ \bibnamefont {Campos}}, \bibinfo
  {author} {\bibfnamefont {D.~A.}\ \bibnamefont {Muller}}, \bibinfo {author}
  {\bibfnamefont {J.}~\bibnamefont {Guo}}, \bibinfo {author} {\bibfnamefont
  {P.}~\bibnamefont {Kim}}, \bibinfo {author} {\bibfnamefont {J.}~\bibnamefont
  {Hone}}, \bibinfo {author} {\bibnamefont {Shepard}},\ and\ \bibinfo {author}
  {\bibfnamefont {C.~R.}\ \bibnamefont {Dean}},\ }\bibfield  {title} {\bibinfo
  {title} {{One-Dimensional Electrical Contact to a Two-Dimensional
  Material}},\ }\href {https://doi.org/27110.21371/journal.pone.0027147.SM2.}
  {\bibfield  {journal} {\bibinfo  {journal} {Science}\ }\textbf {\bibinfo
  {volume} {342}},\ \bibinfo {pages} {614} (\bibinfo {year}
  {2013}{\natexlab{b}})}\BibitemShut {NoStop}%
\bibitem [{\citenamefont {Moon}\ and\ \citenamefont
  {Koshino}(2014)}]{moon2014electronic}%
  \BibitemOpen
  \bibfield  {author} {\bibinfo {author} {\bibfnamefont {P.}~\bibnamefont
  {Moon}}\ and\ \bibinfo {author} {\bibfnamefont {M.}~\bibnamefont {Koshino}},\
  }\bibfield  {title} {\bibinfo {title} {Electronic properties of
  graphene/hexagonal-boron-nitride moir\'e superlattice},\ }\href
  {https://doi.org/10.1103/PhysRevB.90.155406} {\bibfield  {journal} {\bibinfo
  {journal} {Phys. Rev. B}\ }\textbf {\bibinfo {volume} {90}},\ \bibinfo
  {pages} {155406} (\bibinfo {year} {2014})}\BibitemShut {NoStop}%
\bibitem [{\citenamefont {Chen}\ and\ \citenamefont {Chrzan}(2011)}]{Chen2011}%
  \BibitemOpen
  \bibfield  {author} {\bibinfo {author} {\bibfnamefont {S.}~\bibnamefont
  {Chen}}\ and\ \bibinfo {author} {\bibfnamefont {D.~C.}\ \bibnamefont
  {Chrzan}},\ }\bibfield  {title} {\bibinfo {title} {{Continuum theory of
  dislocations and buckling in graphene}},\ }\bibfield  {journal} {\bibinfo
  {journal} {Physical Review B - Condensed Matter and Materials Physics}\
  }\textbf {\bibinfo {volume} {84}},\ \href
  {https://doi.org/10.1103/PhysRevB.84.214103} {10.1103/PhysRevB.84.214103}
  (\bibinfo {year} {2011})\BibitemShut {NoStop}%
\bibitem [{\citenamefont {Lai}\ \emph {et~al.}(2016)\citenamefont {Lai},
  \citenamefont {Zhang}, \citenamefont {Zhou}, \citenamefont {Zeng},\ and\
  \citenamefont {Tang}}]{Lai2016}%
  \BibitemOpen
  \bibfield  {author} {\bibinfo {author} {\bibfnamefont {K.}~\bibnamefont
  {Lai}}, \bibinfo {author} {\bibfnamefont {W.-B.}\ \bibnamefont {Zhang}},
  \bibinfo {author} {\bibfnamefont {F.}~\bibnamefont {Zhou}}, \bibinfo {author}
  {\bibfnamefont {F.}~\bibnamefont {Zeng}},\ and\ \bibinfo {author}
  {\bibfnamefont {B.-Y.}\ \bibnamefont {Tang}},\ }\bibfield  {title} {\bibinfo
  {title} {{Bending rigidity of transition metal dichalcogenide monolayers from
  first-principles}},\ }\href@noop {} {\bibfield  {journal} {\bibinfo
  {journal} {Journal of Physics D: Applied Physics}\ }\textbf {\bibinfo
  {volume} {49}},\ \bibinfo {pages} {185301} (\bibinfo {year}
  {2016})}\BibitemShut {NoStop}%
\bibitem [{\citenamefont {Zhou}\ \emph {et~al.}(2015)\citenamefont {Zhou},
  \citenamefont {Han}, \citenamefont {Dai}, \citenamefont {Sun},\ and\
  \citenamefont {Srolovitz}}]{Zhou2015e}%
  \BibitemOpen
  \bibfield  {author} {\bibinfo {author} {\bibfnamefont {S.}~\bibnamefont
  {Zhou}}, \bibinfo {author} {\bibfnamefont {J.}~\bibnamefont {Han}}, \bibinfo
  {author} {\bibfnamefont {S.}~\bibnamefont {Dai}}, \bibinfo {author}
  {\bibfnamefont {J.}~\bibnamefont {Sun}},\ and\ \bibinfo {author}
  {\bibfnamefont {D.~J.}\ \bibnamefont {Srolovitz}},\ }\bibfield  {title}
  {\bibinfo {title} {{Van der Waals bilayer energetics: Generalized
  stacking-fault energy of graphene, boron nitride, and graphene/boron nitride
  bilayers}},\ }\href {https://doi.org/10.1103/PhysRevB.92.155438} {\bibfield
  {journal} {\bibinfo  {journal} {Physical Review B}\ }\textbf {\bibinfo
  {volume} {92}},\ \bibinfo {pages} {155438} (\bibinfo {year}
  {2015})}\BibitemShut {NoStop}%
\bibitem [{\citenamefont {Kresse}\ and\ \citenamefont
  {Furthm\"uller}(1996)}]{Kresse1996}%
  \BibitemOpen
  \bibfield  {author} {\bibinfo {author} {\bibfnamefont {G.}~\bibnamefont
  {Kresse}}\ and\ \bibinfo {author} {\bibfnamefont {J.}~\bibnamefont
  {Furthm\"uller}},\ }\bibfield  {title} {\bibinfo {title} {Efficient iterative
  schemes for ab initio total-energy calculations using a plane-wave basis
  set},\ }\href {https://doi.org/10.1103/PhysRevB.54.11169} {\bibfield
  {journal} {\bibinfo  {journal} {Phys. Rev. B}\ }\textbf {\bibinfo {volume}
  {54}},\ \bibinfo {pages} {11169} (\bibinfo {year} {1996})}\BibitemShut
  {NoStop}%
\bibitem [{\citenamefont {Peng}\ \emph {et~al.}(2016)\citenamefont {Peng},
  \citenamefont {Yang}, \citenamefont {Perdew},\ and\ \citenamefont
  {Sun}}]{Peng2016}%
  \BibitemOpen
  \bibfield  {author} {\bibinfo {author} {\bibfnamefont {H.}~\bibnamefont
  {Peng}}, \bibinfo {author} {\bibfnamefont {Z.-H.}\ \bibnamefont {Yang}},
  \bibinfo {author} {\bibfnamefont {J.~P.}\ \bibnamefont {Perdew}},\ and\
  \bibinfo {author} {\bibfnamefont {J.}~\bibnamefont {Sun}},\ }\bibfield
  {title} {\bibinfo {title} {Versatile van der waals density functional based
  on a meta-generalized gradient approximation},\ }\href
  {https://doi.org/10.1103/PhysRevX.6.041005} {\bibfield  {journal} {\bibinfo
  {journal} {Phys. Rev. X}\ }\textbf {\bibinfo {volume} {6}},\ \bibinfo {pages}
  {041005} (\bibinfo {year} {2016})}\BibitemShut {NoStop}%
\bibitem [{\citenamefont {Kresse}\ and\ \citenamefont
  {Joubert}(1999)}]{Kresse1999}%
  \BibitemOpen
  \bibfield  {author} {\bibinfo {author} {\bibfnamefont {G.}~\bibnamefont
  {Kresse}}\ and\ \bibinfo {author} {\bibfnamefont {D.}~\bibnamefont
  {Joubert}},\ }\bibfield  {title} {\bibinfo {title} {From ultrasoft
  pseudopotentials to the projector augmented-wave method},\ }\href
  {https://doi.org/10.1103/PhysRevB.59.1758} {\bibfield  {journal} {\bibinfo
  {journal} {Phys. Rev. B}\ }\textbf {\bibinfo {volume} {59}},\ \bibinfo
  {pages} {1758} (\bibinfo {year} {1999})}\BibitemShut {NoStop}%
\end{thebibliography}%

\end{document}